\pdfoutput=1
\documentclass[12pt]{scrbook}

\usepackage[stable]{footmisc}
\usepackage[ngerman,pdfview=FitH,pdfstartview=FitV]{hyperref}

\usepackage[utf8]{inputenc}

\usepackage{upgreek}
\usepackage{graphicx}
\usepackage{subcaption}
\usepackage{amsmath}
\usepackage{cleveref}
\usepackage{amsfonts}
\usepackage{algorithm}
\usepackage{mathtools}
\usepackage{algpseudocode}  
\usepackage{graphicx}
\graphicspath{ {Figures/} }
\usepackage{subcaption}

\usepackage[T1]{fontenc} 
\usepackage{bbm} 
\usepackage{cite}

\makeatletter
\algnewcommand\Input{\item[\textbf{Input:}]}%
\algnewcommand\Output{\item[\textbf{Output:}]}%
\algnewcommand\Init{\item[\textbf{Init:}]}%
\makeatother

\usepackage{amsthm}
\theoremstyle{plain}

\newtheorem{myResult}{Result}
\newtheorem{myLemma}{Lemma}

\theoremstyle{definition}
\newtheorem*{myRemark}{Remark}

\newcommand{\twopartdef}[4]
{
	\left\{
	\begin{array}{lcl}
		#1  &   
		  \hspace{ 100pt } \mbox{if } #2 \eqkomma \\
		#3   & 
		  \hspace{ 100pt} \mbox{if } #4 \eqkomma 
	\end{array}
	\right.
}

\newcommand{\etal}{\textit{et al.}}

\renewcommand{\vec}[1]{\mathrm{#1}}

\newcommand{\indicator}[1]{\mathbbm{1}(#1)}

\newcommand{\E}{\mathrm{E}}
\newcommand{\Eof}[1]{\E[#1]}

\newcommand{\shift}{\mathrm{S}}

\newcommand{\prob}{\mathrm{P}}
\newcommand{\probOf}[1]{\prob(#1)}

\renewcommand{\vec}[1]{{\boldsymbol#1}}

\newcommand{\eqpunkt}{.} 
\newcommand{\eqkomma}{,}
\newcommand{\defeq}{\coloneqq}

\newcommand{\ie}{\textit{i.e.}}
\newcommand{\eg}{\textit{e.g.}}

\newcommand{\viz}{\textit{viz.}}
\newcommand{\vide}{\textit{vide }}

\newcommand{\quotes}[1]{``#1''}

\usepackage{todonotes}
\usepackage{ifthen}
\newboolean{draftversion}
\setboolean{draftversion}{false} 

\newcommand{\tdJulius}[2][]{\ifthenelse{\boolean{draftversion}}{\todo[inline, color=orange!20, caption={2do}, #1]{\begin{minipage}{\textwidth-4pt}\emph{ToDo for Julius:}\\#2\end{minipage}}}{}}
\newcommand{\tdWasiur}[2][]{\ifthenelse{\boolean{draftversion}}{\todo[inline, color=blue!20, caption={2do}, #1]{\begin{minipage}{\textwidth-4pt}\emph{ToDO for Wasiur:}\\#2\end{minipage}}}{}}
\newcommand{\tdGeneral}[2][]{\ifthenelse{\boolean{draftversion}}{\todo[inline, color=green!20, caption={2do}, #1]{\begin{minipage}{\textwidth-4pt}\emph{ToDO:}\\#2\end{minipage}}}{}}
\ifthenelse{\boolean{draftversion}}{\usepackage[firstpage]{draftwatermark}}{}

\usepackage{url}

\usepackage{tabularx} 
\usepackage{booktabs}

\newcommand{\sys}[1]{\textsc{#1}}

\newlength{\longtablewidth}

\title{A Comprehensive Analysis of Swarming-based Live Streaming to Leverage Client Heterogeneity\footnote{Supplementary material to \cite{KhudaBukhsh2016P2P}}}
\author{Wasiur R. KhudaBukhsh\hspace{1.5mm}\href{https://orcid.org/0000-0003-1803-0470}{\includegraphics[width=3mm]{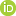}}\\ Julius R\"uckert\\ Julian Wulfheide\\ David Hausheer\\ Heinz Koeppl}

\date{}

\begin{document}
\maketitle

\tableofcontents
\listoffigures
\listoftables


\newpage

 \chapter{Introduction}

Media streaming 
continues dominating the traffic share on
nowadays Internet~\cite{cisco14}.
With the increasing number of content providers that offer their services in an over-the-top (OTT) manner to a world-wide audience, the need for efficient and scalable content delivery mechanisms that are not dependent on special network services continues to be strong.
IP multicast~\cite{deering89} once set off with the objective to provide such a network service and, thus, was hoped to help in the realization of highly efficient one-to-many and many-to-many delivery scenarios.
Yet, due to the inherent limitations of the approach~\cite{diot00}, IP multicast was not adopted in more than network islands and, in particular, is not usable for OTT content deliveries on a large scale.
Due to high demand, multicast functionality was realized at the application
layer  instead.
Content delivery networks (CDNs), such as the biggest one by Akamai~\cite{sitaraman14}, are the most prominent example, where ten thousands of server machines are deployed at strategic points of the Internet and overlay routing networks are established among these nodes for data delivery, including one-to-many multicasting traffic.
To reduce the load on the servers and make the delivery more profitable, an number of peer-to-peer (P2P) and peer-assisted approaches were proposed that shift the data duplication to the clients of the service and, thus, to the very edge of the network~\cite{zhang12,liu08}.
They are used as extension to CDNs~\cite{zhao2013} or even as standalone approaches to realize highly efficient one-to-many multicasting, e.g., of live video streams, to a large number of clients across the public Internet.
Efficiency in this context is understood from the OTT content provider and CDN perspective.
It is important to note that the approaches shift the actual costs of the delivery in terms of network traffic to clients and, thus, mainly the Internet Service Providers that offer broadband access to the clients or function as transport networks.

A large body of research exists on P2P-based live video content delivery.
Over time, different classes of P2P streaming approaches evolved, including single-tree and multi-tree push-based approaches as well as mesh/push-based swarming approaches, and, most recently hybrid approaches~\cite{zhang12, liu08}.
Hybrid streaming systems have shown to exhibit desirable properties and allow to combine advantages of both classes, while reducing the costs in terms of coordination overhead to a minimum~\cite{wichtlhuber14a}.
Here, a common approach is to establish a substrate mesh/pull-based overlay network and augment it with tree structures that manifest data paths that have proven to be stable over the past~\cite{wang10,rueckert15b}.
This way, swarming mechanisms allow for a highly robust delivery whenever tree structures temporarily or permanently fail or are not yet established.
As a result, the ability to achieve a high and stable streaming performance (typically measured by the chunk delivery ratio) using mesh structures only is essential also for hybrid approaches.
Thus, after focusing more on the tree substrate in earlier works~\cite{rueckert15b, wichtlhuber14a}, in this work, we focus on the second important constitute of hybrid streaming systems: the mesh/pull-based swarming substrate.

A key design issue in swarming is the choice for a data scheduling strategy used by the individual peers to select video chunks to be requested from their neighbors.
This strategy, on one hand, has to assure that data is available before the individual peers' playback deadlines to avoid undesired stalling events or video quality degradations.
On the other hand, the strategy has to assure data availability across the peers to allow swarming to take place in the first place~\cite{rejaie14} and, thus, avoid content bottlenecks.

A number of works study data scheduling for swarming and propose sophisticated solutions for it, including adaptive scheduling strategies.
The impact of realistic client populations with heterogeneous resources, however, is not yet fully understood and provides a huge potential to simplify complex scheduling approaches or form completely new ones that avoid this complexity in the first place.
For the latter, we argue that understanding  basic scheduling strategies and
considering their combination has a great potential. From mathematical  perspective too, our understanding of the system in a heterogeneous set-up is far from complete.

In this technical report, therefore, we contribute to closing this gap by mathematically analysing the most basic scheduling mechanisms \emph{latest deadline first} (LDF) and \emph{earliest deadline first} (EDF) in a continuous time Markov chain framework and combining them in a simple, yet powerful, mixed strategy to leverage inherent differences in upload resources of realistic client populations.

Our mathematical framework is a general one and can essentially be interpreted as a contact process (\cite{liggett2013stochastic}) on a random graph, where neighbours contact each other. The purpose could be disparate. It could be an unintentional contagion in the context of infection spread in an epidemic, transmission of a computer virus/malware in a cyber attack.  Or it could be an intentional one. We envisage peer-to-peer live streaming as one such example. Peers contact each other to download chunks. The shifting feature of our model make it particularly interesting on its own as it paves way for plentiful physical interpretations in dissimilar contexts. Another important facet of our model is that it meticulously captures the influence of degree.

The contribution of this technical report consists of two complementary parts:
(1) we present a mathematical framework to study swarming on random graphs as a stochastic model. We complement our theoretical study with an implementation of  a full-stack P2P streaming system based on the \sys{Simonstrator}~\cite{richerzhagen15} evaluation platform. In both theoretical and practical study, we largely focus on the two basic scheduling strategies LDF and EDF in heterogeneous scenarios. 
(2) A new mixed strategy is proposed that leverages peer heterogeneity in the choice of the applied scheduling strategy.


The proposed strategy is shown to outperform the two basic strategies using different abstractions: a mean-field theoretic analysis of buffer probabilities, simulations of the stochastic model on random graphs, and discrete event-based simulations of the full-stack implementation of a P2P streaming system.
This way, we both theoretically and practically show the huge potential of using a simple, yet powerful combination of primitive scheduling mechanisms to improve the overall streaming performance for mesh-/pull-based streaming.
It is shown that a significant gain in delivery ration can be achieved, which is expected to translates to even higher gains for more sophisticated and hybrid approaches.
These results are encouraging to consider using primitive scheduling mechanism combinations in mesh-based and hybrid streaming approaches and enable switching between them.
For the later, the notion of using \emph{transitions} to realize dynamic switching between the individual scheduling mechanisms is considered a promising approach as proposed in~\cite{froemmgen15}.

The remainder of this technical report is structured as follows: 
Section~\ref{sec:model} presents the proposed mathematical framework as well a the full-stack streaming system model.
Section~\ref{sec:mean-field} presents the mean-field analysis of the problem.
Subsequently, Section~\ref{sec:simulation-results} presents the results of both simulations using the stochastic model as well as the full-stack streaming system and Section~\ref{sec:game-theoretic-consideration} discusses the the proposed mixed strategy from a game theoretical view point.
Finally, Section~\ref{sec:related-work} discusses related work in relevant areas and Section~\ref{ref:conclusion} concludes the technical report.

\newpage

\chapter{Model}
\label{sec:model}
We begin this section with some general comments about the way we record our analytic expressions. At the risk of stating the obvious, we use $\mathbb{C},\mathbb{R}, \mathbb{N}$ to denote the sets of all complex, real and natural numbers, respectively. For $s \in \mathbb{C}, \Re(s), \Im(s)$ denote respectively the real and imaginary parts of $s$. 
Unless otherwise mentioned, we shall use $\prob, \E$ to denote  probability measure and expectation, respectively.  For a set $A$, we shall use $|A|$ to denote its cardinality. 

\section{The approach}
\label{sec:model:approach}
First, we briefly explain our modelling strategy. The main idea is to model a swarming-based peer-to-peer live streaming system as a type of  contact process on a random graph (see \cite{liggett2013stochastic}), where the vertices represent the peers. We endow each peer with a buffer of length~$n$ (a vector of $0$'s and $1$'s with $1$'s representing the availability of  chunks). The different possible buffer configurations constitute the \emph{local} states of a peer. The local buffer configuration of a peer changes over time as the peer downloads chunks (from the server or from one of its neighbours following a chunk selection strategy, such as EDF or LDF), or deletes chunks that are already played back. Therefore, the interactions among the peers  define a contact process on the random graph. The matrix with as many rows as the number of peers in the system and  whose $i$-th row  is the buffer configuration of the $i$-th peer,  can then modelled as a CTMC with certain transition intensities or rates (specified in Section~\ref{sec:model:p2pcommsys}). The matrix essentially captures the \emph{global} state of the entire system. The transition intensities between two  different  states  of the matrix naturally depend on the graph structure (specifying which peers can download from which other peers), and the chunk selection policy.  The probability of finding the system, which is now a CTMC,  in a particular state at a particular time is usually found by solving the Kolmogorov forward equation, also known as the (chemical) master equation in the physics/physical chemistry literature. Since this probability is dependent on the chunk selection strategy,  we can now, at least in principle, choose a chunk selection strategy that maximises the probability of the system being in a state that ensures good playback performance, \eg, a state in which the current chunk required for playback is available at every buffer (to ensure playback continuity). In particular, the buffer probabilities (of chunk availability) can be expressed as functions of the chunk selection strategy, and therefore, can  be utilised to improve the chunk selection strategy or devise a new  one. This is precisely our plan.

The major roadblock to executing the plan described above is the fact that the state space of the matrix (comprising all possible buffer configurations of all the peers)  is exponentially large when the number of peers in the system grows arbitrarily. As a result, we can not solve the master equation analytically. Therefore, in order to reduce the state space, we lump the original process and only keep track of some aggregate counts such as the number of peers of degree~$k$ with a particular buffer configuration, instead of the entire matrix. We do lose some information as a consequence of this aggregation. In fact, the aggregated process need not even be Markovian \cite{khudabukhsh2018approximate}. However,  the aggregation or lumping of states  reduces the state space and thereby  greatly simplifies the mathematical analysis. Finally, in order to represent the buffer probabilities as a function of the chunk selection strategy, we carry out a mean-field theoretic analysis of the lumped process  in Section~\ref{sec:mean-field}, assuming the number of peers in the system is large. Based on the resulting insights, we then motivate our mixed strategy \sys{SchedMix}.

We make the ideas discussed above more precise in the following sections.

\section{The network}
\label{sec:model:network}
We describe the underlying network as a random graph. While otherwise allowing for generality  in the choice of candidate random graphs, we assume finiteness of mean of the associated degree distribution. For our simulations, we consider 
Barab\'{a}si-Albert preferential attachment~\cite{BarabasiAlbertScaling} and Watts-Strogatz \quotes{small world}~\cite{WattsStrogatzSmallWorlds} networks. The description, however, remains valid for any  random graph with a finite-mean degree distribution.

Suppose $\mathcal{G}_M$  be the class of all random graphs with $M$ nodes. Although we believe it can cover directed and weighted graphs, to keep our premise uncomplicated, we confine ourselves to simple graphs.  Also, we condition on the event that the resulting graph is connected. Without introducing a new notation, we continue with $\mathcal{G}_M $ to denote the reduced class of random graphs possessing the properties mentioned above. Let $\pi \in \mathcal{P}(\Lambda)$ 
  be the associated degree distribution, where $\mathcal{P}(\Lambda)$ is the class of all probability distributions on measurable space $\Lambda \defeq (\mathbb{N},\mathcal{N})$ and $\mathcal{N}$ is the class of all subsets of $\mathbb{N} $. 
We also define the size-biased degree distribution, $q \in \mathcal{P}(\Lambda) $ as follows:
\begin{equation}
q(k) \defeq \frac{k \pi(k) }{ \sum_k k \pi(k)} , k \in \mathbb{N} \eqkomma
\label{eq:size-biased-degree}
\end{equation}
where $\pi(k), q(k)$ should be interpreted as $\pi(\{k\}), q(\{k\})$ respectively. The denominator is assumed to be finite so as to make $q$ a legitimate probability distribution. The quantity $q(k)$ is the probability that a given edge points to a vertex of degree $k$.

\begin{figure}[htbp]
\centering
	\begin{subfigure}[b]{.3\columnwidth}
	\centering
 		\includegraphics[keepaspectratio,scale=0.25]{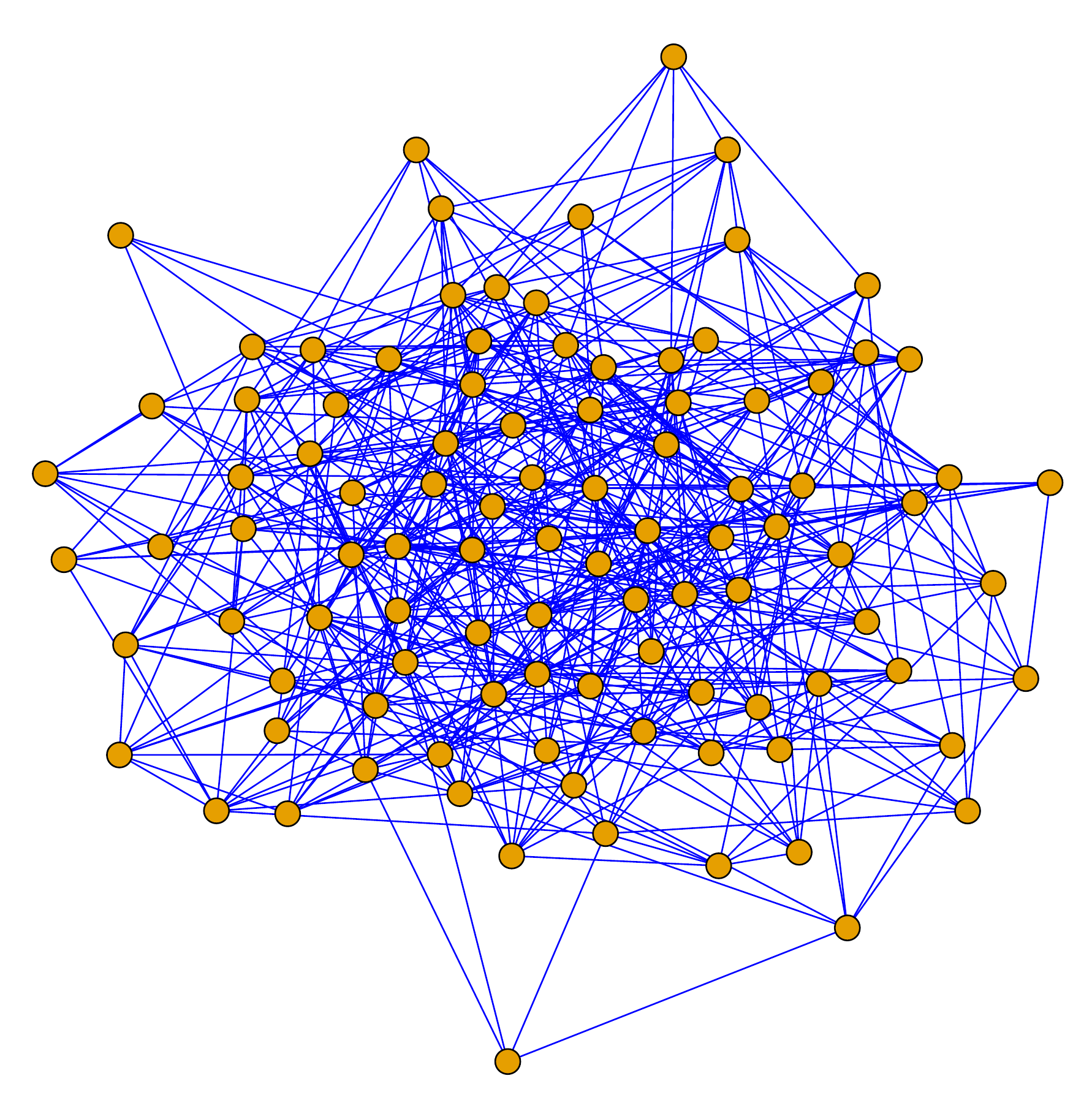}
		\caption{Erd\"{o}s-R\'{e}nyi random graphs}
		\label{fig:ERGraph}
	\end{subfigure}
	\begin{subfigure}[b]{.3\columnwidth}
	\centering
		\includegraphics[keepaspectratio,scale=0.25]{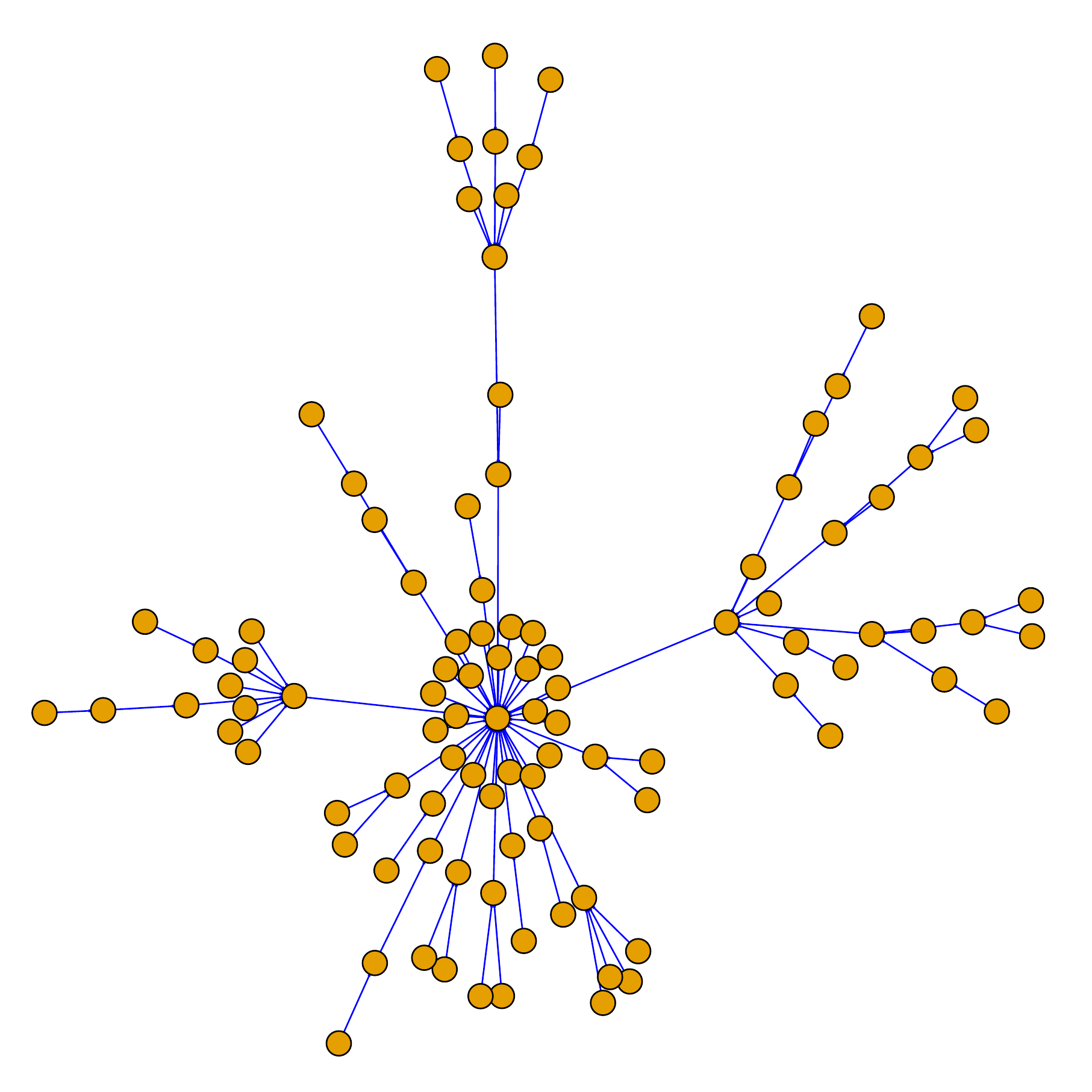}
		\caption{Barab\'{a}si-Albert preferential attachment graph}
		\label{fig:BAGraph}
	\end{subfigure}
	\begin{subfigure}[b]{.3\columnwidth}
	\centering
		\includegraphics[keepaspectratio,scale=0.25]{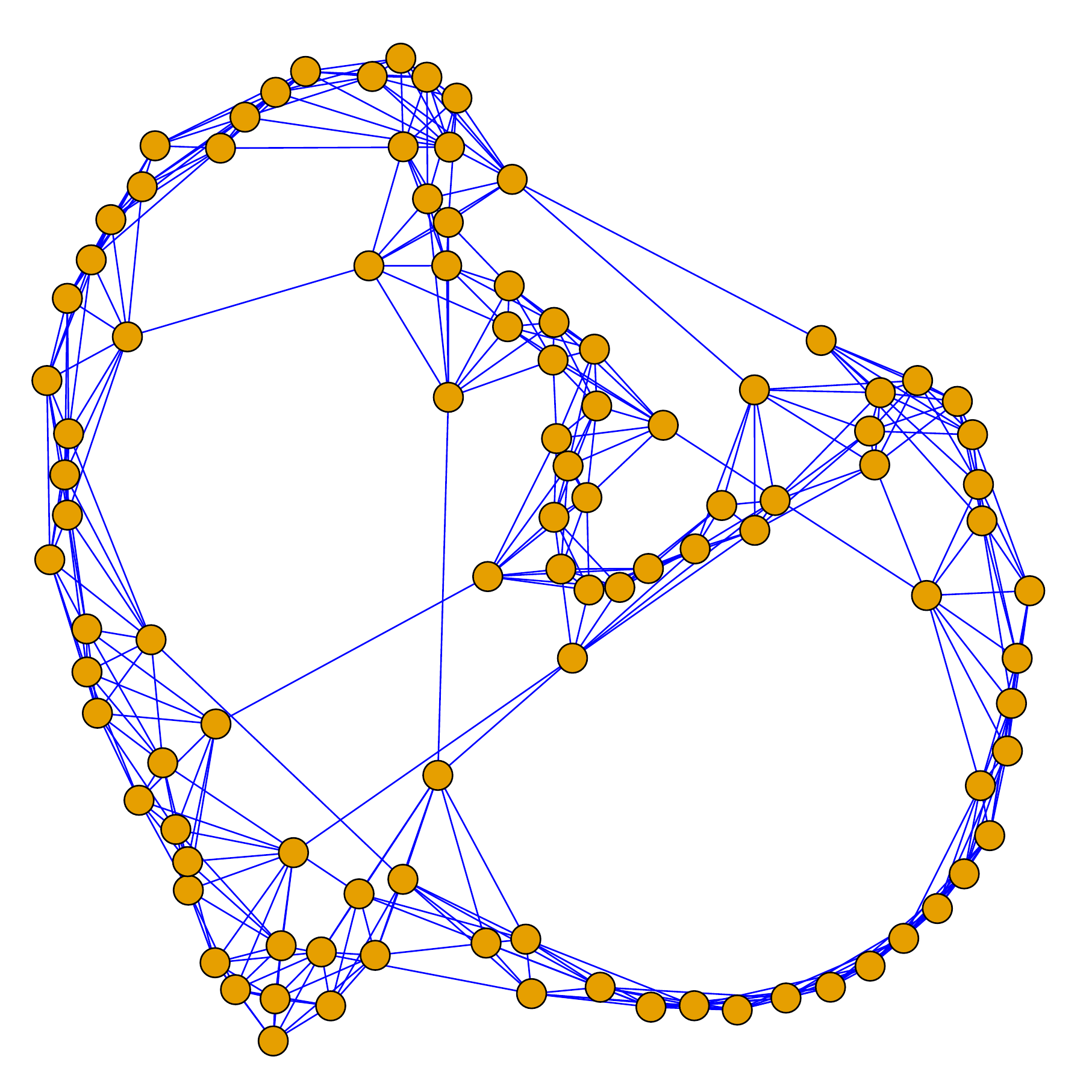}
		\caption{Watts-Strogatz small world graph}
		\label{fig:WSGraph}
	\end{subfigure}
	\caption{Examples of typical random graphs. The figures are created using \cite{igraph}.}
	\label{fig:random-graphs}
\end{figure}

\paragraph{Examples} The asymptotic degree distribution of an Erd\"{o}s-R\'{e}nyi random graph $\mathcal{G}(M, \bar{k} /M)$ is a \emph{Poisson} distribution with mean $\bar{k} $. The associated size-biased degree distribution also turns out to be  (shifted) \emph{Poisson} distribution with the same mean $ \bar{k}  $. For scale-free networks such as Barab\'{a}si-Albert preferential attachment models, the degree distribution $p(.)$ has the property that $p(k) \propto k^{-\gamma}$ for all $k \in \mathbb{N}$ and for some $\gamma \in (2,3]$, usually\footnote{$\gamma >2$ is a technical requirement for finite first order moment.}. It can be shown that the associated size-biased degree distribution $q(.)$ is given by
\begin{equation*}
q(k) = \frac{k^{- (\gamma -1)}}{\zeta(\gamma-1)}, k \in \mathbb{N}  \eqkomma
\end{equation*}
where $\zeta : \{ s \in \mathbb{C} : \Re(s) >1 \} \rightarrow \mathbb{R}$ is the \emph{Riemann zeta function}  defined as $\zeta(s) \defeq \sum_{k \in \mathbb{N} } k^{-s} $. Refer to \cref{fig:random-graphs} for typical examples. 

\section{The peer-to-peer communication system}
\label{sec:model:p2pcommsys}
\begin{figure}[!t]
	\centering
	\includegraphics[scale=0.2]{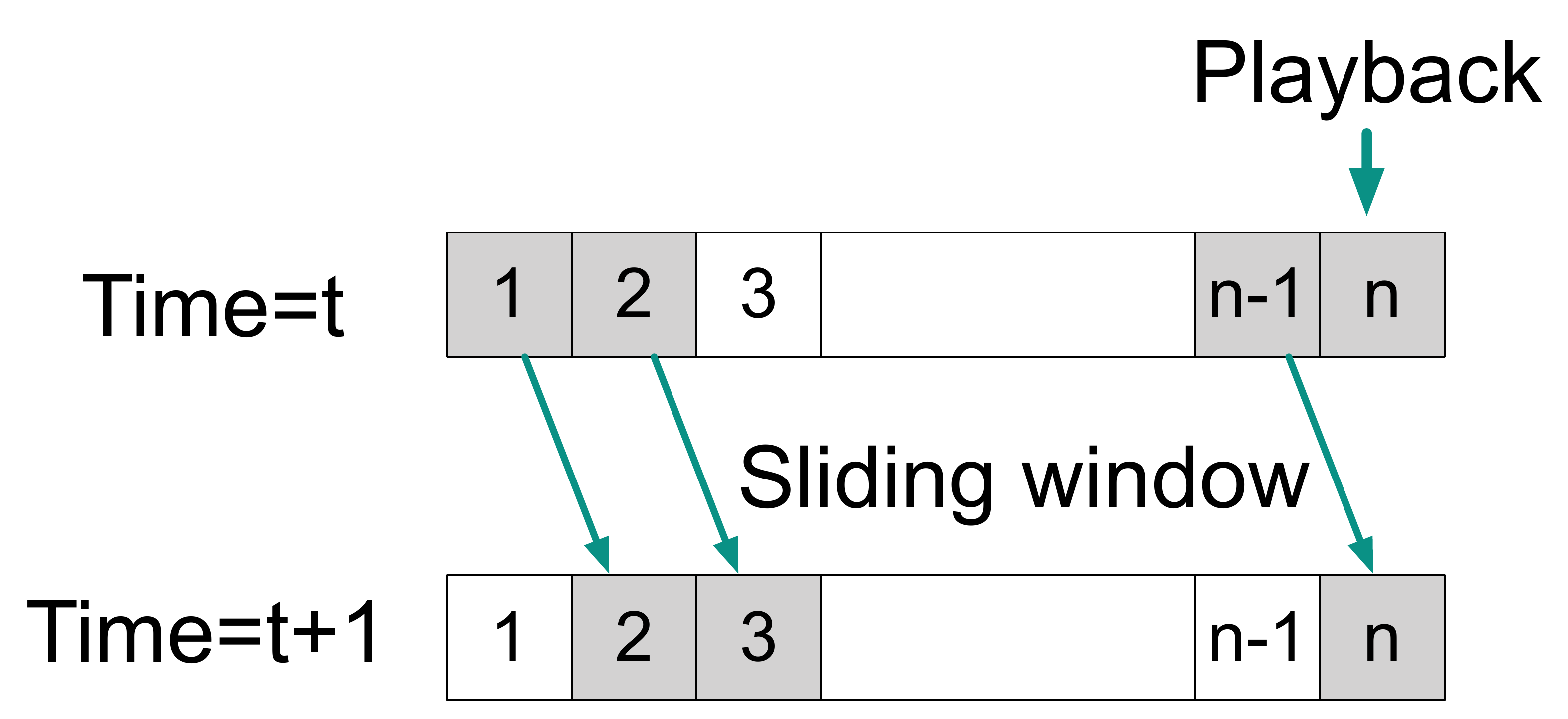}
	\caption{Shifting operation of the buffer}
	\label{fig:SlidingWindow}
\end{figure}

Suppose there are $M$ peers and a single server. Let $n$ denote the buffer length. The server uniformly selects one peer and uploads a chunk at buffer position $1$. The server continues to upload chunks to the chosen peer until there is a connection breakage  (an event that occurs with a small probability, say $\varepsilon \in [0,1]$) in which case the server chooses a peer again uniformly at random (could be the same as before). The chunk at buffer position $n$, if available, is pushed for playback.  After playback, the chunk 
is removed  and all other chunks are shifted one index to the right (\vide \cref{fig:SlidingWindow}). Each peer maintains a Poisson clock with rate proportional to its degree\footnote{That is, we place a Poisson clock on each edge of the graph.}. A peer, if not selected by the server, contacts one of its neighbours uniformly at random at each tick of its Poisson clock and seeks to download a missing chunk. The chunk it downloads from among all downloadable chunks is decided by its chunk selection policy. For simplicity, we assume that playback rate is one chunk per unit of time.

Suppose $\mathcal{G} \defeq \left( \mathcal{V}, \mathcal{E} \right) \in \mathcal{G}_M $ be a given random graph, where $  \mathcal{V} $  and $ \mathcal{E} \subseteq  \mathcal{V} \times  \mathcal{V} $ are the sets of vertices and  edges, respectively. Each node in $\mathcal{G}$ is a peer. Let $\Omega \defeq \{ \omega \in \{0,1\}^{M\times n} \mid \sum_{i=1}^M \omega(i,1) =1 \}$ 
and denote all subsets of $\Omega$ by $ \mathcal{A}$. Then define a stochastic process $\{X_t \}_{t\geq 0}$ on measurable space $\left( \Omega , \mathcal{A} \right)$ as $X_t(i,j) \defeq 1$ if the $i^{th}$ peer has the chunk required to fill the $j^{th}$ buffer location, and $0$ otherwise. The rows $X_t^{1}, X_t^{2},\ldots, X_t^{M}$  of $X_t$ denote the buffer states of peer $1,2,\ldots,M$ respectively. 

Let  $\shift  :\cup_{f,g \in \mathbb{N}} \{0,1\}^{f \times g} \rightarrow \cup_{f,g \in \mathbb{N}} \{0,1\}^{f \times g} $ denote the shifting operator defined as $\shift {Y} \defeq \left( 0, y_1,y_2,\ldots, y_{g-1} \right)$ for $Y=  \left( y_1,y_2,\ldots, y_{g} \right) \in \{0,1\}^{f \times g} $ for some $f, g \in \mathbb{N}$, where $y_1,y_2,\ldots,y_g$ denote the columns of $Y$, the union $\cup_{f,g \in \mathbb{N}}$ is a disjoint union and the product spaces $\{0,1\}^{f \times g} $ carry usual interpretation.
Let us now define the transition rates for a node $v \in \mathcal{V}$ as follows
\begin{equation}
\mu^{v} (u, u+ e_i) =		\left\{
		\begin{aligned}
	\sum_{l \in \mathcal{V} : \left(v, l\right) \in \mathcal{E}}  \varsigma  \indicator{X_t(l,i)=1}  \alpha^{v}  (i,u,X_t^{l} )    \eqkomma & \\
		\hspace{ 100pt } \mbox{if } i \neq 1 \eqkomma  &\\
			\indicator{X_t(v,1)=1} (1- \varepsilon + \varepsilon /M)  &\\
				\hspace{10pt} +   \indicator{X_t(v,1)=0} \varepsilon / M   &\\
		\hspace{ 100pt} \mbox{if } i=1  \eqkomma &
		\end{aligned}
		\right.
\label{eq:FirstRateEqn}
\end{equation}
where $u = ( u_1,u_2,\ldots, u_{n} )  \in \mathcal{T} \defeq  \{0,1\}^n$, $i \in \mathcal{F} \defeq
\{1,2,\ldots, n\}$,  such that $u_i=0$,
$\varsigma >0$ is a constant, $ \indicator{.}$ is the indicator function, 
$e_i$ is the $i^{th}$ unit vector of the $n$-dimensional Euclidean space $\mathbb{E}^n$, and $ \alpha^{v}: \mathcal{F} \times \mathcal{T} \times \mathcal{T} \rightarrow [0,1] $ is the chunk selection function of the peer $v \in \mathcal{V}$. In words, $ \alpha^{v}  (i,u,X_t^{l} )$ is the probability of downloading chunk $i$ when peer $v$ is in buffer state $u$ 
and contacts peer $l$ in buffer state  $X_t^{l}$. For LDF, $ \alpha^{v} $ attaches higher mass to smaller $i$ and for EDF, it attaches higher mass to larger $i$. We defer an elaborate discussion of the chunk selection policy to a later section. 
The system is described by the following \emph{master equation}
\begin{equation}
\begin{split}
	\frac{\, d \probOf{X} }{\, dt} = & - \probOf{X}
	  +  \sum_{v' \in \mathcal{V}}    \indicator{X(v',1)=1}\bigg[    \sum_{Y \in \Omega : \shift Y=X - \Delta(v',1)}
	   \mu^{v'} (Y^{v'} -e_1, Y^{v'} )  \bigg\{ \probOf{Y} \\
	  & + \sum_{i \in \mathcal{F} \setminus \{1\}} \sum_{v \in \mathcal{V}
	  \setminus \{v'\}}  \bigg(  \sum_{   Z=Y - \Delta(v,i) } \mu^{v} (Y^{v} -e_i, Y^{v} ) \probOf{Z}
	    - \mu^{v} (Y^{v} , Y^{v} +e_i ) \probOf{Y}  \bigg)  \bigg\} \bigg]
	   \eqkomma
  \end{split}
\label{eq:exact-master-equation}
\end{equation}
for $X \in \Omega$ where $\Delta(v,i)$ is an $M \times n$ matrix of all zeroes except a unity at position $(v,i)$. 

We can not solve the above \cref{eq:exact-master-equation} analytically. The reason we furnish this description of the process is to motivate our mean field analysis in \cref{sec:mean-field} where we gradually make a number of simplifying assumptions to obtain analytic insights.

We propose an aggregation of the chain in order to simplify our study. Define $\deg (v) \defeq \sum_{l \in \mathcal{V} } \indicator{(v, l) \in \mathcal{E}} \forall v \in \mathcal{V}$, $\mathcal{D} \defeq  \{d \mid  \exists v \in \mathcal{V}, \deg(v)=d  \}$ and $M_0 \defeq \{0,1,\ldots, M\}$.  Consider the measurable map $ \mathrm{T} : ( \Omega ,  \mathcal{A}) \rightarrow (\Upsilon, 
\mathcal{M} )$ defined by $\mathrm{T}(X) \defeq (t^k(x) : x \in \mathcal{T}, k \in \mathcal{D})$ where $ t^k(x) \defeq \sum_{v \in \mathcal{V}} \indicator{X^v =x} \indicator{\deg(v)=k }$,
$\Upsilon \defeq \{ \upsilon \in M_0^ {|\mathcal{T}| \times \mathcal{D}} : \sum_{k \in \mathcal{D}} \sum_{x \in \mathcal{T}: x_1=1} \upsilon^k(x) =1 , \sum_{k \in \mathcal{D}} \sum_{x \in \mathcal{T}: x_1=0} \upsilon^k(x) =M-1 ,  \sum_{x \in \mathcal{T} } \upsilon^k(x) =n_k \} $, $n_k$ is the number of peers of degree~$k$ and $\mathcal{M}$ is the $\sigma$-field generated by all subsets of $\Upsilon$. 
Define a binary relation $\overset{\mathrm{T}}{\sim}$ on $\Omega$ as $X \overset{\mathrm{T}}{\sim} Y \iff \mathrm{T}(X) =\mathrm{T}(Y)$ and $\Omega_t \defeq \{ X : \mathrm{T}(X) =t \}$ for each $t \in \Upsilon$. 
Then, $\{ \Omega_t  : t \in \Upsilon  \} $  
is a partition of $\Omega$ and each $\Omega_t$ is an equivalence class. The induced probability is given by
\begin{equation}
\probOf{\mathrm{T}(X) =t} = \sum_{X \in \Omega : \mathrm{T}(X) =t} \probOf{X} \eqpunkt
\label{eq:aggregation-prob}
\end{equation}

Such an aggregation into a population model is useful in reducing the state space. 
Since there is no \emph{perfect} graph\footnote{A
	graph with two or more nodes is called perfect if for each pair of distinct vertices $u$ and $v$, $\deg(u) \ne \deg(v)$,	\ie, no two vertices have the same degree. }(see \cite{behzad1967no} for proof), we are certain that $|\mathcal{D}| <M$. However, for $M \geq 2$, we can construct a unique \emph{quasiperfect} graph\footnote{A graph 	with 	at 	least 	two nodes 	is 	quasiperfect 	if 	there 	are 	precisely 	two 	vertices with 	the same 	degre} (unique upto isomorphism) that is connected and entails  $|\mathcal{D}| =M-1$,  worst case scenario. We attempt to derive conditions for  such an aggregation to actually reduce  state space. Before presenting our result in this context, let us define some necessary quantities.

Define $R\defeq (n_k : k \in \mathcal{D})$ and $\mathcal{C} \defeq \{ C \in M_0^ {|\mathcal{T}|}: \sum_{x \in \mathcal{T}: x_1=1} c(x) =1 ,\sum_{x \in \mathcal{T}: x_1=0} c(x) =M-1 \}$. Given $R$ and a $C \in \mathcal{C}$, define the function $F : (0,1)^ {|\mathcal{D}| \times 2^n} \rightarrow \mathbb{R}$ as
\begin{equation}
F(x,y) \defeq \left( \prod_{i \in \mathcal{D}} x_i^{-n_i} \right) \left(  \prod_{j \in \mathcal{T}} y_j^{-c(j)}  \right) \left(   \prod_{i \in \mathcal{D}, j \in \mathcal{T}} \frac{1}{1- x_i y_j} \right) \eqkomma
\end{equation}
where $x =(x_i : i \in \mathcal{D}) \in (0,1)^ {|\mathcal{D}|} $ and $y= (y_j : j \in \mathcal{T}) \in (0,1)^ { 2^n}$. Also define its minimum on the open ball $(0,1)^ {|\mathcal{D}| \times 2^n} $ as follows
\begin{equation}
\chi (R,C) \defeq \min_{x_i,y_j \in (0,1) \forall i \in \mathcal{D}, j \in \mathcal{T}} F(x,y) \eqpunkt
\end{equation}
Now we present our result regarding state space reduction.
\begin{myResult}
	For $\mathcal{G} \in \mathcal{G}_M $, a necessary condition for $\mathrm{T}$ to engender state space reduction is
	\begin{equation}
	M 2^{(M-1) (n-1)}   \geq  \binom{M-2 + 2^{n-1}}{M-1}  \min_{C \in \mathcal{C}}   M^{- a_0 (|\mathcal{D}| + 2^n)} 	\chi (R,C)  \eqkomma
	\end{equation}
	for an absolute constant $a_0 >0$. The following gives us a sufficient condition,
	\begin{equation}
	M 2^{(M-1) (n-1)}  \geq  \binom{M-2 + 2^{n-1}}{M-1}  \max_{C \in \mathcal{C}}  	\chi (R,C) \eqpunkt
	\end{equation}
	\label[result]{result:state-space-reduction}
\end{myResult}


\begin{proof}[Proof of \cref{result:state-space-reduction}]
	Given $\mathcal{G}$, we try to find the size of $\Upsilon$. Suppose  $\upsilon \in \Upsilon$. Elements of $\upsilon$ must satisfy three sets of constraints, \viz,
	\begin{eqnarray*}
		\sum_{x \in \mathcal{T}} \upsilon^k(x) =n_k  \forall k \in \mathcal{D} \eqkomma \\
		\sum_{k \in \mathcal{D}} \sum_{x \in \mathcal{T}: x_1=1} \upsilon^k(x) =1 \eqkomma \\
		\sum_{k \in \mathcal{D}} \sum_{x \in \mathcal{T}: x_1=0} \upsilon^k(x) =M-1 \eqpunkt \\
	\end{eqnarray*}

	We treat this as a combinatorial problem of finding the number of contingency tables of nonnegative elements, satisfying given row and column sums. In this context, regard the first set of equations as row constraints. These are fixed, given $\mathcal{G}$. Now set column constraints as  
	\begin{equation*}
	\sum_{k \in \mathcal{D}} \upsilon^k(x)  = c(x) \forall x \in \mathcal{T} \eqkomma
	\end{equation*}
	where the column constraints are further constrained as follows
	\begin{eqnarray}
		\sum_{x \in \mathcal{T}: x_1=1} c(x) =1 \eqkomma \\
		\sum_{x \in \mathcal{T}: x_1=0} c(x)=M-1 \eqpunkt
	\end{eqnarray}

	Before we proceed further, let us define certain quantities that would come into play later on. Let $R\defeq (n_k : k \in \mathcal{D}), C\defeq (c(x) : x \in \mathcal{T})$. Notice that $R\vec{1}=C\vec{1}=M$, the number of peers. Define the function $F : (0,1)^ {|\mathcal{D}| \times 2^n} \rightarrow \mathbb{R}$ as
	\begin{equation}
	F(x,y) \defeq \left( \prod_{i \in \mathcal{D}} x_i^{-n_i} \right) \left(  \prod_{j \in \mathcal{T}} y_j^{-c(j)}  \right) \left(   \prod_{i \in \mathcal{D}, j \in \mathcal{T}} \frac{1}{1- x_i y_j} \right) \eqkomma
	\end{equation}
	where $x =(x_i : i \in \mathcal{D}) \in (0,1)^ {|\mathcal{D}|} $ and $y= (y_j : j \in \mathcal{T}) \in (0,1)^ { 2^n}$. Also define its minimum on the open ball $(0,1)^ {|\mathcal{D}| \times 2^n} $ as follows
	\begin{equation}
	\chi (R,C) \defeq \min_{x_i,y_j \in (0,1) \forall i \in \mathcal{D}, j \in \mathcal{T}} F(x,y) \eqpunkt
	\end{equation}

	Elements of the vector $C$ can be partioned into two equal halves. Each half can be thought of as a solution in non-negative integers to a linear Diophantine equation (the constraints in the definition of $\Upsilon$). The first constraint is
	\begin{equation*}
	\sum_{x \in \mathcal{T}: x_1=1} c(x) =1 ,
	\end{equation*}
	which allows $2^{n-1}$ solutions in non-negative integers. The second constraint is,
	\begin{equation*}
	\sum_{x \in \mathcal{T}: x_1=0} c(x) =M-1 \eqpunkt
	\end{equation*}
	The above has $\binom{M-1 +2^{n-1}-1 }{2^{n-1}-1} = \binom{M-2 + 2^{n-1}}{M-1}$ solutions in non-negative integers. Define $\mathcal{C} \defeq \{ C : \sum_{x \in \mathcal{T}: x_1=1} c(x) =1 ,\sum_{x \in \mathcal{T}: x_1=0} c(x) =M-1 \}$. Since the above two equations can be solved independently, the total number of admissible $C$ is, therefore, $|\mathcal{C}|= 2^{n-1} \binom{M-2 + 2^{n-1}}{M-1} $.

	Fix a $C \in \mathcal{C}$. Let $\#(R,C)$ denote the number of $|\mathcal{D}| \times 2^n$ matrices (contingency tables) with nonnegative  elements satisfying row sum $R$ and column sum $C$.
	Then, following \cite{barvinok2009asymptotic}, we get
	\begin{equation}
	\chi (R,C) \geq \#(R,C) \geq M^{- a_0 (|\mathcal{D}| + 2^n)} 	\chi (R,C) \eqkomma
	\end{equation}
	for an absolute constant $a_0 >0$. It should be noted that the quantity $a_0$ depends on both $R$, and $C$ and hence, should be written as $a_0(R,C)$. However, for simplicity, we omit the arguments $R$, and $C$ and just write $a_0$.
	Please refer to \cite{barvinok2009asymptotic} for proof. Since any $C \in \mathcal{C}$ is a valid choice for $\Upsilon$, we must have
	\begin{equation*}
	|\Upsilon| = 	\sum_{C \in \mathcal{C}}  \#(R,C) \geq |\mathcal{C}|	\min_{C \in \mathcal{C}}   M^{- a_0 (|\mathcal{D}| + 2^n)} 	\chi (R,C)  \eqpunkt
	\end{equation*}	Similarly, we get an upper bound as follows
	\begin{equation*}
	|\Upsilon| \leq 	|\mathcal{C}|	\max_{C \in \mathcal{C}}  	\chi (R,C)  \eqpunkt
	\end{equation*}
	Combining the above two, we get
	\begin{equation}
	|\mathcal{C}|	\min_{C \in \mathcal{C}}   M^{- a_0 (|\mathcal{D}| + 2^n)} 	\chi (R,C)  \leq 		|\Upsilon| \leq 	|\mathcal{C}|	\max_{C \in \mathcal{C}}  	\chi (R,C)  \eqpunkt
	\end{equation}

	Now, see that $|\Omega|= M 2^{M (n-1)}$.  We seek to find $n \in \mathbb{N}$ such that $|\Omega| \geq |\Upsilon| $.

	\paragraph{\underline{Necessary condition:}}
	\begin{align*}
	|\Omega| &\geq |\Upsilon| \\
	\implies	|\Omega| & \geq  |\mathcal{C}|	\min_{C \in \mathcal{C}}   M^{- a_0 (|\mathcal{D}| + 2^n)} 	\chi (R,C)  \\
	\iff  M 2^{M (n-1)} &  \geq 2^{n-1} \binom{M-2 + 2^{n-1}}{M-1}
	   \min_{C \in \mathcal{C}}   M^{- a_0 (|\mathcal{D}| + 2^n)} 	\chi (R,C)  \\
	\iff  M 2^{(M-1) (n-1)} &  \geq  \binom{M-2 + 2^{n-1}}{M-1}  \min_{C \in \mathcal{C}}   M^{- a_0 (|\mathcal{D}| + 2^n)} 	\chi (R,C)  \\
	\end{align*}

	\paragraph{\underline{Sufficient condition:}}
	Set
	\begin{align*}
	|\Omega| &\geq  |\mathcal{C}|	\max_{C \in \mathcal{C}}  	\chi (R,C) \\
	\iff M 2^{M (n-1)} & \geq 2^{n-1} \binom{M-2 + 2^{n-1}}{M-1} 	\max_{C \in \mathcal{C}}  	\chi (R,C) \\
	\iff M 2^{(M-1) (n-1)} &  \geq  \binom{M-2 + 2^{n-1}}{M-1}  \max_{C \in \mathcal{C}}  	\chi (R,C) \\
	\end{align*}

	\begin{myRemark}
		Note that, with the substitution $x_i=e^{-x_i'}$ and $y_j= e^{-y_j'}$, the task of evaluating the minimum of $F$ on the open ball $(0,1)^ {|\mathcal{D}| \times 2^n} $, \ie, $\chi (R,C)$ reduces to the problem of minimising the convex function

		\begin{equation}
		F'(x',y') \defeq \sum_{i \in \mathcal{D}} x'_i {n_i} +  \sum_{j \in \mathcal{T}} y'_j {c(j)}   -  \prod_{i \in \mathcal{D}, j \in \mathcal{T}} \ln(1- e^{-x_i' -  y_j'}) \eqkomma
		\end{equation}
		on the positive, open orthant $x_i', y_j' > 0$. This makes possible the use of convex optimization methods to compute $\chi$ in polynomial time (see \cite{nesterov1994interior}).
		\end{myRemark}

\end{proof}



We  emphasize that we do lose  information in the process of aggregation. Also, the aggregated process, or the \quotes{lumped} process is not necessarily Markovian\footnote{Please refer to \cite{kemeny1960finite} for a leisurely read on this topic.}. However, if we impose that  peers having  the same degree play  the same chunk selection strategy, 
we expect peers having same degree to behave indistinguishably in mean field. We discuss these arguments in \cref{sec:mean-field} and strive to derive master equation for the population model.
Of particular interest to us is the scenario where  peers with higher degree play LDF and others play EDF. We show that this heterogeneous setup has a number of benefits compared to homogeneous ones. We elucidate this in later sections.

\newpage
\chapter{Mean-field theoretic analysis}
\label{sec:mean-field}

\section{Mean-field master equations}
\label{sec:mean-field:master-equations}
In this section, we impose simplifying assumptions on the general description of the process developed in \cref{sec:model}, in an attempt to facilitate furtherance of analytic treatment when the number of peers in the system, $ M$, is large. As a first step, peers are assumed to be independently interacting with a \quotes{mean} environment (\quotes{mean-field}), nullifying the complexity of interactions. This essentially engenders independence among the rows of $\{X_t \}_{t\geq 0}$. Peers having  same degree  play  same chunk selection policy and thus, behave indistinguishably in an infinitely large random graph.  This insinuates that such a mean-field behaviour can very well be described by a population model, where we just count the number of peers of each buffer configuration $x \in \mathcal{T}$, for each degree $k \in \mathbb{N}$, instead of recording each peer separately. Thus, instead of indexing by peers as done in \Cref{sec:model}, we shall index all the relevant quantities by degree. This approach has been extensively followed in statistical physics and probability literature, \eg, infection models \cite{pastor2002epidemic,durrett2007random}.

Consider the Markov chain $\{Z_t  \}_{t\geq 0} $ on measurable space $( \mathbb{N}_0^{|\mathcal{T}| \times \mathbb{N}} ,\mathcal{N}_0)$  defined as
$ Z_t \defeq ( z_x^{k} (t) : x \in \mathcal{T}, k \in \mathbb{N} ) $ where
$z_x^{k} (t) $ is the number of degree-$k$ peers at buffer configuration $x \in \mathcal{T}$ at time $t$,  $\mathcal{N}_0$ is the $\sigma$-field generated by all subsets of $ \mathbb{N}_0^{|\mathcal{T}|  \times \mathbb{N}}$ and $ \mathbb{N}_0 \defeq \mathbb{N} \cup \{0\}$. We  omit  time index whenever   dependence is unambiguous. 
Our mean field assumption allows us to treat each neighbour of a degree-$k$ peer as an independent sample from a \quotes{mean} environment (\quotes{mean-field}). Therefore, we get our mean-field transition
rates for a degree-$k$ peer as follows, for each $k \in \mathbb{N}, u \in \mathcal{T}$ and $i \in \mathcal{F} \setminus \{1\}$ such that $u_i=0$
\begin{align*}
	\beta^{k} (u,u+e_i) &= \sum_{l=1}^{k} \varsigma \Eof{ \indicator{Y_l(i)=1} \alpha^{k}  (i,u,Y_l)  }  \\
	&= k \varsigma \Eof{ \indicator{Y_1(i)=1} \alpha^{k}  (i,u,Y_1)  }  \eqkomma
	\end{align*}
where~$\{(Y_l,d_l) \mid Y_l = \left(Y_l(1),Y_l(2), \ldots, Y_l(n) \right) \in \mathcal{T}, d_l \in \mathbb{N} \}_{l=1}^k$ is a set of $k$  independent samples from the mean environment of a degree-$k$ peer. The first component of each neighbour is the buffer state and the second component, its degree. Note that $d_l$'s are distributed according to $q$ of \cref{eq:size-biased-degree}. The expectation above is found in a straightforward way.
\begin{align*}
	& \Eof{ \indicator{Y_1(i)=1}  \alpha^{k}  (i,u,Y_1)  } \\
	=&  \sum_{v \in \mathcal{T}} \sum_{m \in \mathbb{N}} \Eof{ \indicator{Y_1(i)=1}  \alpha^{k}  (i,u,Y_1) \mid Y_1=v, d_1=m }
	 \probOf{Y_1=v, d_1=m} \\
	=&  \sum_{v \in \mathcal{T} : v_i=1} \sum_{m \in \mathbb{N}} \alpha^{k}  (i,u,v)  \probOf{Y_1=v \mid  d_1=m} \probOf{ d_1=m}\\
	=& \sum_{v \in \mathcal{T} : v_i=1} \sum_{m \in \mathbb{N}}  q(m) \frac{\Eof{z_v^{m}}}{n_m} \alpha^{k}  (i,u,v)   \eqkomma
	\end{align*}
where $n_m$ is the number of peers of degree $m$. 
Having found the expectation, we now turn to buffer index $1$. Since only the server can upload chunks at buffer index $1$, we need to consider this case separately. As we assumed the server selects a peer uniformly at random, the probability of a degree-$k$ peer of being served directly by the server is $1/M$. Therefore, 
we get
\begin{equation}
\beta^{k} (u,u+e_i) = \twopartdef{k \varsigma \sum_{v \in \mathcal{T} : v_i=1} \sum_{m \in \mathbb{N}}  q(m) \frac{\Eof{z_v^{m}}}{n_m} \alpha^{k}  (i,u,v)}{i \neq 1}{1/ M}{i=1} \eqkomma
\label{eq:per-peer-rate}
\end{equation}
for each $k \in \mathbb{N}, u \in \mathcal{T}$ and $i \in \mathcal{F} $ such that $u_i=0$. Let us define $\varrho : \mathbb{N} \times \mathcal{T} \times \mathcal{F} \rightarrow  \{-1,0,1\} ^{|\mathcal{T}| \times \mathbb{N}} $  such
that $Y =Z - \varrho (k,u,i) \implies y_u^{k} = z_u^{k} +1, y_{u+e_i}^{k} = z_{u+e_i}^{k} -1 , y_x^{l} =z_x^{l} \forall l \in \mathbb{N} \setminus \{k\}, x \in \mathcal{T} \setminus \{u\}$. Broadening the scope of definition of $ \beta$ by setting it to $0$ for all arguments not covered in \cref{eq:per-peer-rate}, we have the following \emph{mean-field master equations}
\begin{equation}
\begin{split}
\frac{\,d \probOf{Z}}{\,dt} =& - \probOf{Z} + \sum_{\substack{Y : \sum_{ \shift  v= u} y_v^{l} = z_u^{l} \\ \forall u,v \in \mathcal{T}, l \in
		\mathbb{N}}} \Bigg[ \probOf{Y}
 +  \sum_{l \in \mathbb{N}, u \in \mathcal{T}, i \in \mathcal{F}} (y_u^{l} +1) \beta^{l} (u,u+e_i)
\probOf{Y -\varrho(l,u,i )}  \\
  &  \hspace{100pt } -  \sum_{l \in \mathbb{N}, u \in \mathcal{T}, i \in \mathcal{F}} y_u^{l} \beta^{l} (u,u+e_i) \probOf{Y}
    \Bigg] \eqpunkt
\end{split}
\label{eq:master-equation}
\end{equation}

In pursuance of the mean dynamics, we begin by first setting $\prob$ to zero outside its domain of definition, and then by defining, for each $l \in \mathbb{N}, u \in \mathcal{T}, i \in \mathcal{F}$, the following
quantity $\gamma_{l,u,i} (Z) \defeq  z_u^{l} \beta^{l} (u,u+e_i) $. Next, we note that, in mean field, we can write $\Eof{\gamma_{l,u,i} (Z)}$
as $\Eof{z_u^{l}}\beta^{l} (u,u+e_i) $. The following result encapsulates the mean dynamics of the system.

\begin{myResult}
		\label[result]{result:mean-dynamics}
	The process $\{Z_t  \}_{t\geq 0} $ defined on measurable space $(
	\mathbb{N}_0^{|\mathcal{T}| \times \mathbb{N}} ,\mathcal{N}_0)$  admitting master equation \cref{eq:master-equation} satisfies
	\begin{equation}
	\label{eq:mean-dynamics}
	\frac{\,d \Eof{Z}}{\,dt} = - \Eof{Z} + \Eof{Y}
	+ \sum_{l \in \mathbb{N}, u \in \mathcal{T}, i \in \mathcal{F} } \varrho(l,u,i ) \Eof{\gamma_{l,u,i} (Y)} \eqkomma
	\end{equation}
	where $Y \in  \mathbb{N}_0^{|\mathcal{T}| \times \mathbb{N}} $ is such that $y_u^{l} = \sum_{\shift v=u} z_v^{l}  \forall l \in \mathbb{N}$.
\end{myResult}
We make use of the following lemma to base our proof of \cref{result:mean-dynamics} upon.

\begin{myLemma}
	\label[lemma]{lemma:for-mean-dynamics}
For $Y$ as defined in \cref{result:mean-dynamics}, the following identity holds true,
for all $k \in \mathbb{N} $, \begin{equation*}
	\sum_{Z \in  \mathbb{N}_0^{|\mathcal{T}| \times \mathbb{N}} } z_u^{k} \sum_{\substack{Y : \sum_{ \shift  v= u} y_v^{l} = z_u^{l} \\ \forall u,v \in \mathcal{T}, l \in
			\mathbb{N}}} \probOf{Y} = \sum_{v \in \mathcal{T} : \shift v =u } \Eof{z_v^{k} }
	\end{equation*}

\end{myLemma}

The proof is provided in \cref{sec:appendix}.
Now, we provide a sketch of proof of \cref{result:mean-dynamics}.
\begin{proof}[Proof of \cref{result:mean-dynamics}]
From \cref{eq:master-equation} and applying  \cref{lemma:for-mean-dynamics}, we get
\begin{align*}
\frac{\,d \Eof{Z}}{\,dt} =& - \Eof{Z} + \Eof{Y} +
	\sum_{Z \in  \mathbb{N}_0^{|\mathcal{T}| \times \mathbb{N}} } Z \sum_{\substack{Y : \sum_{ \shift  v= u} y_v^{l} = z_u^{l} \\ \forall u,v \in \mathcal{T}, l \in
		\mathbb{N}}} \bigg[  	\sum_{l \in \mathbb{N}, u \in \mathcal{T}, i \in \mathcal{F} }
  \gamma_{l,u,i} (Y - \varrho(l,u,i))   \probOf{Y - \varrho(l,u,i)}  	\\
  &   \hspace{100pt} - \sum_{l \in \mathbb{N}, u \in \mathcal{T}, i \in \mathcal{F} }  \gamma_{l,u,i} (Y ) \probOf{Y} \bigg] \\
=&  - \Eof{Z} + \Eof{Y} + 	\sum_{l \in \mathbb{N}, u \in \mathcal{T}, i \in \mathcal{F} }  \sum_{Z \in  \mathbb{N}_0^{|\mathcal{T}| \times \mathbb{N}} }
  \sum_{\substack{Y : \sum_{ \shift  v= u} y_v^{l} = z_u^{l} \\ \forall u,v \in \mathcal{T}, l \in
				\mathbb{N}}}   \varrho(l,u,i) \gamma_{l,u,i} (Y - \varrho(l,u,i))
		  \probOf{Y - \varrho(l,u,i)}  \\
=&  - \Eof{Z} + \Eof{Y} + \sum_{l \in \mathbb{N}, u \in \mathcal{T}, i \in \mathcal{F} } \varrho(l,u,i ) \Eof{\gamma_{l,u,i} (Y)}
\end{align*}
The second line is arrived at by addition and subtraction of $\varrho(l,u,i)$  and rearrangement of summands.
\end{proof}

Now, looking closely at \cref{eq:mean-dynamics} and
 recalling the definition of $\varrho(l,u,i)$, we write down explicitly
 \begin{equation}
 \begin{split}
 	\frac{\,d \Eof{z_u^{k}}}{\,dt} &= - \Eof{z_u^{k}}  +
 	\sum_{v \in \mathcal{T} : \shift v=u} \Bigg[ \Eof{z_v^{k}} +  \sum_{i \in \mathcal{F}} \Eof{z_{v-e_i}^{k}} \beta^{l} (v-e_i,v)
    - \sum_{i \in \mathcal{F}} \Eof{z_{v}^{k}} \beta^{l} (v,v+e_i) \Bigg] \eqkomma
 \end{split}
 \end{equation}
for each $u \in \mathcal{T}, k \in \mathbb{N}$.

It is convenient to work with proportions to study the mean dynamics. Therefore,
consider the associated Markov chain $\{W_t  \}_{t\geq 0} $ on measurable space
$( [0,1]^{|\mathcal{T}|  \times \mathbb{N} } , \mathcal{B})$  defined as $ W_t
\defeq ( w_x^{k} (t) : x \in \mathcal{T}, k \in \mathbb{N} ) $ where $w_x^{k} (t) \defeq \frac{z_x^{k}}{n_k}$ 
and  $\mathcal{B}$ is the Borel $\sigma$-field on $[0,1]^{|\mathcal{T}|  \times \mathbb{N} }$.
Let us define,
\begin{equation}
	\lambda^{k}(u,u+e_i) = w_u^{k} 	\beta^{k} (u,u+e_i)   \eqkomma
	\label{eq:lambda-1}
	\end{equation}
	for each $k \in \mathbb{N}, u \in \mathcal{T}$ and $i \in \mathcal{F} $ such that $u_i=0$.
\Cref{eq:lambda-1} quantifies the contribution of  $u \rightarrow u+e_i$
transitions among  degree-$k$ peers to the rate of change of $w_x^{k}$. Apart from  transitions that are due to downloading of chunks by the peers from among themselves, the only other source of transition is shifting after playback, an event we assume to take place at rate unity. Then the total influx into a buffer state $u \in \mathcal{T}$ is~$  \sum_{v \in \mathcal{T} : \shift v=u} \bigg( w_v^{k} +\sum_{i \in \mathcal{F} } \lambda^{k}(v-e_i,v) \bigg)$ , while the total outflux is~$\sum_{v \in \mathcal{T} : \shift v=u} \sum_{i \in \mathcal{F}} \lambda^{k}(v,v+e_i) +w_u^{k}$. 

We argue that, when the number of peers is large, it sufficies to study the mean dynamics of the proportions, for the fluctuation around  mean is expected to be negligible for infinitely large systems. Therefore, denoting $\Eof{w_x^{k}}$, with abuse of notation, by $w_x^{k}$ itself, we are in a position to write down the following \emph{rate equations} to capture the mean dynamics of the system in the  form of an ordinary differential equation (ODE),
\begin{equation}
\begin{split}
\frac{\,dw_u^{k}}{\,dt} & = -w_u^{k} +
 \sum_{v \in \mathcal{T} : \shift v=u} \Bigg[ w_v^{k} + \sum_{i \in \mathcal{F}} \bigg( \lambda^{k}(v-e_i,v)  -\lambda^{k}(v,v+e_i) \bigg) \Bigg]  \eqkomma
\end{split}
\label{eq:rate-equation}
\end{equation}
for each $u \in \mathcal{T}, k \in \mathbb{N}$. We find the stationary distribution by setting $\frac{\,dw_u^{k}}{\,dt} =0$, giving rise to  following fixed point equations at stationarity,
\begin{equation}
w_u^{k} = \sum_{v \in \mathcal{T} : \shift v=u} \Bigg[w_v^{k} + \sum_{i \in \mathcal{F}} \bigg( \lambda^{k}(v-e_i,v)  -\lambda^{k}(v,v+e_i) \bigg) \Bigg] \eqkomma
\label{eq:stationary-distribution}
\end{equation}

Observe that
\begin{align*}
	\sum_{u \in \mathcal{T}}\frac{\,dw_u^{k}}{\,dt}  &= 	- \sum_{u \in \mathcal{T}} w_u^{k}  + \sum_{u \in \mathcal{T}} \sum_{v \in \mathcal{T} : \shift v=u} w_v^{k}
	+  \sum_{u \in \mathcal{T}} \sum_{v \in \mathcal{T} : \shift v=u}  \sum_{i \in \mathcal{F}} \bigg( \lambda^{k}(v-e_i,v)
	  -\lambda^{k}(v,v+e_i) \bigg)  \\
	&= - \sum_{u \in \mathcal{T}} w_u^{k} +
     \sum_{u \in \mathcal{T}} w_u^{k} +
      \sum_{u \in \mathcal{T}}   \sum_{i \in \mathcal{F}} \bigg( \lambda^{k}(u-e_i,u)  -\lambda^{k}(u,u+e_i) \bigg)  \\
	&= 0 \eqkomma
	\end{align*}
for all $k \in \mathbb{N}$. This is  because of the fact that proportions sum up to 1, \ie, ~$	\sum_{u \in \mathcal{T}} w_u^{k} =1 \forall k \in \mathbb{N}$. Notice that, by definition,  for $u,v \in \mathcal{T},  \shift v=u \implies u_1=0$ and for $u, u', v, v' \in \mathcal{T} :  \shift v=u ,  \shift v'=u', u \ne u' \implies v \ne v'$. Therefore, in the second term on the right hand side, the first summation accounts for exactly half of $u \in \mathcal{T}$, \viz, with $u_1=0$ and the second summation runs over exactly two distinct $v \in \mathcal{T} : \shift v=u$ for each $u$ in the first summation. So, we get $\sum_{u \in \mathcal{T}} \sum_{v \in \mathcal{T} : \shift v=u} w_v^{k} =  \sum_{u \in \mathcal{T}} w_u^{k} $. The third term vanishes by simple rearrangement of summands.

It does merit some attention that the population model presented here can be thought of as an infection model with $2^n$ distinct levels of a disease, each level being represented by  $u \in \mathcal{T}$ and (gradual) recovery being represented by the shifting of buffer state after playback. This amounts to saying, a peer with all buffer positions filled is infected to the highest extent of a disease and if it does not download any chunk,  \ie, if it does not get infected, within the next $n$-time units ($n$ chunk-shifting operations), it will  gradually recover to a state of complete susceptibility (no chunk available). Another analogy that we would like to draw here is that as the chances of being infected increases with the number of infected neighbours in contact, so improves the playback experience. Higher the degree, better are the chances of downloading a piece from among neighbours.

Let $p_k : \{1,2,\ldots,n \} \rightarrow [0,1] $ denote the buffer probability of a peer of degree $k \in \mathbb{N}$. Then,
\begin{equation}
p_k(i)= \sum_{u \in \mathcal{T} : u_i=1} w_u^{k} \eqpunkt
\end{equation}
The global performance of the network is linked to these degree-specific buffer probabilities through the associated degree distribution of $\mathcal{G}$ as follows
\begin{equation}
p(i) = \sum_{k \in \mathbb{N}} \pi(k) p_k(i) \eqpunkt
\label{eq:global-definition}
\end{equation}
Now, we try to derive a recurrence relation among $p_k(.)$'s (and then, in turn, among $p(i)$'s) by means of  \cref{eq:stationary-distribution}. We have the following result in that direction.
\begin{myResult}
		\label[result]{result:recurrence-from-master}
	The process $\{W_t\}_{t \geq 0}$ of proportions obeying rate equation \cref{eq:rate-equation}, admits  the following recursion relation among its buffer probabilities at stationarity
	\begin{align}
	p_k(i+1) &= p_k(i) + \sum_{u \in \mathcal{T} : u_{i}=1} \lambda^{k}( u-e_{i},u) \\
	p(i+1) &= p(i) +  \sum_{k \in \mathbb{N}} \pi(k)\sum_{u \in \mathcal{T} : u_{i}=1} \lambda^{k}( u-e_{i},u)  \eqkomma
	\end{align}
	for all $i,k \in \mathbb{N}$. Moreover, the buffer probabilities are nondecreasing functions of their arguments, \ie, buffer indices.
\end{myResult}

Before providing a proof of \cref{result:recurrence-from-master}, let us first prove the following lemma.
\begin{myLemma}
	\label[lemma]{lemma1}
 For the Markov chain $\{W_t  \}_{t\geq 0} $ obeying rate equation
 \cref{eq:rate-equation}, for each $i,k \in \mathbb{N}$, we have the following two identities
 \begin{align}
	\sum_{u \in \mathcal{T}: u_{i+1}=1} \sum_{v \in \mathcal{T}: \shift v=u} w_v^{k} &= p_k(i) \\
	 \sum_{u \in \mathcal{T}: u_{i}=1}  \sum_{j \in \mathcal{F}} \left[  \lambda^{k}(u-e_j,u)  -\lambda^{k}(u,u+e_j) \right]
	&  =  \sum_{u \in \mathcal{T}: u_{i}=1} \lambda^{k}( u-e_{i},u)
 \end{align}

\end{myLemma}

The proof of \cref{lemma1} is provided in \cref{sec:appendix}.
Now we furnish a proof of  \cref{result:recurrence-from-master}.
\begin{proof}[Proof of \cref{result:recurrence-from-master}]
Summing both sides of \cref{eq:stationary-distribution} and using \cref{lemma1}, we get
\begin{align*}
	\sum_{u \in \mathcal{T}: u_{i+1}=1} w_u^{k} = & 	\sum_{u \in \mathcal{T}: u_{i+1}=1}  \sum_{v \in \mathcal{T} : \shift v=u} w_v^{k}
	 + 	\sum_{u \in \mathcal{T}: u_{i+1}=1}  \sum_{v \in \mathcal{T} : \shift v=u} \sum_{j \in \mathcal{F}} \bigg( \lambda^{k}(v-e_j,v)  -\lambda^{k}(v,v+e_j) \bigg) \\
	\implies p_k(i+1) &= p_k(i) + \sum_{u \in \mathcal{T} : u_{i}=1} \lambda^{k}( u-e_{i},u)  \eqpunkt
\end{align*}

Summing the above according to \cref{eq:global-definition}, we get the other recurrence relation pertaining to global performance. The fact that buffer probabilities are nondecreasing in buffer indices follows from the nonnegativity of $\lambda^{k}$'s.
\end{proof}

\paragraph{Interpretation of \cref{result:recurrence-from-master}:}

The recurrence relation among  buffer probabilities succintly captures a peer's performance aspect in the live video streaming framework. The left hand side gives the probability that the chunk required to fill buffer location $i+1$ is present. The right hand side virtually tells us that there are two possible ways to have the chunk at buffer location $i+1$ present. One, it could already be there at buffer location $i$, with  probability of buffer index $i$, and was made available at location $i+1$ due to shifting. Two, the chunk was not there, but the peer could download it in the mean time. Somewhat imprecisely speaking, this takes place with probability~$\sum_{u \in \mathcal{T} : u_{i}=1} \lambda^{k}( u-e_{i},u) $ for a degree-$k$ peer and $\sum_{k \in \mathbb{N}} \pi(k)\sum_{u \in \mathcal{T} : u_{i}=1} \lambda^{k}( u-e_{i},u) $ for the global performance, respectively. This forms the basis for our further analysis of the buffer probabilities.


With the recurrence relation in place, we make another simplifying assumption. We assume that the  chunk selection policy of a degree-$k$ peer, $ \alpha^{k} ( i, u, v) $ for buffer index $i$, own buffer state $u$ and contacted peer's buffer state $v$ is agnostic of both $u$ and $v$, but rather attaches mass to buffer indices according as their relative importance. 
Call this simplified policy $s_k$, instead of $\alpha^{k}$. This assumption simplifies the problem to a great extent.
\begin{align*}
		\lambda^{k}(u,u+e_i) &= k \varsigma w_u^{k} \sum_{v \in \mathcal{T} : v_i=1} \sum_{l \in \mathbb{N}} q(l) w_v^{l} \alpha^{k} ( i, u, v) \\
		&=k \varsigma w_u^{k}  s_k ( i)  \sum_{l \in \mathbb{N}} q(l)  \sum_{v \in \mathcal{T} : v_i=1}w_v^{l}  \\
		&= k \varsigma w_u^{k}  s_k ( i)  \sum_{l \in \mathbb{N}} q(l)  p_l(i) \\
		&=k \varsigma w_u^{k}  s_k ( i) \theta_i \eqkomma
	\end{align*}
	where $i \in \mathcal{F} \setminus \{1\}$ and  $ \theta_i \defeq  \sum_{l \in \mathbb{N}} q(l)  p_l(i)$ encapsulates the probability that \quotes{an arbitrarily given edge points to a node 
		where chunk $i$ is available}.

Let us now revisit the recurrence relation and plug in the above simplified quantities. In order to do so, note that, for all $i \in \mathcal{F} \setminus \{1\}$,
\begin{align*}
	 \sum_{u \in \mathcal{T} : u_{i}=1} \lambda^{k}( u-e_{i},u)  & =   \sum_{v \in \mathcal{T} : v_{i}=0}  \lambda^{k}( v,v+e_{i})  \\
	&= k \varsigma \theta_i  s_k( i) \sum_{v \in \mathcal{T} : v_{i}=0}  w_v^{k}  \\
	&= k \varsigma \theta_i  (1- p_k(i) )s_k(i)
	\end{align*}

The recursion relation in  \cref{result:recurrence-from-master} then reads:
\begin{equation}
\label{equation:marginalProbRecursion}
p_k(i+1) = \twopartdef{p_k(i) +  k \varsigma \theta_i  (1- p_k(i) )s_k(i) }{i \neq 1}{p_k(1) +  \frac{1}{M} }{i=1}
\end{equation}
where  $i=1,2,\ldots,n-1 \text{ and } k \in \mathbb{N} $ and $\sum_{u \in \mathcal{T} : u_{1}=1} \lambda^{k}( u-e_{1},u) =  \frac{1}{M} $.

Such a recurrence relation in the special case of a homogeneous system has  served as a starting point for study of buffer probabilities in a number of articles in the literature, \eg, \cite{Zhou:P2Pmodel,zhou2007simple,ying2010asymptotic}. In fact, by choosing $\pi(k)=\indicator{k=k^*}, \varsigma= \frac{1}{k^*}$ for some $k^* \in \mathbb{N}$, we retrieve from \cref{equation:marginalProbRecursion} the corresponding recurrence relation in the homogeneous setup, as found in \cite{Zhou:P2Pmodel,zhou2007simple,ying2010asymptotic}. Our endeavour was to provide a principled approach to derive such a recurrence relation in a more general heterogenous setup maintaining degree dependence of peers.

An attentive reader might have noticed from \cref{result:recurrence-from-master} that 
$p_k(1)$ is $0$ for all $k \in \mathbb{N}$, an artefact of the way we have built the model with shifting of chunks at rate unity\footnote{ Refer to \cref{eq:stationary-distribution} to see why this is true. In reality, chunks are pushed onto the player by means of the shifting mechanism.}. This is because buffer index $1$ is solely reserved for the server and at each shifting, the chunk at buffer index $1$ is pushed to buffer index $2$ and buffer index $1$ is emptied. Therefore, $p_k(2)= \sum_{u \in \mathcal{T} : u_{1}=1} 	\lambda^{k} (u-e_1,u)  
=\frac{1}{M} $, total input to the system by the  server. 
Since the relevant buffer index where  peers can download and fill chunks starts from $2$, we rename this as index $1$ and set  $p_k(1)= \frac{1}{M}$. Notice that the quantity $p_k(1)= \frac{1}{M} \rightarrow 0$ as $M \rightarrow \infty$. 
This will be the case if  there are infinitely many peers and the server can serve only  finitely many of them. The probability that a peer has a chunk available at buffer index $1$ is linked to the capacity of server. In practice, this probability, however small,  can be assumed non-zero. We symbolically represent  this small non-zero probability by  $p_k(1)$ and use it as  boundary condition  in \cref{result:recurrence-from-master}.

		We do realise that these simplifying assumptions are not tenable in most finite-sized real systems and hence, the model is not an accurate one. However, we emphasize that the purpose of this exercise is to get some analytic intuition, rather than accurate computation of equilibrium buffer probabilities.

We shall now focus on the two popular chunk selection strategies, \viz, rarest first or the latest deadline first (LDF) and greedy strategy or the earliest deadline first (EDF). We follow the interpretations  laid down in \cite{Zhou:P2Pmodel}.

\section{Chunk selection function}
\subsection{Rarest-first strategy}
\label{subsubsection:RFstrategyChunkSelection}
This strategy aims to download the rarest piece first. The priority is thus on the initial buffer indices. Therefore, for rarest-first strategy, $s_k(i)$ can be written as
\begin{equation}
\label{equation:LDFchunckSelectionFunction}
s_k(i)= \left[  1- p_k(1) \right]   \prod_{j=1}^{i-1} \left[   p_k(j) + (1- p_k(j))(1-k \varsigma \theta_j) \right] \eqpunkt
\end{equation}
The explanation for the above is as follows: the event that a peer of degree $k$ in need of and having found chunk $i$ selects it from among all downloadable chunks is tantamount to the joint occurance of the events that the peer is not being served directly by the fixed source (call this event $A$), that chunk selection does not take place for all buffer locations prior to $i$, \ie, for all buffer locations $1,2,\ldots,i-1$ (call this event $B$) and that chunk selection  does take place at buffer location $i$ (call  this event $C$). As the server picks up a peer for direct upload unbiasedly, event $A$ takes place with probability $1-p_k(1)=1- \frac{1}{M}$. The event $B$ necessitates that for all buffer indices $1,2,\ldots,i-1$, the peer is either in possession of the chunk at that buffer index or  when in need of the chunk, it can not find it among its neighbours. Due to the independence of the buffer states, the probability of this event can be expressed as a product over the buffer indices, as shown in the above equation. The event  $C$ takes place with probability $1$.

\begin{myResult}
	\label[result]{result:LDFchunkSelectionFunction}
	\begin{enumerate}
		\item	The chunk selection function for rarest-first strategy can be expressed as
		\begin{equation}
		s_k(i) =1- p_k(i) \eqpunkt
		\end{equation}
		\item The recursion relation for buffer probabilities for rarest-first strategy has the following form
		\begin{equation}
		p_k(i+1)= p_k(i) + k \varsigma \theta_i (1-p_k(i))^2 \eqkomma
		\end{equation}
		for $ i=1,2,\ldots,n-1 \text{ and } k \in \mathbb{N} $.
	\end{enumerate}
\end{myResult}
The proof is similar to \cite{Zhou:P2Pmodel}, however, for the sake of completeness, it is provided in \Cref{appendix:chunkSelectionFunction}.

\begin{figure}
	\centering
	\caption{LDF and EDF strategies}
	\includegraphics[scale=0.2]{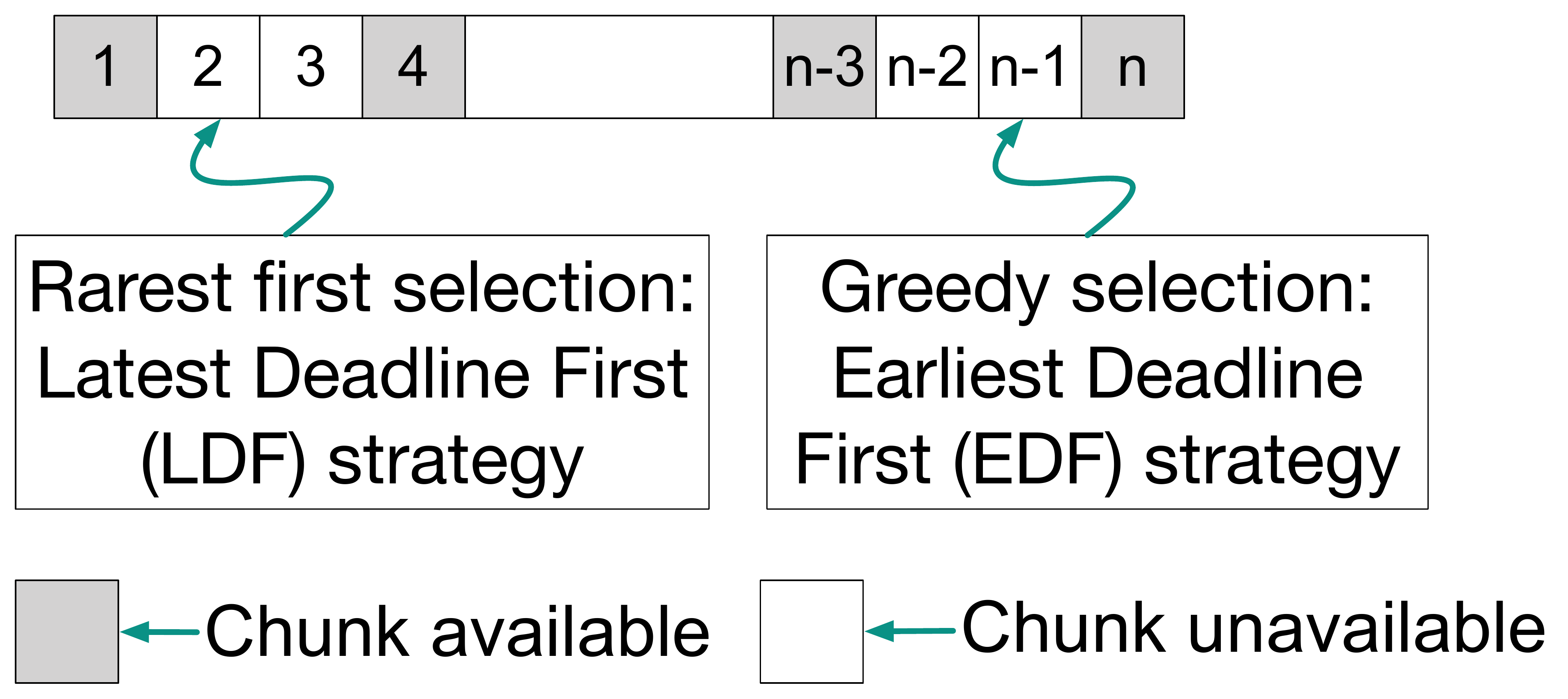}
\end{figure}

\subsection{Greedy strategy}
\label{subsubsection:EDFstrategyChunkSelection}
The greedy strategy seeks to download pieces that are close to playback. The priority is thus on playback urgency and hence towards the final buffer indices. Therefore, for the greedy strategy, the chunk selection function can be expressed as
\begin{equation}
\label{equation:EDFchunckSelectionFunction}
s_k(i)= \left[  1- p_k(1) \right]  \prod_{j=i+1}^{n-1}  \left[   p_k(j) + (1- p_k(j))(1-k \varsigma \theta_j)  \right] \eqpunkt
\end{equation}
The explanation is similar to the case of rarest-first strategy, with the notable exception that now we require to search buffer index $n$ first, then $n-1$ and so on.

\begin{myResult}
	\label[result]{result:EDFchunkSelectionFunction}
	\begin{enumerate}
		\item	The chunk selection function for greedy strategy can be expressed as
		\begin{equation}
		s_k(i) =1- p_k(1) -p_k(n) +p_k(i+1) \eqpunkt
		\end{equation}
		\item The recursion relation for buffer probabilities for greedy strategy has the following form
		\begin{align}
			\begin{split}
				p_k(i+1) &= p_k(i) + k \varsigma \theta_i (1-p_k(i))
				 \left[  1- p_k(1) -p_k(n) +p_k(i+1) \right]\eqkomma
			\end{split}
		\end{align}
		for $ i=1,2,\ldots,n-1 \text{ and } k \in \mathbb{N} $.
	\end{enumerate}
\end{myResult}
The proof is provided in \Cref{sec:appendix}.

\begin{myRemark}
	A typical EDF buffer probability curve exhibits a late sharp increase, contrary to an LDF curve (\cite{Zhou:P2Pmodel,zhou2007simple}). However, when $M$ is large, EDF hinders propagation of new chunks.	While LDF is known to possess good scalability, EDF outperforms LDF  when  $M$ is small. We wish to exploit this feature of EDF even when $M$ is large. In order to do so, we must devise a way to ensure circulation of new chunks. We surmise this can be done by employing strong peers to play LDF so as to act as \emph{pseudo-servers} in the system. Next we pursue this supposition by studying different strategy profiles.
	\label[remark]{remark:supposition}
\end{myRemark}

\par
To keep our premise simple, we consider a 2-degree system. Suppose there are only two degrees $k_1,k_2 \in \mathbb{N}$ in the system where $k_1 < k_2$. For typographical convenience, we shall subscript all the relevant variables with only $1,2$ instead of $k_1, k_2$ respectively, whenever the degree of a vertex appears as a subscript or as an argument to a function, \eg, $\pi_1,\pi_2$ in place of $\pi(k_1),\pi(k_2)$ respectively and $p_1(i),p_2(i)$ in place of $p_{k_1}(i), p_{k_2}(i)$ respectively. Also, to set a convention, call the peers of higher degree ($k_2$ in this case) \quotes{strong peers}, and peers of lower degree ($k_1$ in this case), \quotes{weak peers}.

\section{Pure LDF strategy}
\begin{figure}[!t]
\centering
	\begin{subfigure}[b]{.45\columnwidth}
		\includegraphics{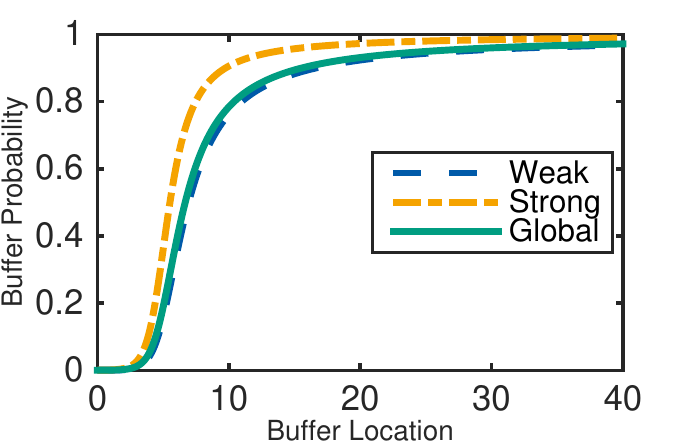}
		\caption{Buffer probabilities under  LDF }
	\end{subfigure}
	\hfill
	\begin{subfigure}[b]{.45\columnwidth}
		\includegraphics{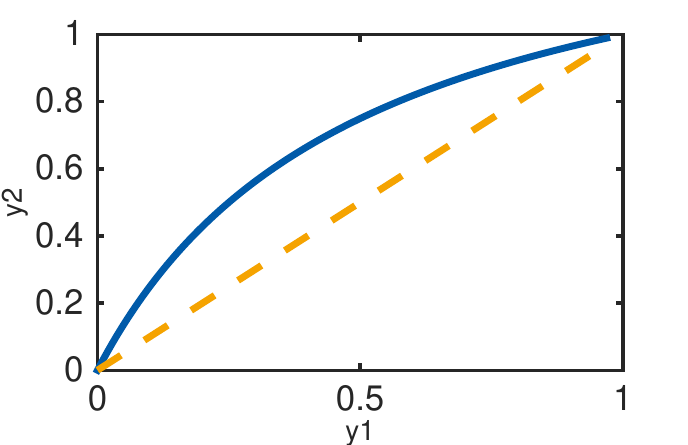}
		\caption{Weak versus strong peers under LDF}
		\label{fig:y1vsy2LDFODE}
	\end{subfigure}
	\caption{Performance under pure LDF strategy}
\end{figure}

As seen in \cref{subsubsection:RFstrategyChunkSelection}, the buffer probabilities for the two degrees $k_1,k_2$ in the system when everybody plays LDF, are given by the following recursion relations:
\begin{equation}
	\begin{split}
p_1(i+1) = p_1(i) + k_1 \varsigma \theta_i (1-p_1(i))^2  \eqkomma \\
p_2(i+1) =  p_2(i) + k_2 \varsigma \theta_i (1-p_2(i))^2 \eqkomma
	\end{split}
\end{equation}
	for $ i=1,2,\ldots,n-1$. To study their behaviour, we adopt continuous approximation of the above two difference equations. Treating the buffer index $i$ as a continuous variable and writing $y, y_1,y_2,\theta$ for $p(i),p_1(i),p_2(i) $ and $ \theta_i$ respectively, we have the following  differential equations:
\begin{align}
	\label[equation]{eq:pureLDF_ODE}
	\begin{split}
\frac{\, dy_1}{\, dx} &= k_1 \varsigma \theta (1-y_1)^2 \eqkomma \\
\frac{\, dy_2}{\, dx} & = k_2 \varsigma \theta (1-y_2)^2 \eqkomma \\
\frac{\, dy}{\, dx} &=  \pi_1 \frac{\, dy_1}{\, dx}+ \pi_2\frac{\, dy_2}{\, dx} \eqpunkt
	\end{split}
\end{align}
The above luckily allows an exact solution which we present in the next result.
\begin{myResult}
	\label[result]{result:comparisonLDF}
	For the pure LDF strategy and large systems, \ie, when $M$ is large, 
	the two buffer probabilities are related according to the following equation
	\begin{equation}
	y_2= \frac{y_1}{1-(1-r)(1-y_1)} \eqkomma
	\label{eq:y1vsy2relationPureLDF}
	\end{equation}
	where $r=\frac{k_1}{k_2}$. 
\end{myResult}
The proof is given in \Cref{proof:comparisonLDF}. As an immediate observation, we see that $y_2>y_1$, \ie, the stronger peers have better performance irrespective of buffer length. This gain in buffer probability is due to their  greater rate of interaction. 
However, this difference in performance for the weak peers due to degree disparity can be made arbitrarily small if sufficiently large buffer length is made available. \Cref{fig:y1vsy2LDFODE} shows a plot of $y_1$ and $y_2$. Another interesting consequence is that the above can now be used to derive an expression for buffer-size requirements and facilitate sensitivity analysis therefrom. We have the following result in that direction.

\begin{myResult}
	\label[result]{result:LDFbuffersizeRequirement}
	For large systems playing pure LDF strategy, the buffer size requirement for the weaker peer corresponding to a desired level of continuity probability $1-\epsilon_1$, where $\epsilon_1=1-p_1(n)$, is given by
		\begin{align*}
	n_1 &= \frac{A}{k_1} \ln\left(\frac{1-\epsilon_1}{p(1)}\right)+ B  \ln\left( \frac{1-p(1)}{\epsilon_1} \right) + \frac{C}{\epsilon_1}
 + \frac{D}{q_1(1-r)} \ln\left( \frac{1+E (1-\epsilon_1)}{1+E (1-p(1))} \right)\\
 &{}\quad  - \frac{C-1+p(1)}{1-p(1)}   \eqkomma
	\end{align*}
where $A=\frac{r}{q_1r+q_2}, B= \frac{r}{k_1 \varsigma (q_1r+q_2)} \left( \frac{1}{1-r} + \frac{q_1^2 q_2 r(1-r)}{1+q_1r}\right) + \frac{1}{k_1 \varsigma r}\left( 1- \frac{1-r^2}{1+q_1r} \right) , C= \frac{1}{k_1 \varsigma }, D= \frac{q_1^2 q_2 (1-r)^3}{k_1 \varsigma (q_1r+q_2)}$ and $E=\frac{q_1(1-r)}{q_1r+q_2}$.
\end{myResult}
Proof is shown in \Cref{proof:LDFbuffersizeRequirement}. The above can also be used when we intend to achieve a prespecified level of global performance, $1-\epsilon= p(n)$. Recall that the global performance is functionally related to $\epsilon_1$ by:
\begin{equation*}
1-\epsilon= \pi_1 (1-\epsilon_1) + \pi_2 \frac{1-\epsilon_1}{1- (1-r) \epsilon_1} \eqpunkt
\end{equation*}
The above expression for buffer-size requirement also paves way for a number of sensitivity analyses, \eg, by means of $\frac{\, dn_1}{\, d\epsilon_1}$
we gauge how much additional buffer space would be required if we wanted a marginal increase in performance. Similarly, $\frac{\, dn_1}{\, dp(1)}$ can be used to infer how much additional buffer space would be required to maintain the same continuity as population size increases\footnote{$p(1)=\frac{1}{M}$ enables us to compute $ \frac{\, dn_1}{\, dM}$ from  $\frac{\, dn_1}{\, dp(1)}$} and hence, about the scalability of the system.

\section{Mixed strategy: \sys{SchedMix}}
\begin{figure}[htbp]
\centering
	\begin{subfigure}[b]{.48\textwidth}
	\centering
			\includegraphics{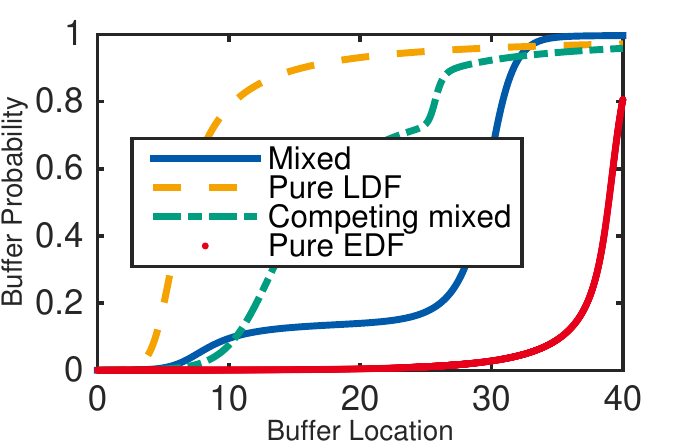}
		\caption{Global performance comparison under LDF and mixed strategy}
		\label{fig:LDFvsMixedODE}
	\end{subfigure}
	\begin{subfigure}[b]{.48\textwidth}
	\centering
			\includegraphics{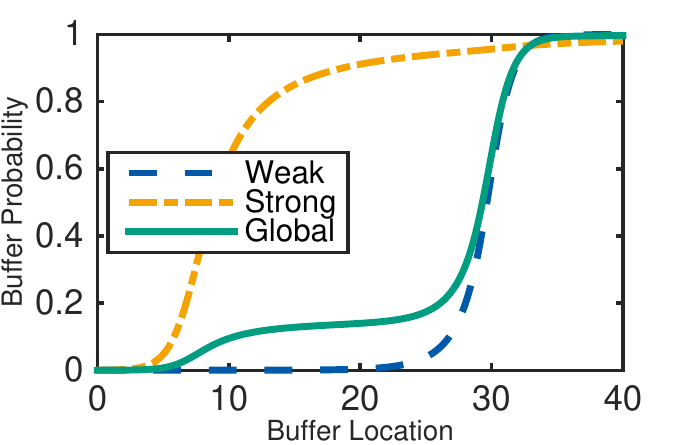}
		\caption{Performance comparison of weak and strong peers under mixed strategy}
		\label{fig:WeakvsStrongMixedODE}
	\end{subfigure}
	\caption{Performance comparison}
\end{figure}

Now we turn to the mixed strategy. Suppose the weaker peers of degree $k_1$ adopt EDF and the stronger peers of degree $k_2$, LDF. As shown in \Cref{subsubsection:RFstrategyChunkSelection,subsubsection:EDFstrategyChunkSelection}, 
the equilibrium probabilities are given by the following recursion relations:
\begin{equation}
	\begin{split}
p_1(i+1) &= p_1(i) + k_1 \varsigma \theta_i (1-p_1(i))
 \left[ 1- p_1(1) -p_1(n) +p_1(i+1) )\right] \eqkomma \\
p_2(i+1) &= p_2(i) + k_2 \varsigma \theta_i (1-p_2(i))^2 \eqkomma
  \end{split}
\end{equation}
	for $ i=1,2,\ldots,n-1$. As before, we shall use continuous approximation to study their behaviour. So, writing $\epsilon_1=1-p_1(n)$, we have the following differential equations:
\begin{equation}
	\label[equation]{eq:mixed_ODE}
		\begin{split}
		\frac{\, dy_1}{\, dx} &= \frac{k_1 \varsigma \theta (1-y_1) (y_1-p_1(1) + \epsilon_1)}{1-k_1 \varsigma \theta (1-y_1)  } \eqkomma \\
		\frac{\, dy_2}{\, dx} &= k_2 \varsigma \theta (1-y_2)^2 \eqkomma \\
		\frac{\, dy}{\, dx} &=\pi_1 \frac{\, dy_1}{\, dx}+ \pi_2\frac{\, dy_2}{\, dx} \eqpunkt
		\end{split}
\end{equation}
The above equations, unfortunately, do not yield an exact solution. So, an attempt to derive closed-form expression for buffer-size requirement  is fruitless. Therefore, we resort to numerical solution to compare  global performance of the system under pure LDF and mixed strategies. It turns out that performance under the mixed startegy is indeed better than that under pure LDF strategy (\vide  \Cref{fig:LDFvsMixedODE}), which also corroborates our claim.

When we compared the performance of the weak peers \emph{vis-\`{a}-vis} the strong ones, an interesting phenomenon was observed. The buffer probability curve for weak peers expectedly remained below the curve for strong peers for the initial buffer indices, but the weak peers could eventually manage to outperform the strong ones even for moderate buffer-lengths, buoyed by a sharp increase in buffer probabilities that a typical \quotes{EDF curve} enjoys and what we call the \quotes{boon of heterogeneity} (\vide \Cref{fig:WeakvsStrongMixedODE}). This phenomenon is in agreement with our supposition in \cref{remark:supposition} and can be explained intuitively. Both strong and weak peers playing LDF and EDF respectively benefit immensely from being exposed to a heterogeneous environment. As a peer trying to fill the buffer as much as possible, when exposed to a homegeneous environment, be it LDF or EDF, one would expect somewhat similar availability of chunks among all its neighbours. Indeed, in a homogeneous environment, all the peers have the same buffer probabilities. On the contrary, when exposed to a hetergoneous environment, if a peer requires a chunk in one of the initial buffer locations, it has a higher probability of downloading it from an LDF-playing neighbour. Similarly, for a missing chunk close to playback, one would preferably contact an EDF-playing neighbour. Thus, a peer can fill larger part of the buffer when exposed to a hetergeneous environment, for it makes available a diverse set of chunks. We call this phenomenon the \quotes{boon of heterogeneity}. This 
prepones the steep rise that a typical \quotes{EDF curve} enjoys. Since an EDF curve has a greater growth-rate in the neighbourhood of $1$ (see \cite{Zhou:P2Pmodel,zhou2007simple}), the weaker peers eventually outperform the stronger peers playing LDF even for moderate buffer-lengths.

To get an analytic understanding of this phenomenon, we make a very crude, but  \quotes{conservative} approximation\footnote{Conservative in the sense that the approximate differential equation would underestimate the buffer probabilities of the weak peers, \ie, $ \frac{\, dy_1}{\, dx} > \frac{\, dy_1^{approx}}{\, dx}$. }. Ignoring all second and higher order terms of $k_1 \varsigma \theta (1-y_1)$ in the expansion of $1/{(1- k_1 \varsigma \theta (1-y_1))}$ that appears in $\frac{\, dy_1}{\, dx}$ yields the following exact solution
\begin{equation}
y_2= \frac{\frac{1}{r(1-p(1)+\epsilon_1)} \ln \left( \frac{y_1-p(1) + \epsilon_1}{1-y_1} \right) -C- 1}{\frac{1}{r(1-p(1)+\epsilon_1)} \ln \left( \frac{y_1-p(1) + \epsilon_1}{1-y_1} \right) -C} \eqkomma
\label{eq:y2vsMixed}
\end{equation}
where $C= \frac{1}{(1-p(1)+\epsilon_1)} \ln \left( \frac{\epsilon_1}{1-p(1)} \right) - \frac{1}{1-p(1)} $ and $r=\frac{k_1}{k_2}$ (proof shown in \Cref{derivation:y2vsy1Mixed}). As seen in the pure LDF case, the apparent shortcoming of degree-disparity can be overcome if suffieciently large buffer is made available. Rewriting \Cref{eq:y2vsMixed} and setting $\epsilon_1=p(1)$, a legitimate choice given sufficiently large buffer, we get the following simplified expression
\begin{equation}
y_1= \frac{1}{1+ e^{-r (\frac{1}{1-y_2} + C_0)}} \eqkomma
\label{eq:y2vsy1MixedSimplified}
\end{equation}
where $C_0= \frac{1}{r} \ln \left( \frac{p(1)}{1-p(1)} \right) -\frac{1}{1-p(1)}  $.  Fortunately  \Cref{eq:y2vsy1MixedSimplified} is simple enough to derive analytic intuition into its behaviour. Consider two functions $f_1(z)=z, f_2(z)= 1+ e^{-r (\frac{1}{1-z} + C_0)} $ for $z \in (0,1)$. As $z \uparrow 1$, we observe $f_1 \uparrow 1$ and $f_2 \downarrow 1$. Moreover, expanding $\frac{1}{1-z}=1+z+z^2+\ldots$, we see that $f_2 \downarrow 1$ exponentially fast as $z \uparrow 1$. Since the decay of $f_2$ is faster than the growth of $f_1$, there must be a point $z$ in the neighbourhood of $1$ such that $f_1(z)f_2(z) <1$, \ie, $f_1(z)<1/f_2(z) $. This precisely implies that there must be a point where the weak peers outplay the strong peers.

\begin{figure}[!t]
\centering
		\includegraphics{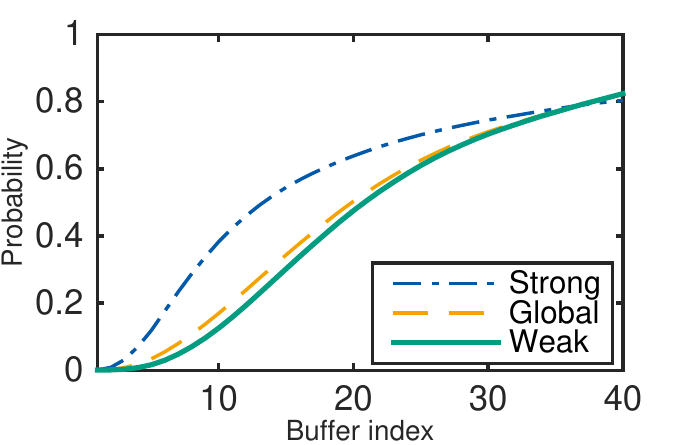}
	\caption{$y_1$ vs. $y_2$ under mixed strategy from the continuous time model \Cref{sec:model} on Watts-Strogatz graph. Relevant parameters are: $M=2000, n=40, \varsigma= 0.25 \eqpunkt$}
	\label[figure]{fig:y1vsy2MixedActual}
\end{figure}

We verified this phenomenon of the weaker peers outperforming the stronger ones (see \Cref{fig:y1vsy2MixedActual}) in the \quotes{more} realistic continuous time model described in \Cref{sec:model}.

\section{Stability analysis}
Here we briefly touch upon the stability considerations for the system of differential equations developed in earlier sections, \viz,
\cref{eq:pureLDF_ODE} and \cref{eq:mixed_ODE}. This exercise is only intended to be elucidative of the system's behaviour when sufficiently large buffers are made available. Please refer to \cite{strogatz2014nonlinear}  for a leisurely read.

Let us first take up the pure LDF strategy, \cref{eq:pureLDF_ODE}. Rewriting $	\frac{\, dy_1}{\, dx} = f_1(y_1,y_2), 	\frac{\, dy_2}{\, dx} = f_2(y_1,y_2)$, we evaluate the \emph{Jacobian} matrix $\mathrm{J}=  \bigg( \bigg( \frac{\, df_i}{\, dy_j}  \bigg) \bigg)_{i,j=1,2}$  and study its eigen values. 
At $(1,1)$, both the eigen values of $\mathrm{J}$ are $0$, implying \emph{doubly degenerate equilibrium} of the system around $(1,1)$.

The more interesting case is the mixed strategy, \cref{eq:mixed_ODE}. Proceeding as before, we get
\begin{equation}
\mathrm{J} \Big|_{(1,1)}=  \left( \begin{matrix}
-k_1 \varsigma (1-y_0) & 0\\
0 & 0 \\
\end{matrix} \right)  \eqkomma
\end{equation}
with eigen values $0$ and $-k_1 \varsigma (1-y_0)$. This implies a singly degenerate equilibrium around $(1,1)$. Moreover, this tells us that the EDF-peers will continue to grow at an exponential rate, while the LDF-peers will not have any growth around $(1,1)$. Thus, the EDF-peers must outsmart the LDF-peers eventually. This further corroborates our suppsotion regarding the weak peers eventually outperforming the strong ones under the mixed strategy.



\newpage
\chapter{Game theoretic argument}
\label{sec:game-theoretic-consideration}
In this section, we attempt to bring in a game theoretic perspective for the strategies by pitting  weak peers against  strong ones. Let $\mathcal{S}_1=\mathcal{S}_2=\{\text{LDF},\text{EDF}\}$ be the strategy profiles of  weak and  strong peers, respectively and $\mathcal{S} \defeq \mathcal{S}_1 \times \mathcal{S}_2$ denote the set of all possible strategy vectors. Also define utility functions $u_i : \mathcal{S} \rightarrow \mathbb{R}$ as the playback continuity probability for a fixed buffer length $n \in \mathbb{N}$, \ie, $u_i(\varphi) \defeq p_i^\varphi(n) \text{ for } \varphi \in \mathcal{S} \text{ and } i=1,2$, where~$p_i^\varphi : \mathbb{N} \rightarrow ~[0,1]$ is the buffer probability under strategy vector $\varphi$. As a convention, for $\varphi \in \mathcal{S}$, denote by $\varphi_i $ the strategy of player $i$ and by $\varphi_{-i}$, the vector of strategies of all other players except $i$. A strategy vector $\varphi \in \mathcal{S}$ is defined to be a \quotes{Nash equilibrium} \cite{nisan2007algorithmic} if for all players $i$ and each alternate strategy $\varphi_i' \in \mathcal{S}_i$, we have that
\begin{equation}
\label{eq:Nash}
u_i(\varphi_i, \varphi_{-i}) \geq u_i(\varphi_i', \varphi_{-i} ) \eqpunkt
\end{equation}
We claim that (EDF, LDF) is a Nash equilibrium.

\begin{figure}[htbp]
	\centering
		\begin{subfigure}[b]{.48\textwidth}
			\includegraphics[scale=0.35]{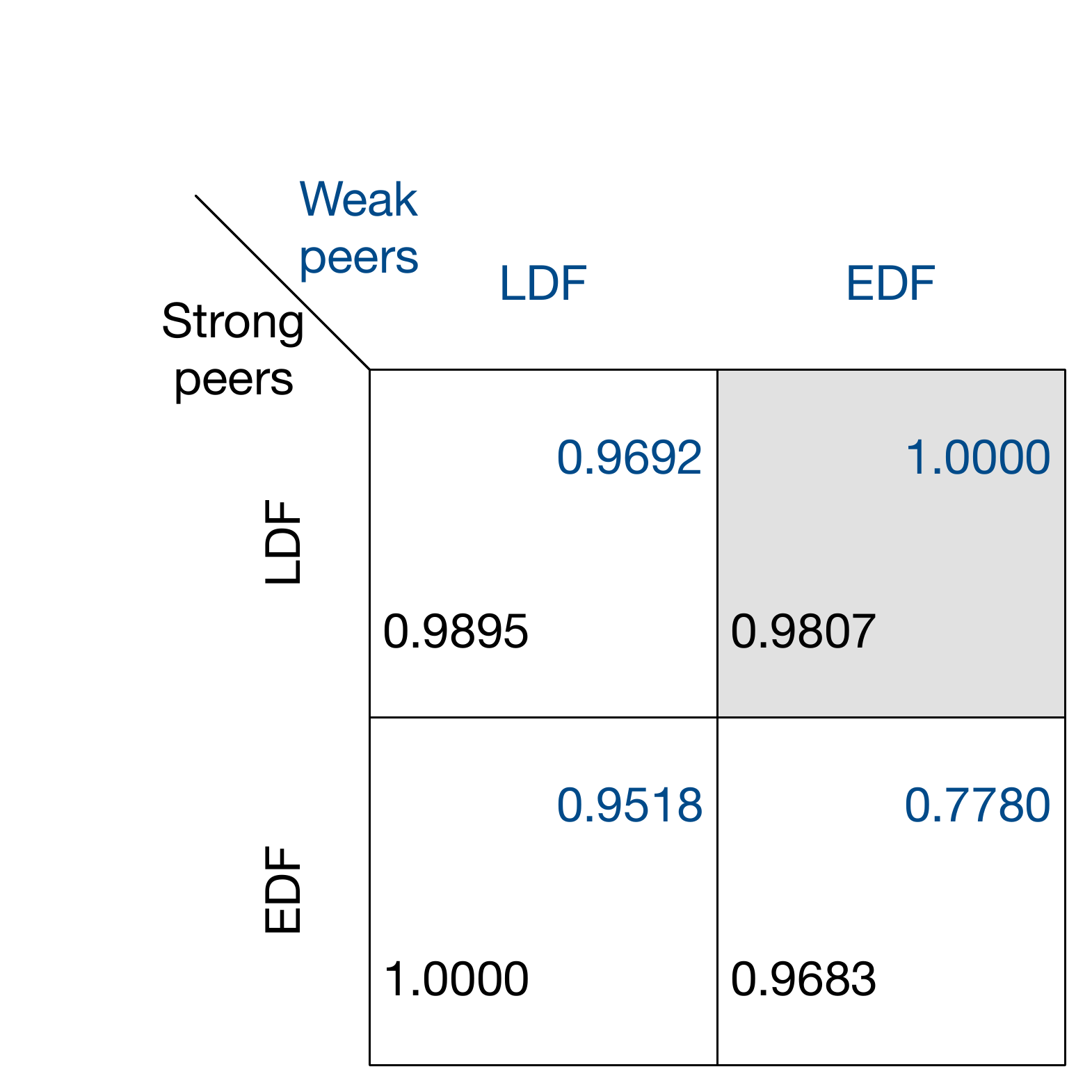}
			\caption{Payoff table, calculated from exact ODEs for buffer length $n=40, k_1=25, k_2=55, \pi_1=0.85=1-\pi_2, M=1000$.}
			\label{fig:PlayoffTable}
		\end{subfigure}
	\begin{subfigure}[b]{.48\textwidth}
			\includegraphics{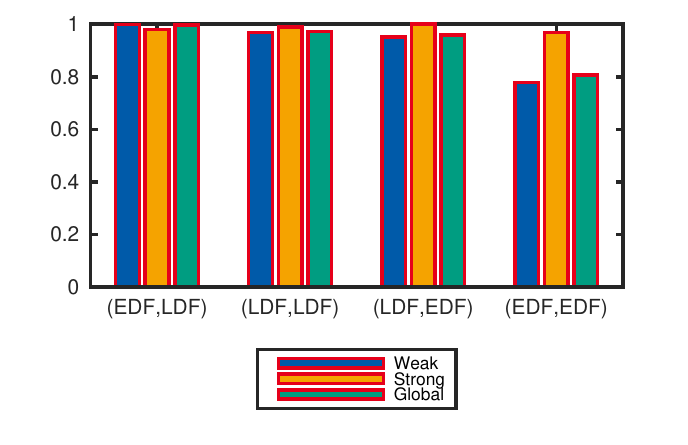}
		\caption{Bar plot of performance of weak, strong and global performance}
		\label{fig:BarPlot}
		\end{subfigure}
	\caption{Comparison of all possible strategy vectors in $\mathcal{S}$ }
\end{figure}

Let us first argue about the weak peers' strategy against  LDF-playing strong peers. We have seen that, from the perspective of weak peers, LDF is not a \quotes{rational strategy}, because under the pure LDF strategy, the weak remains weak. On the other hand, weak peers can eventually outperform the LDF-playing strong peers if they are greedy. From the perspective of strong peers,  they are better-off playing LDF against EDF-playing weak peers, establishing that (EDF, LDF) is a \quotes{Nash equilibrium}. For an illustration of these arguments, refer to \Cref{fig:PlayoffTable,fig:BarPlot} and verify that \Cref{eq:Nash} is indeed satisfied for (EDF, LDF) $\in \mathcal{S}$.

It does merit some attention that the strategy vector (LDF, EDF) is also \emph{Nash}. This explains
the \quotes{boon of heterogeneity} phenomenon that we explained earlier. However, this is a suboptimal strategy
vector (\vide the yellow bars in \Cref{fig:BarPlot}). When the strong peers play LDF, they act as \emph{pseudo-sources} and facilitate propagation of
rarest chunks. That is why (EDF, LDF) is more beneficial from the perspective of global performance. It must also be noted
that the utility functions $u_i$'s depend on different choices of $n, k_1, k_2, \pi_1=1-\pi_2, M$, therefore, \Cref{fig:PlayoffTable,fig:BarPlot} should only
be deemed as an illustration of the game theoretic arguments in a realistic setup with moderate buffer size. Extreme parameter choices, \eg, $n\rightarrow \infty, \frac{\pi_1}{\pi_2} \rightarrow \infty $ do not catch our fancy and hence, are excluded from consideration.

\newpage

\chapter{Simulation of the stochastic model}
\label{sec:simulation-results}

Simulation of the stochastic model as described in  \cref{sec:model} is carried out in two steps: first, generation of a random graph and second, simulation of the content delivery process. We confine ourselves to Barab\'{a}si-Albert preferential attachment~\cite{BarabasiAlbertScaling} and Watts-Strogatz \quotes{small world}~\cite{WattsStrogatzSmallWorlds} networks for the purpose of simulation.

\begin{figure}[htbp]
	\centering
	\begin{subfigure}[t]{.3\textwidth}
		\centering
		\includegraphics{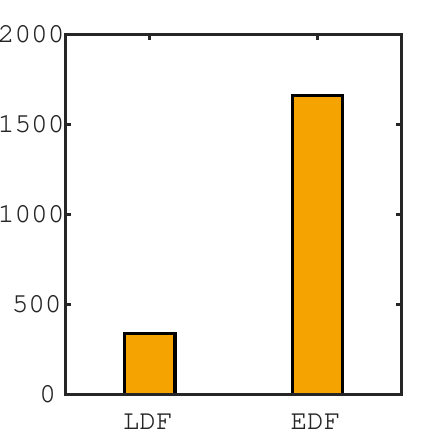}
		\caption{Distribution of EDF and LDF peers}
		\label{fig:BA_EDF_LDF_dist_exp}
	\end{subfigure}
	\begin{subfigure}[t]{.3\textwidth}
		\centering
		\includegraphics{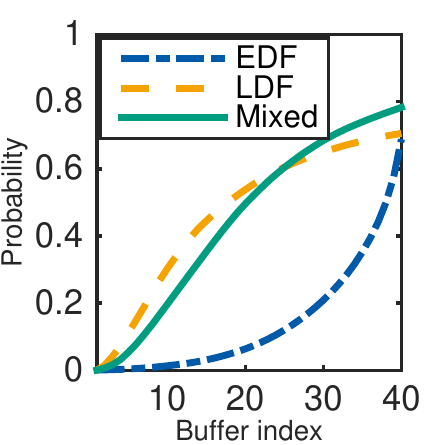}
		\caption{Buffer probablities}
		\label{fig:BABufferProbabilities_exp}
	\end{subfigure}
	\begin{subfigure}[t]{.3\textwidth}
		\centering
		\includegraphics{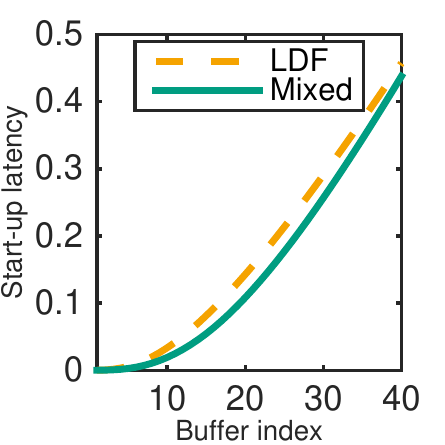}
		\caption{Start-up latency}
		\label{fig:BAStartup_latency_exp}
	\end{subfigure}

	\caption{Performance on a Barab\'{a}si-Albert preferential attachment graph. Here shifting takes place at each tick of a Poisson clock with rate unity. Relevant parameters are: $M=2000, n=40, \varsigma=0.25$.}
	\label{fig:performance-on-BA_exp}
\end{figure}

We begin with a random sample from $\mathcal{G}_M$. A node with degree $k$ maintains a Poisson clock with rate $ k \varsigma $. Given this graph, we fix the buffer size $n$ and simulate the content delivery process in accordance with \cref{sec:model:p2pcommsys} separately for the three chunk selection policies, \viz,  pure LDF, pure EDF and mixed strategy. We initialise $X$ randomly. The server selects one of the peers uniformly at random to upload a chunk at buffer index $1$. It continues to do so, until there is a link breakage in which case it again picks one of the peers uniformly at random (could be the same as before) to upload chunks directly. Link breakage takes place with a small probability. At every tick of the Poisson clocks, a peer, not being served by the server, contacts one of its neighbours uniformly at random and seeks to download a missing chunk, as dictated by its policy.  To carry out this step, we first draw a random sample from a Poisson distribution with mean $k\varsigma$. This gives the number of times a degree-$k$ interacts with its neighbours in a unit amount of time. We repeat this for all peers except the one being directly served by the server. Finally all these interactions are randomly permuted. After unit amount of time, all the chunks are shifted one place to the right. This completes one step of the live video streaming process. We repeat the procedure a large number of times. After discarding adequate amount of initial simulations (burn-in phase) to ensure stationarity, we record the availability of chunks at each buffer index of each peer to compute buffer probabilities. 

The second metric that we look at is the start-up latency. It is the time a peer should wait before starting playback. While there is no unanimity as to how one should define the quantity in question, it seems reasonable to wait until a newly arrived peer's buffer attains a steady state. In a homogeneous set-up where everybody plays the same policy and has the same buffer probabilities, as argued in \cite{zhou2007simple}, this is well represented by $\sum_i p(i)$, the average number of available chunks at each peer. In our heterogeneous model, a higher degree peer interacts more often than a lower degree peer. Therefore, a newly arrived degree-$k$ peer should have start-up latency of $k \varsigma  \sum_i p(i)$ in the mean-field. The corresponding global metric is $\Eof{k} \varsigma \sum_i p(i)$. For aesthetic reasons, we normalise this quantity to $(0,1)$.


\begin{figure}[htbp]
	\centering
	\begin{subfigure}[t]{.3\textwidth}
		\centering
		\includegraphics{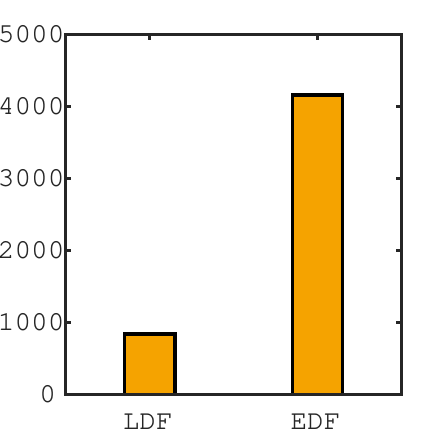}
		\caption{Distribution of EDF and LDF peers}
		\label{fig:BA_EDF_LDF_dist}
	\end{subfigure}
	\begin{subfigure}[t]{.3\textwidth}
		\centering
		\includegraphics{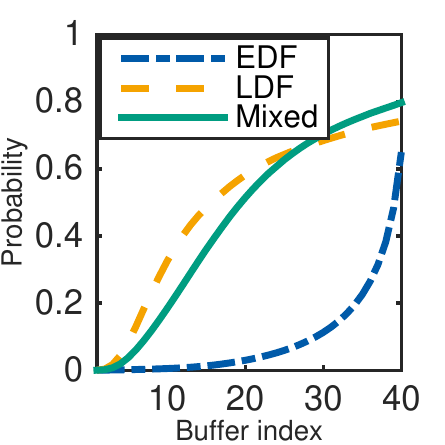}
		\caption{Buffer probablities}
		\label{fig:BABufferProbabilities}
	\end{subfigure}
	\begin{subfigure}[t]{.3\textwidth}
		\centering
		\includegraphics{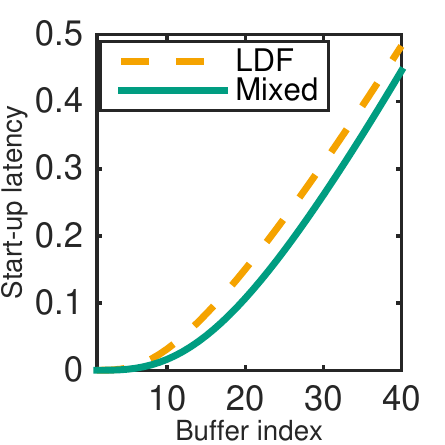}
		\caption{Start-up latency}
		\label{fig:BAStartup_latency}
	\end{subfigure}
	
	\caption{Performance on a Barab\'{a}si-Albert preferential attachment graph.
	Relevant parameters are: $M=5000, n=40, \varsigma=0.25$.}
	\label{fig:performance-on-BA}
\end{figure}

\section{Barab\'{a}si-Albert network}
Simulation results on Barab\'{a}si-Albert network are depicted in \cref{fig:performance-on-BA}. In \cref{fig:BA_EDF_LDF_dist}, we show the distribution of EDF and LDF-playing peers. The performance of mixed strategy pitted against pure LDF and pure EDF strategies are shown in \cref{fig:BABufferProbabilities}. The mixed strategy indeed gives a better performance, corroborating our claim. However, the gain in performance is not significantly higher than pure LDF strategy which is known to be a good strategy. The mixed strategy, however, effectuates handsome reduction in start-up latency (\vide \cref{fig:BAStartup_latency}).

\section{Watts-Strogatz network}
\begin{figure}[htbp]
	\centering
	\begin{subfigure}[t]{.3\textwidth}
		\centering
		\includegraphics{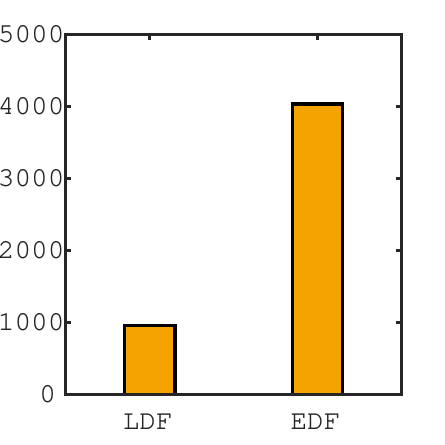}
		\caption{Distribution of EDF and LDF peers}
		\label{fig:WS_EDF_LDF_dist}
	\end{subfigure}	
	\begin{subfigure}[t]{.3\textwidth}
		\centering
		\includegraphics{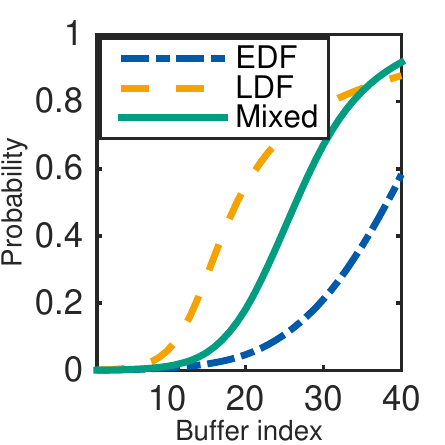}
		\caption{Buffer probablities}
		\label{fig:WSBufferProbabilities}
	\end{subfigure}
	\begin{subfigure}[t]{.3\textwidth}
		\centering
		\includegraphics{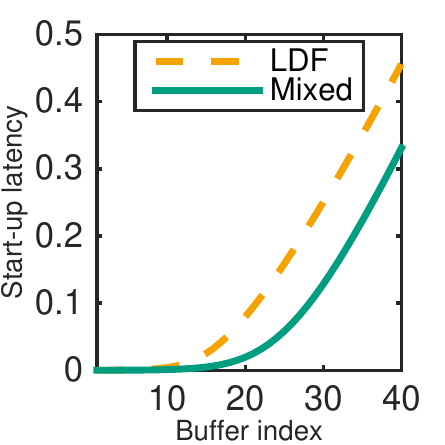}
		\caption{Start-up latency}
		\label{fig:WSStartup_latency}
	\end{subfigure}
	
	\caption{Performance on a Watts-Strogatz small world graph. Relevant parameters
	are: $M=5000, n=40, \varsigma=0.25$.}
	\label{fig:performance-on-WS}
\end{figure}

\Cref{fig:performance-on-WS} limns our findings on Watts-Strogatz network. In \cref{fig:WS_EDF_LDF_dist}, we show the distribution of EDF and LDF-playing peers. \Cref{fig:WSBufferProbabilities} portrays  performance of the three strategies. As earlier, giving credence to our claim, the mixed strategy indeed outperforms the other two, although the gain in performance is nominal. However, as seen in the case of Barab\'{a}si-Albert network, it does, to its credit, beget  weighty reduction in start-up latency (\vide \cref{fig:WSStartup_latency}).


\section*{Remarks}
Although both \cref{fig:performance-on-BA,fig:performance-on-WS} stand affirmatory to the fact that mixed strategy does outperform pure LDF and pure EDF strategies, even if marginally, the crux of employing the mixed strategy \sys{SchedMix} remains in letting most peers play \quotes{greedy}. This is tentamount to saying that the mixed strategy necessitates much smaller start-up latency to ensure good playback performance for everyone (at least as good as pure LDF strategy). This seems a significant benefit.

\begin{figure}[htbp]
	\centering
	
	\begin{subfigure}[t]{.3\textwidth}
		\centering
		\includegraphics{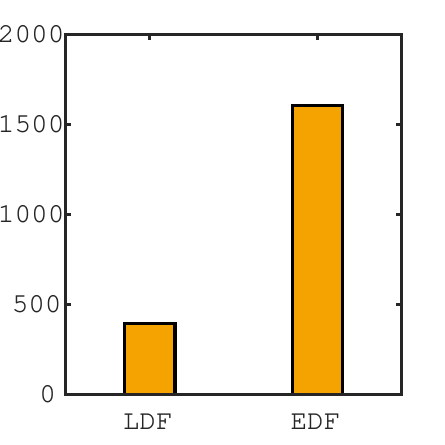}
		\caption{Distribution of EDF and LDF peers}
		\label{fig:WS_EDF_LDF_dist_exp}
	\end{subfigure}	
	\begin{subfigure}[t]{.3\textwidth}
		\centering
		\includegraphics{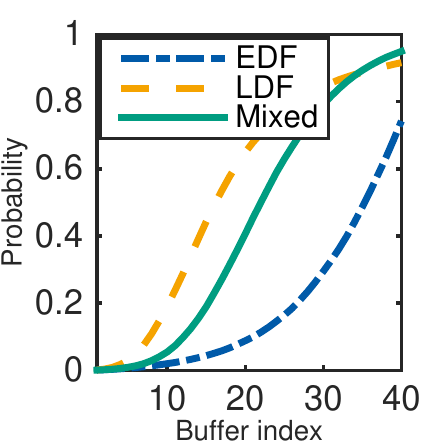}
		\caption{Buffer probablities}
		\label{fig:WSBufferProbabilities_exp}
	\end{subfigure}
	\begin{subfigure}[t]{.3\textwidth}
		\centering
		\includegraphics{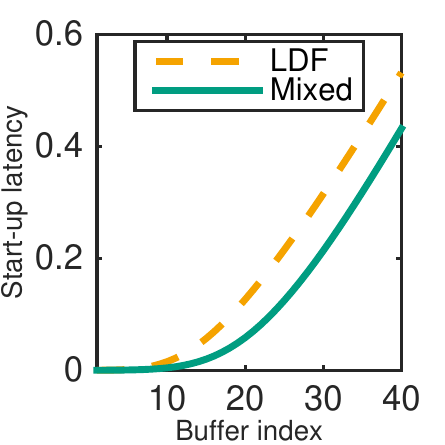}
		\caption{Start-up latency}
		\label{fig:WSStartup_latency_exp}
	\end{subfigure}
	
	\caption{Performance on a Watts-Strogatz small world graph. Here shifting takes place at each tick of a Poisson clock with rate unity. Relevant parameters are: $M=2000, n=40, \varsigma=0.25$.}
	\label{fig:performance-on-WS_exp}
\end{figure}

\section{Effect of shifting after exponential time}
In this section, we investigate the effect of assuming exponential shifting times instead of a deterministic shifting process. We run the system exactly as before except for the shifting which is also a stochastic process now, \viz, a Poisson process of rate unity. \Cref{fig:performance-on-BA_exp,fig:performance-on-WS_exp} respectively depict our findings on Barab\'{a}si-Albert and Watts-Strogatz graph. As can be seen from the figures, this assumption does not disturb the overall behaviour of buffer probabilities. Our claim about the mixed strategy also remains unaffected. 

\newpage
\chapter{Full-stack implementation}
\label{sec:full-stack}

\section{Practical system model}
Besides the previously introduced stochastic model, we additionally designed and implemented a more practical simulation model of the P2P streaming system to study the impact of a full-stack implementation of the proposed mechanism as well as communication network-related factors.
The simulation model is implemented based on the \sys{Simonstrator} API as proposed in~\cite{richerzhagen15}, making it compatible with the most recent version of the discrete-event-based network simulation framework \sys{PeerfactSim.KOM}~\cite{stingl2011}.

The simulation model implements a full-stack version of the proposed mechanism, in the sense that all mechanisms required for a real P2P-based stream delivery are realized.
This includes all protocols involved in the establishment and maintenance of a mesh-based streaming overlay, the implementation of the scheduling mechanisms themselves, the buffer management at individual clients as well as the streaming server, and the modeling of actual data transmissions over a resource-constraint network.
Due to space limitations, only key parts of the simulation model are presented in the following.
All other parts are based on standard approaches used in state-of-the-art P2P-based streaming systems as presented by the authors in previous works~\cite{wichtlhuber14a,rueckert15b}.

\subsection{Mesh establishment}
In contrast to the model described in Section~\ref{sec:model:p2pcommsys}, in a real implementation it is important to also provide mechanisms for the actual establishment of the mesh structure over time.
For this, a join procedure is implemented that uses a BitTorrent-like tracker as central node registry, which is queried for initial neighbor lists by new peers on joining the system.
The tracker selects a maximum of $30$ neighbors uniformly at random from the set of previously joined peers and sends the list to the joining peer.
The peer now strives to establish as many incoming connections to other peers as possible (the maximum number of active connections is a system parameter).
For this, the peer contacts peers from the initial list, where a maximum of $10$ parallel open connection requests are allowed.
Peers that receive a connection request, accept the request depending on their available resources.
For this, the available up- and download bandwidth of a peer is used to define a maximum number of up- and download connections respectively.
In combination with the used message-based communication and video data delivery, this approach limits the maximum number of parallel data transfers and thus the time for an individual video chunk to be delivered.
The maximum number of connections is calculated by dividing $90\%$ of the respective bandwidth by the video bitrate and flooring the result to the next integer.
In case a peer denies a connection request, the requesting peer blacklists the candidate for $60$ seconds, selects an alternative candidate from its local peer list, and tries to connect to it.
This process is repeated until either enough in-connections where established or no more candidates are available.
In the latter case, the peer retrieves a fresh list of potential neighbors from the tracker.

Peers in general accept all incoming connection requests by other peers, unless they do not have free out-connection slots.
In the latter case, requests are not simply denied but can still be accepted with a small probability to achieve two goals:
(1) making the established mesh topology more random as otherwise especially the very early peers tend to be blocked by other peers that joined shortly afterwards;
(2) to ensure that peers with more bandwidth to eventually establish more in-connections than weaker peers.
The second aspect showed to play an important role for applying the proposed mixed scheduling strategy that otherwise would not be able to leverage degree heterogeneity.
In case of no free out-connection slots and if a request is originated from a strong peer, the receiving peer replaces with a probability of $1/16$ one of its existing connections to a weaker peer.
If the request is received from a peer that does not belong to the strong peer class, a randomly chosen connection to any other peer is replaced with a probability of $1/64$ to achieve the first of the above goals.
The right choice for the two probabilities are to be evaluated as part of future work.
Yet, introducing them and assuring that string peers are accepted with a higher probability showed to result in the desired mesh connectivity as discussed based on the observed degree distributions in Section~\ref{subsec:eval-fullstack}.

\subsection{Scheduling and data exchange}
The actual scheduling of data transmissions is done by each peer individually based on its local clock and buffer status.
To account for the heterogeneity of peers and in line with the mathematical model, the rate at which chunks are requested from neighbors is proportional to the in-degree (i.e. the number of in-connections) of a peer.
This is reasonable as peers with higher degrees also have more capacity and thus can request chunks at a higher rate.
The delay between individual requests is calculated using \ref{eq:fullstack-sched-delay}, where $|\text{conn}_\text{in}|$ depicts the number of in-connections of the peer and $t_\text{base}$ is the \emph{request base interval}.

\begin{equation}
\text{delay}_\text{sched} = 1 / |\text{conn}_\text{in}| * t_{base}
\label{eq:fullstack-sched-delay}
\end{equation}

On each $\text{delay}_\text{sched}$ tick, a peer uses the scheduling strategy (e.g. EDF or LDF) to select the next chunks to be requested from a defined request window on the local buffer.
The size of this window ($|\text{req}_\text{win}|$) is used as upper limit for the number of chunks requested in a single tick, which showed to be necessary to account for the fact that chunk transmissions happen over a bandwidth-limited medium and previously requested chunks might still be delivered while new ones are requested.
In our simulation settings, $|\text{req}_\text{win}|$ is set to a default value of $20$, starting either at the beginning of the buffer for EDF or the end for LDF.
Only chunks that are locally unavailable and are not pending for delivery are considered.
Each of the chunks is assigned uniformly at random to one of the peer's in-connections.
Once all selected chunks of a tick are assigned, they are batched on a per-neighbor basis into a single chunks request message and send out.
The buffer has a length of $\text{bl}=4$ seconds, which translates to $50$ chunks at a rate of $8$ chunks per second.



\subsection{Playback policy}
For the playback of the video, a simplified policy was realized in this first version of the simulation model.
A peer learns about the current broadcasting position of the server when it initially contacts the tracker.
Due to the transmission delays of messages, this information is not assumed to be precise but close to the real position.
On receiving this information, the peer starts a timer that triggers its playback after $\text{bl}$ seconds.
In the meantime, the peer starts establishing connections to peers and requesting chunks as described above.
Once the playback was started, it constantly proceeds following the local clock of the peer, furthering the buffer chunk by chunk at the rate defined by the video bitrate.
In case the currently first chunk in the buffer is not available on playback, the playback is nevertheless assumed to proceed.
Playback deadline misses are recorded and reported in terms of the achieved \emph{playback continuity}.

\section{Simulation of the full-stack simulation model}
\label{subsec:eval-fullstack}
A simulative sensitivity analysis was conducted, covering key system and environment parameters.
The system parameters are listed in Table~\ref{tab:eval-system-parameters}, where default values are underlined.
Parameters were varied one at a time, while fixing the other parameters to their respective defaults.
The default system parameter settings were derived by carefully conducting calibration runs for several parameter combinations.
Finding the overall optimal configuration is hard to achieve as it would require exploring the overall configuration space, which is generally not feasible.
We acknowledge this fact and focus on studying the potential of the proposed scheduling mechanisms in a realistic setting.
All simulations presented in this section were repeated at least $30$ times with different random seeds and $95\%$-confidence intervals are reported for all mean values.

\begin{table}[ht]
	\centering
	\caption{Used system parameters and applied variations (default value is underlined).}
	\begin{tabularx}{0.5\textwidth}{p{5.25cm}p{3cm}}
		\toprule
		\textbf{Parameter} & \textbf{Variations} \\
		\midrule
		Request base interval($t_\text{base}$ [s])				& \underline{1}, 2, 3\\
		Request window size ($|\text{req}_\text{win}|$)			& 10, \underline{20}, 30, 40 \\

		Source upload capacity [Mbit/s]							& 2.5, 3.5, 6.5, \underline{12.5}, 24.5 \\
		\bottomrule
	\end{tabularx}
	\label{tab:eval-system-parameters}
\end{table}

\subsection*{Workload models}
For all workloads, the peers are divided into three resource classes based on the bandwidth distribution reported in~\cite{oecd14} for broadband access users (cf. Table~\ref{tab:eval-bw-distribution}).
We acknowledge that these bandwidths are rather high in comparison to configurations used in related works.
Yet, we intended to reflect a realistic setting in that the delivery of the video streams is not simply limited by the peer bandwidths but rather the content bottlenecks resulting from the scheduling strategy~\cite{guo08}.
As observed in~\cite{liang08}, the bandwidth of the source node plays an important role and, thus, its effect is studied in more detail to show its general effect in combination with the different scheduling strategies.

\begin{table}[ht]
	\centering
	\caption{Used peer bandwidth distribution based on~\cite{oecd14}.}
	\begin{tabularx}{0.5\textwidth}{p{1.7cm}p{1.25cm}p{1.26cm}p{1.3cm}p{1.5cm}}
		\toprule
		\textbf{Class} 	& \textbf{Number} 	& \textbf{Share} 	& \textbf{UL BW (Mbps)}	& \textbf{DL BW (Mbps)} \\ \midrule
		Source 			& $1$ 			& $-$ 				& $12.5$						& $12.5$ \\
		Low 			& $50$ 			& $50\%$ 			& $5$							& $26$ \\
		Medium 			& $30$ 			& $30\%$ 			& $4.5$							& $60$ \\
		High 			& $20$ 			& $20\%$ 			& $56$							& $134$ \\
		\bottomrule
	\end{tabularx}
	\label{tab:eval-bw-distribution}
\end{table}

The video bitrate was configured to be $1,500$~Kbps, which is a commonly observed bitrate as recently reported in~\cite{krishnappa15}.
At the beginning of the simulation scenario, the peers join the system with in a random order and a constant arrival rate until all peers are online.
After all peers joined the system, we leave the system some time to stabilize and then start recording a number of different performance and cost metrics.
They are obtained periodically every $60$ seconds of simulation time and on a per-peer basis.
Peers are assumed to stay in the system and not leave it until the end of the simulation.

In the following, first, results for the default configuration are presented, followed by a study of different essential system parameters.

\subsection{Results for default configuration}
\begin{figure}[htbp]
	\centering
	\begin{subfigure}[b]{.48\textwidth}
		\centering
		\includegraphics[height=150pt]{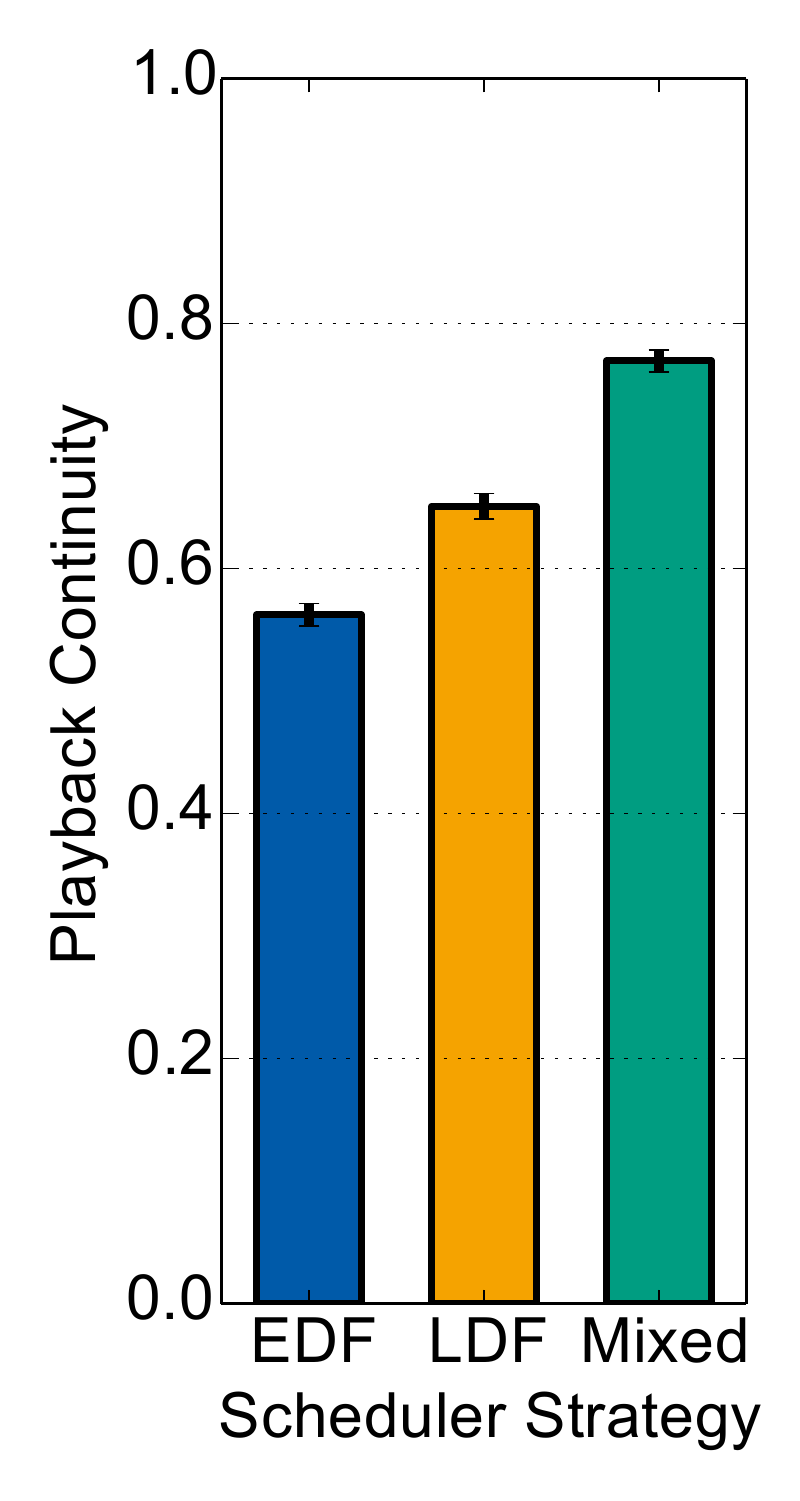}
		\includegraphics[height=150pt]{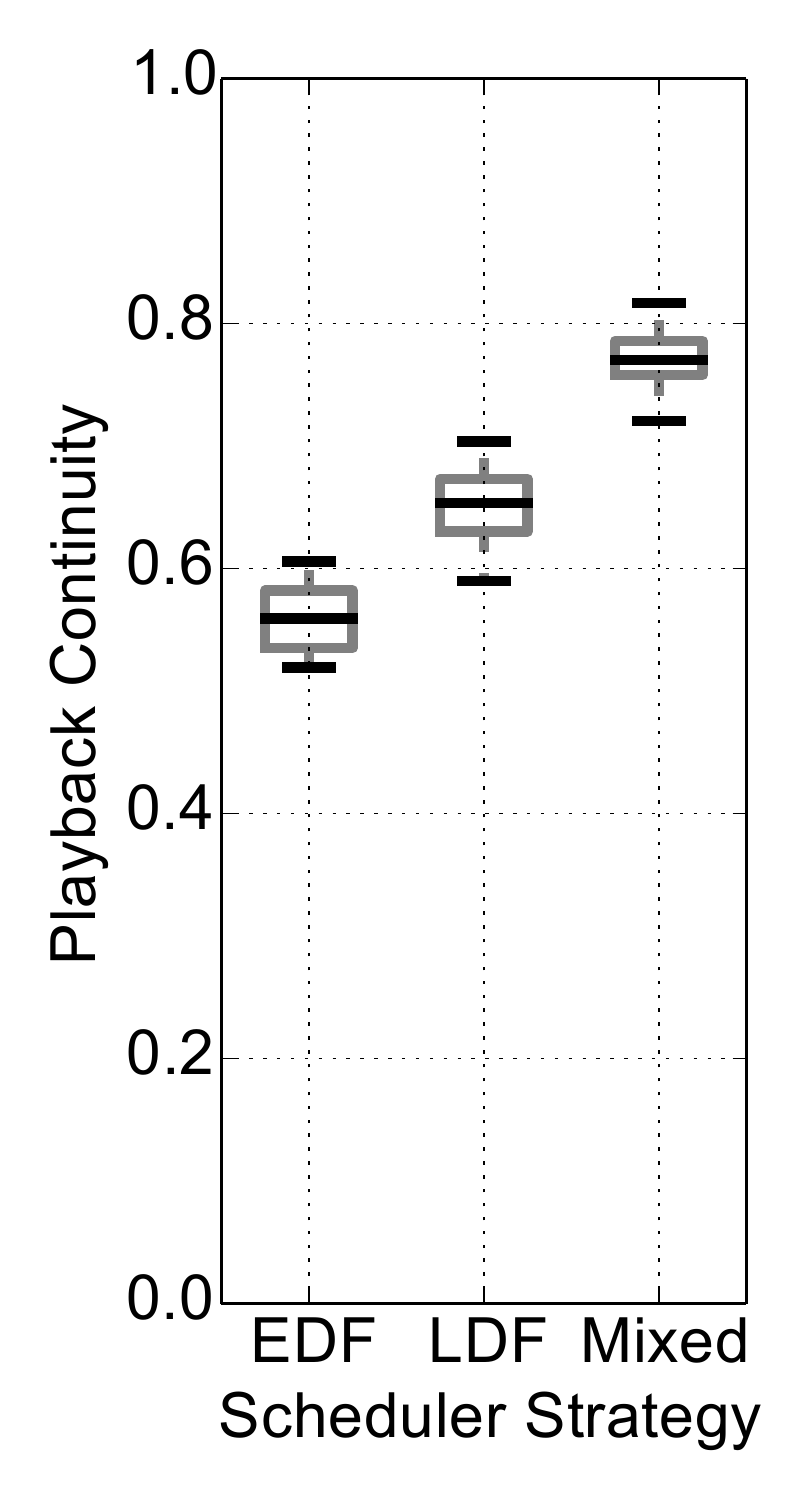}
		\caption{Playback continuity.}
		\label{fig:scheduler-a}
	\end{subfigure}
	\begin{subfigure}[b]{.48\textwidth}
		\centering
		\includegraphics[height=150pt]{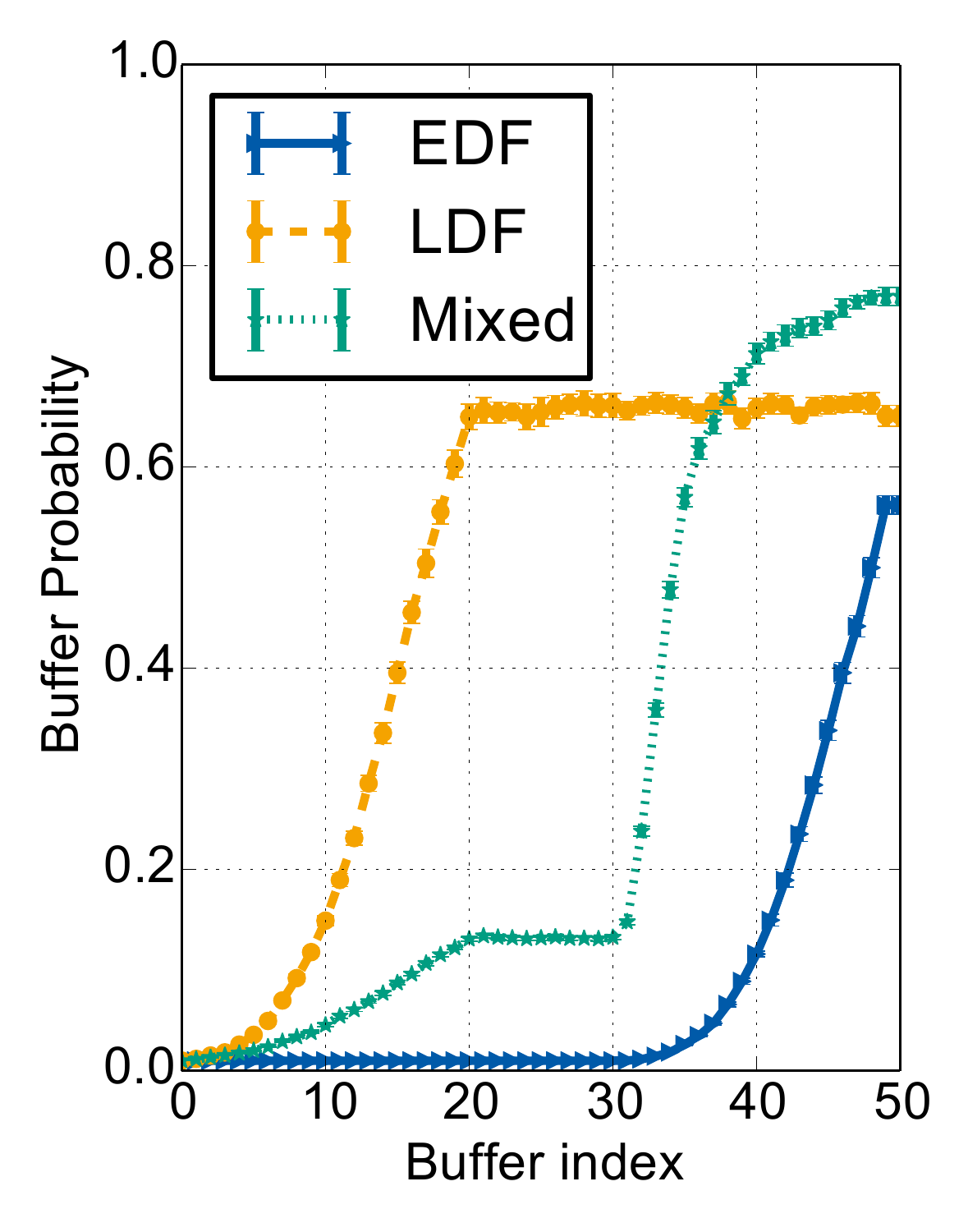}
		\caption{Buffer probability.}
		\label{fig:scheduler-b}
	\end{subfigure}

	\caption{Streaming performance and buffer characteristics.}
	\label{fig:scheduler-performance}
\end{figure}

Figure~\ref{fig:scheduler-performance} shows the streaming performance of the used default configuration.
Here, it is to note that using a request window of size 20 is considered an extreme case as it artificially limits the request rate by localizing the chunks to be selected to the beginning or the end of the buffer.
This is done to highlight the key difference between the chunk selection strategies.
Other configurations of this parameter are described later on.
The metric \emph{playback continuity} describes the relative average availability of the playback chunk (at buffer index 49), while the \emph{buffer probability} depicts the observed probability of all individual chunks in the buffer.
Here as well, buffer index 49 is the next playback chunk and index 0 depicts the end of a peer's local buffer.
It is clearly visible that the mixed strategy achieves a significantly higher playback continuity as both EDF and LDF.
The buffer probability shows that the early replication of newly broadcasted chunks greatly supports the greedy replication by the EDF peers once they enter their request window.

\begin{figure}[htbp]
	\centering
	\begin{subfigure}[b]{.45\textwidth}
		\centering
		\includegraphics[height=150pt]{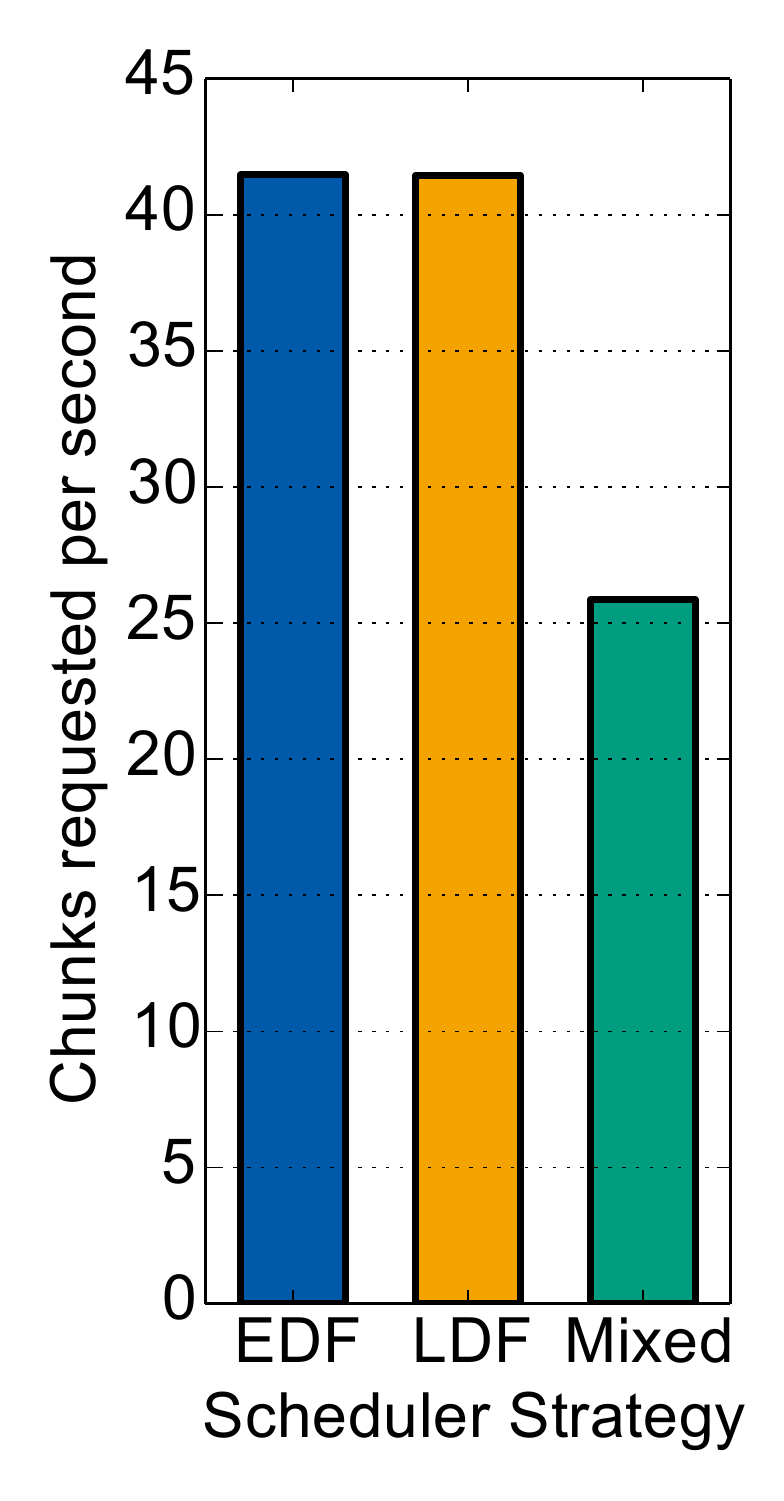}
		\includegraphics[height=150pt]{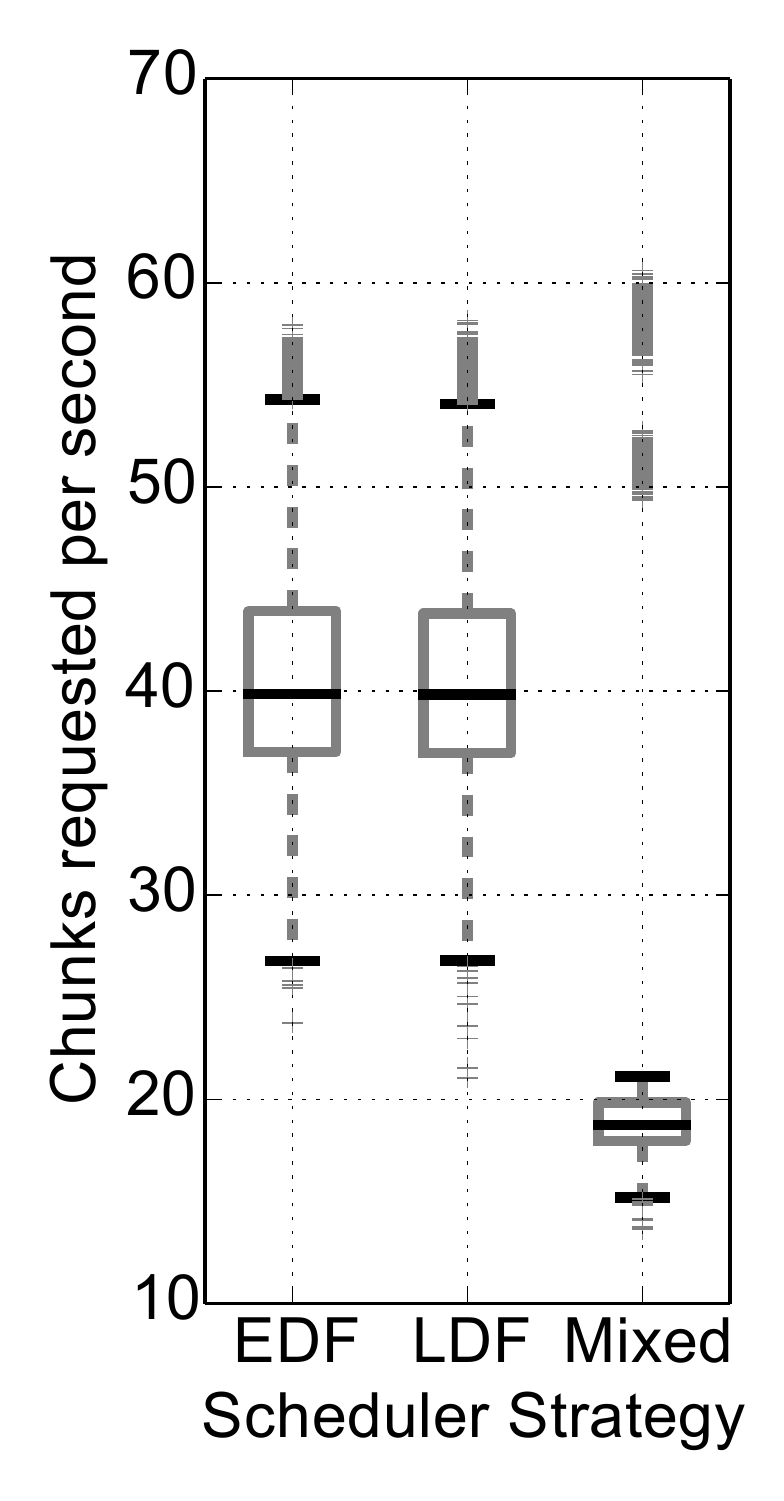}
		\caption{All peers.}
		\label{fig:scheduler-c}
	\end{subfigure}
	\begin{subfigure}[b]{.24\textwidth}
		\centering
		\includegraphics[height=150pt]{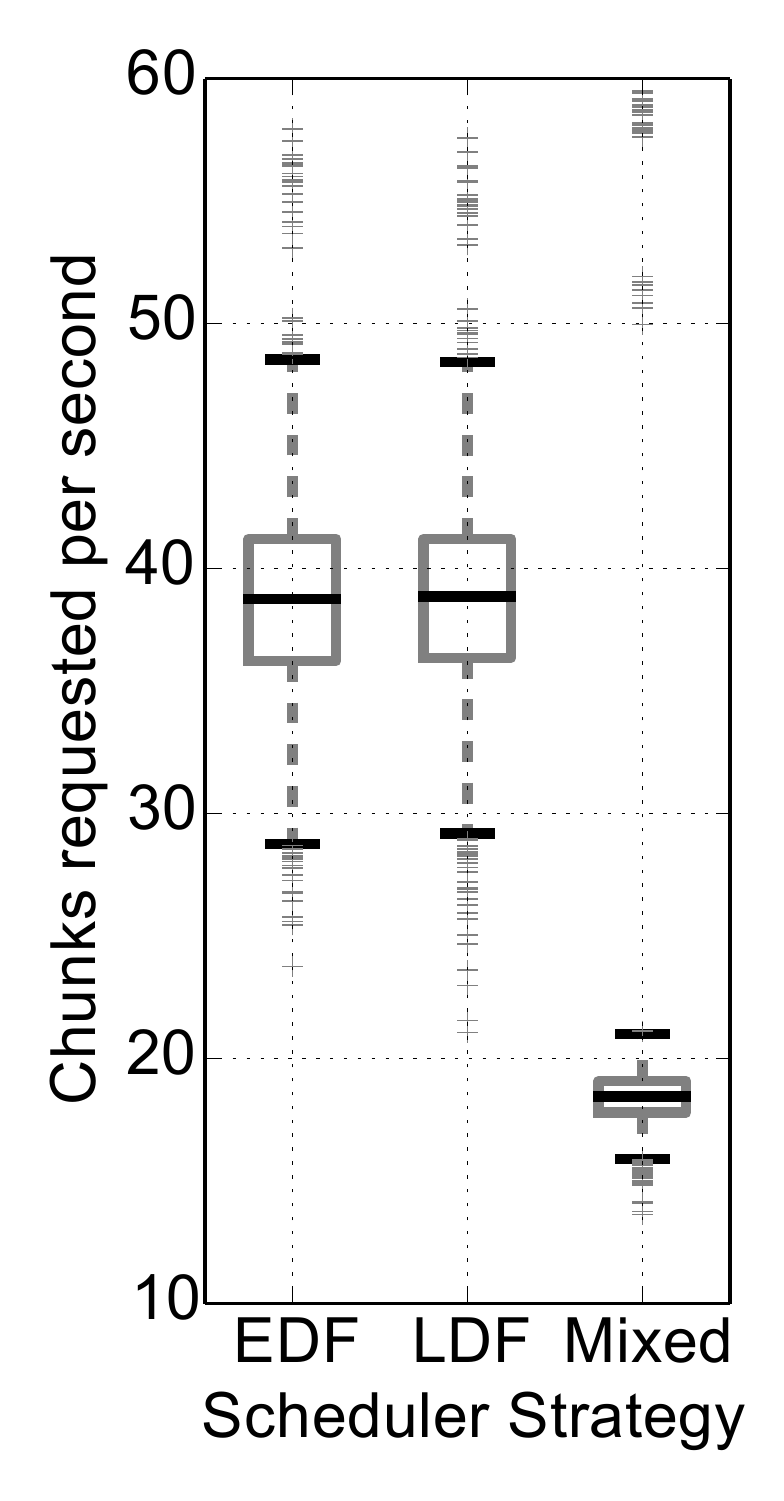}
		\caption{w/o LDFs.}
		\label{fig:scheduler-c}
	\end{subfigure}
	\begin{subfigure}[b]{.24\textwidth}
		\centering
		\includegraphics[height=150pt]{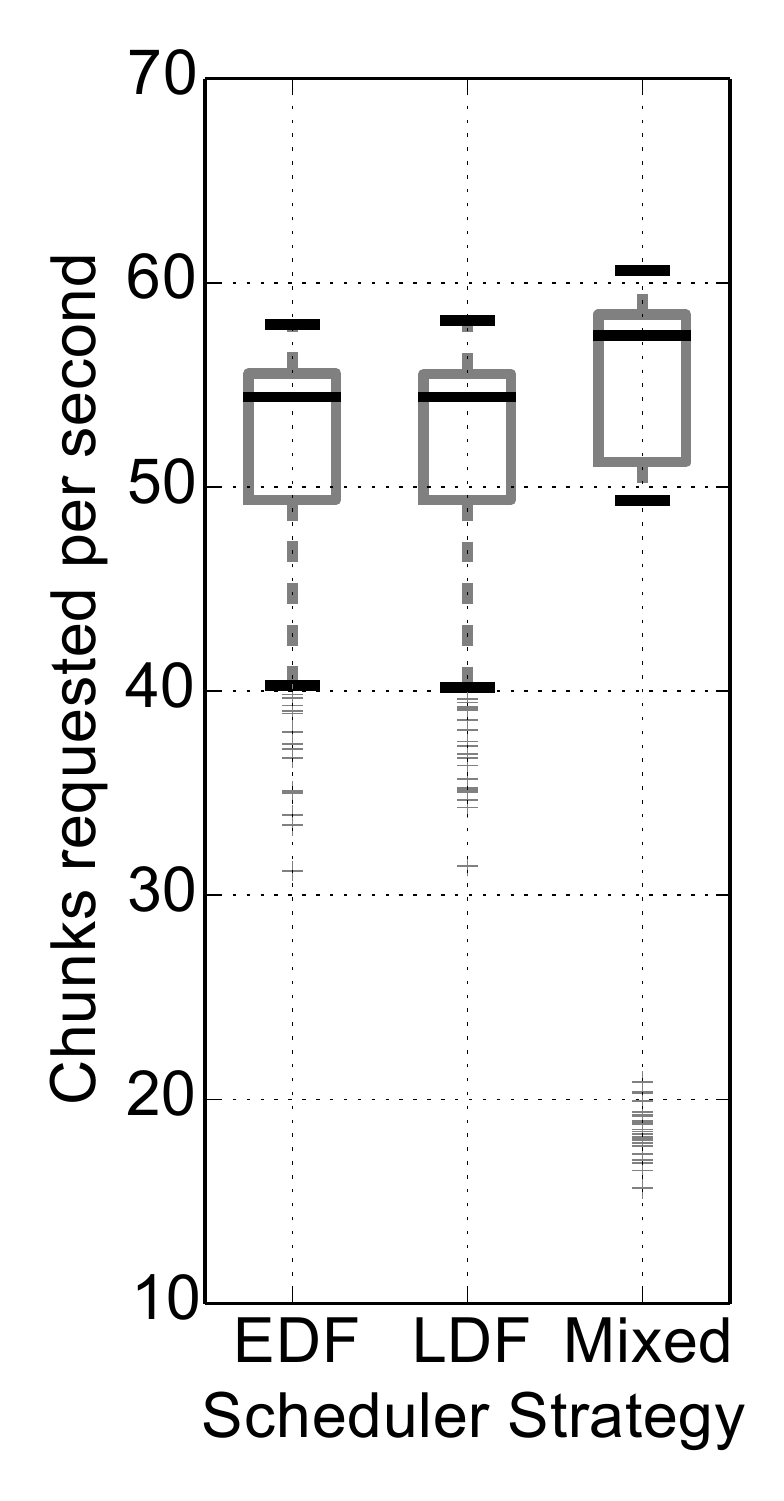}
		\caption{LDF peers.}
		\label{fig:scheduler-c}
	\end{subfigure}
	\caption{Number of requests}
	\label{fig:scheduler-costs}
\end{figure}

Figure~\ref{fig:scheduler-costs}, in addition, shows the resulting request rates for the individual chunk selection strategies.
It shows that for the overall population, the request rate drastically drops using the mixed strategy.
When separating the EDF from the LDF sub-population, it becomes apparent that this reduction is caused by the peers running EDF, which is roughly $50\%$ lower in the mixed configuration.
For the LDF peers, the average request rate is only slightly increased, showing the rather small additional overhead that these peers are penalized with.
At the same time (figures not shown here) the average playback continuity rate across the sub-populations does not show any difference.
This supports the argumentation that there is a high incentive for strong peers to run LDF instead of using EDF as the other peers do.

\subsection{Results of parameter study}
\begin{figure}[htbp]
	\centering
	\begin{subfigure}[b]{.45\textwidth}
		\centering
		\includegraphics[height=150pt]{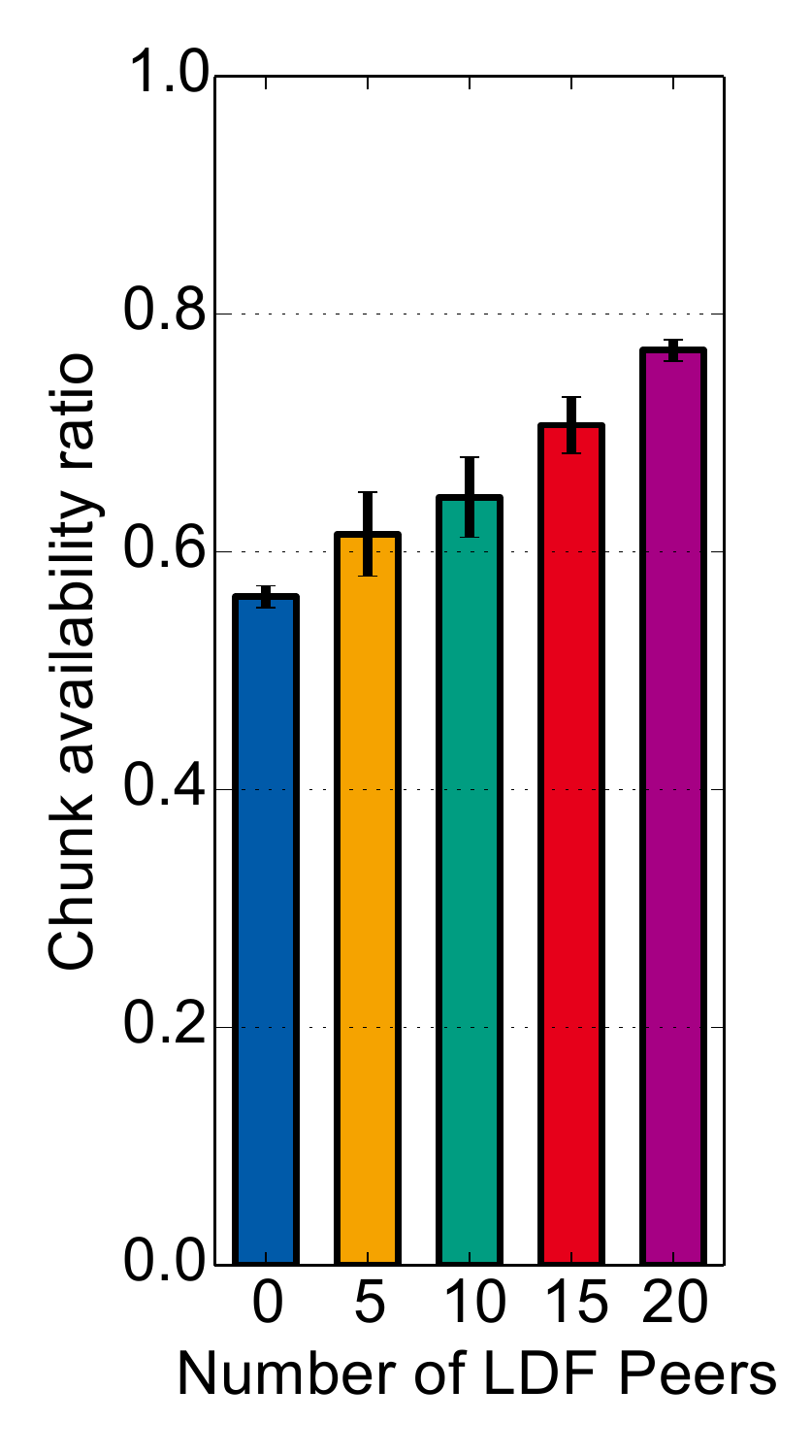}
		\includegraphics[height=150pt]{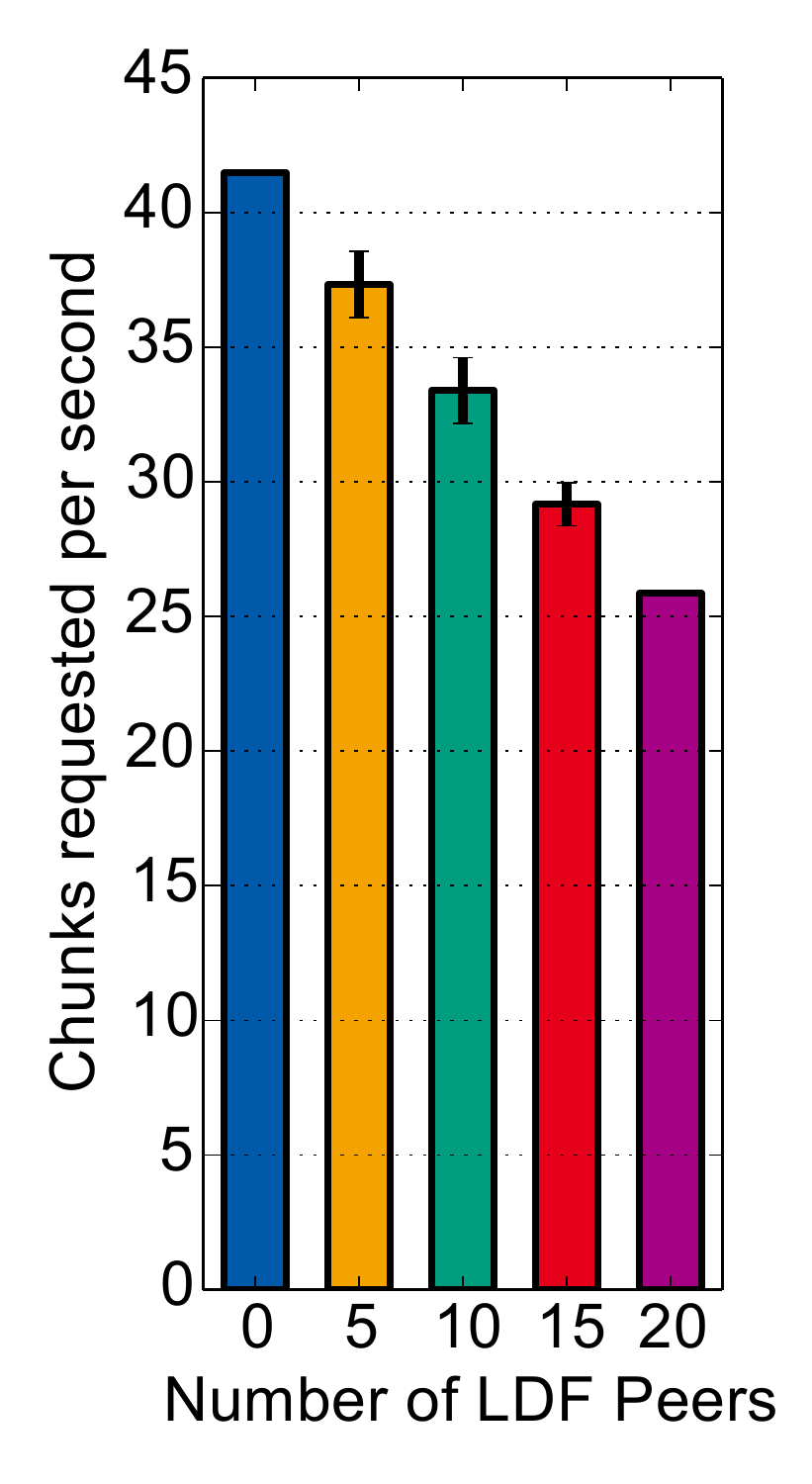}
		\caption{Performance vs. costs.}
		\label{fig:num-ldf-b}
	\end{subfigure}
	\begin{subfigure}[b]{.45\textwidth}
		\centering
		\includegraphics[height=150pt]{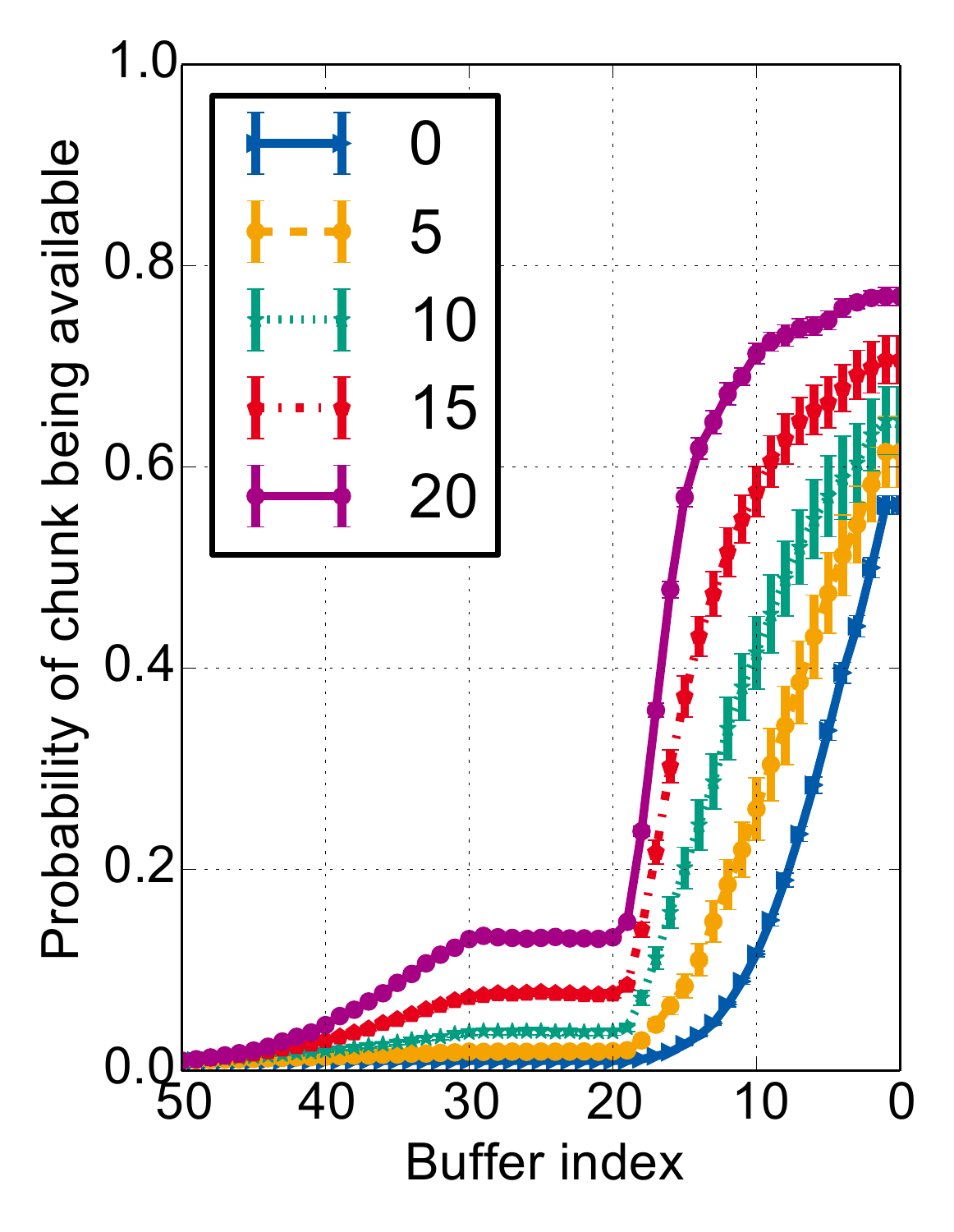}
		\caption{Buffer probability.}
		\label{fig:num-ldf-c}
	\end{subfigure}
	\caption{Number of LDF peers using mixed strategy.}
	\label{fig:num-ldf}
\end{figure}

To understand the impact of key system parameters on the performance and costs of the different chunks selection strategies, a parameter study on these parameters was conducted.
First, the number of peers that run LDF in the mixed configuration is varied between (i.e. all peers run EDF) and $20$.
LDF peers are assumed to be strong peers and thus $20$ LDF peers is the maximum configuration for an overall population size of $100$ peers and using the peer classes as described in Tab.~\ref{tab:eval-bw-distribution}.
The average playback continuity rate as well as the request rate are are depict in Figure~\ref{fig:num-ldf}.
With more strong peers running LDF, the playback continuity steadily increases, where with $20\%$ of the overall population running LDF, the streaming performance is increases by about $20\%$.
At the same time, the average chunk request rate is reduced by about $36\%$.
As discussed before, this reduction only affects the EDF peers, while the LDF peers experience a slightly increased request rate.

\begin{figure}[htbp]
	\centering
	\includegraphics[height=150pt]{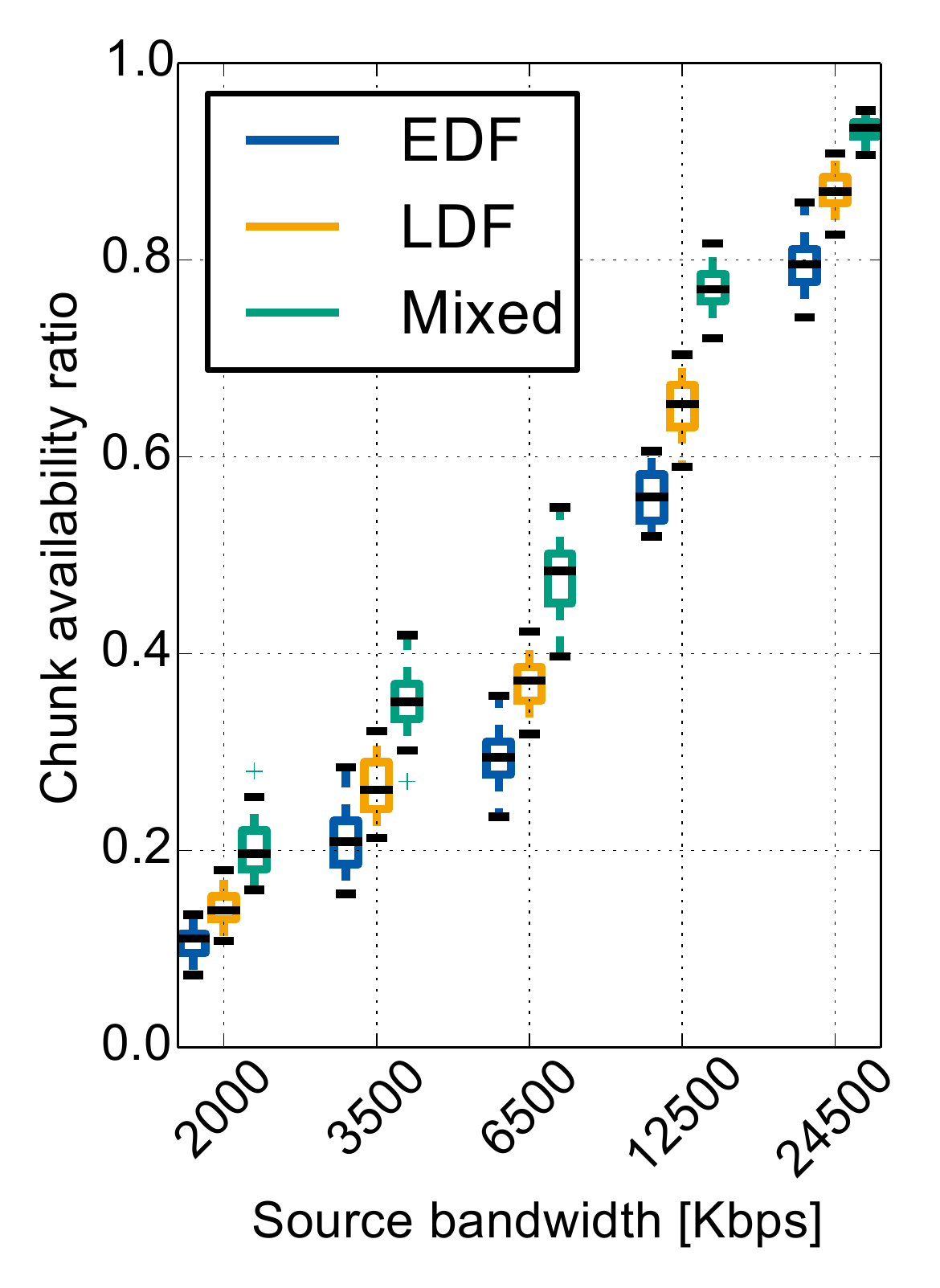}
	\includegraphics[height=150pt]{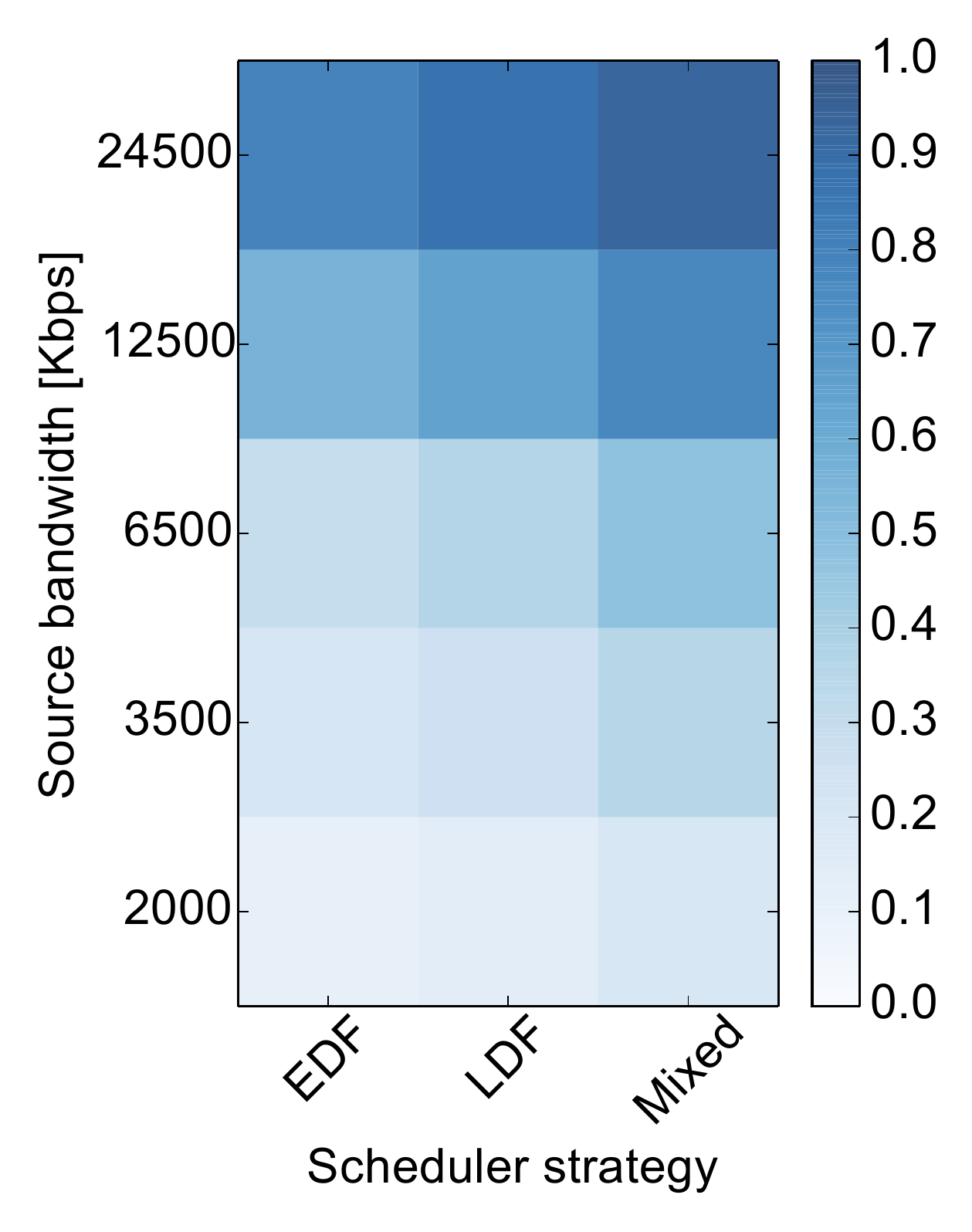}
	\caption{Source bandwidth: Playback continuity.}
	\label{fig:source-bw-a}
\end{figure}

\begin{figure}[htbp]
	\centering
	\includegraphics[height=150pt]{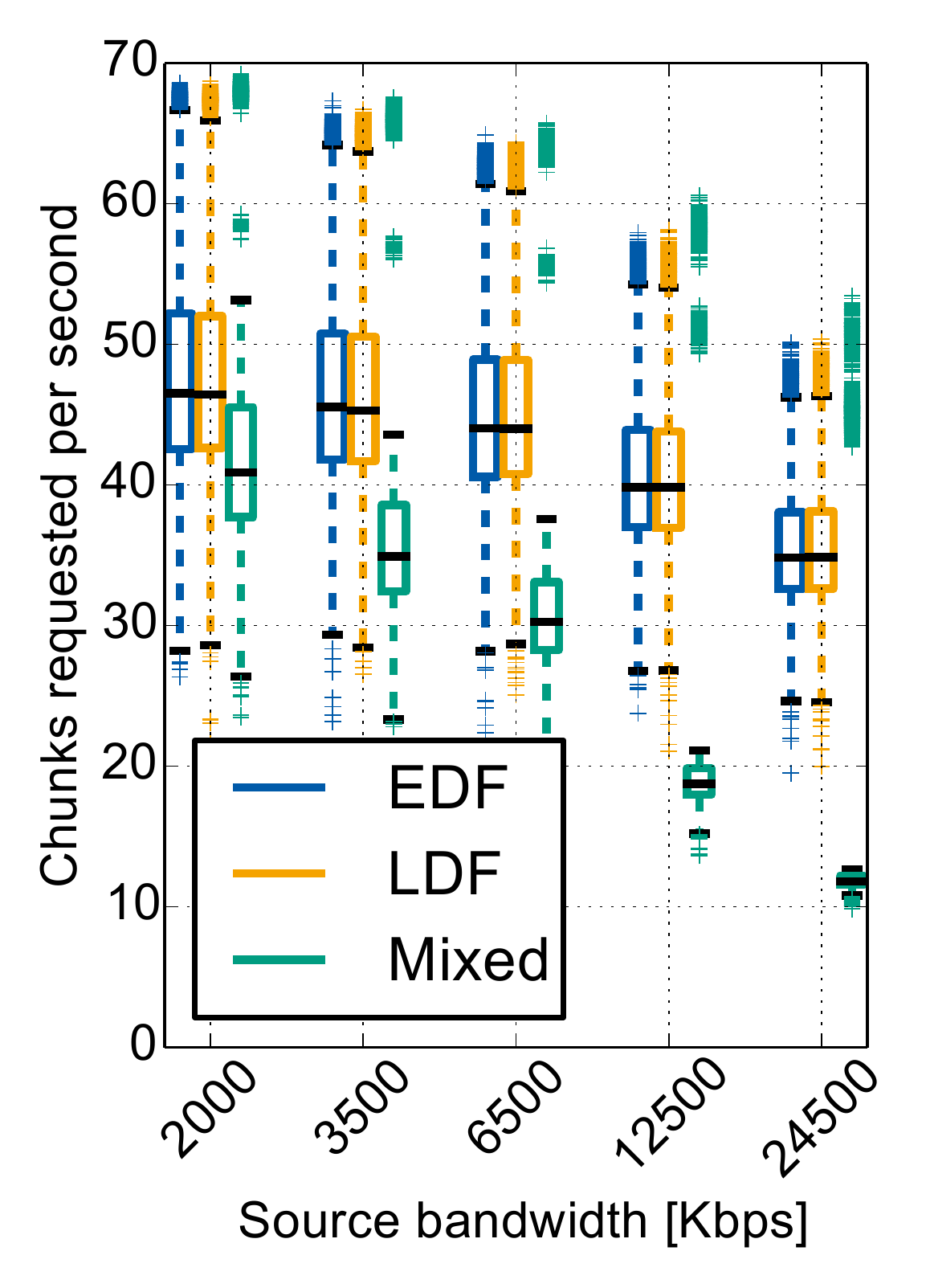}
	\includegraphics[height=150pt]{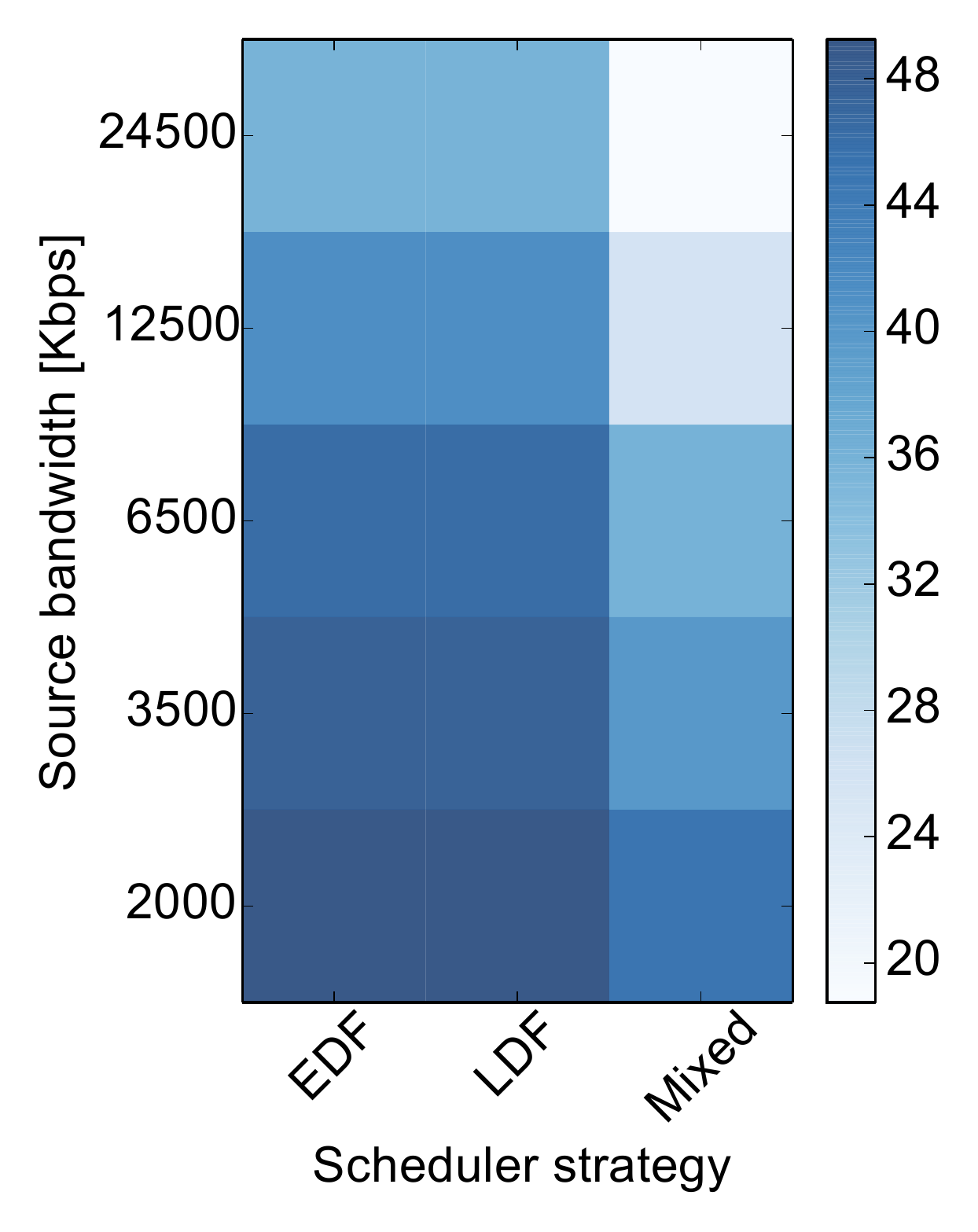}
	\caption{Source bandwidth: Chunk requests per second.}
	\label{fig:source-bw-b}
\end{figure}

Figure~\ref{fig:source-bw-a} and Figure~\ref{fig:source-bw-b} present the performance and costs for both changing scheduling strategies and source upload capacity.
The chosen source bandwidths are motivated by multitudes of the video bitrate plus an small buffer of $500$ Kbps to account for the system overhead.
In line with the observations in~\cite{zhang07,liang08}, increasing the source bandwidth has a huge impact on the overall system performance.
This is true across all scheduling strategies, where a steady increase in playback continuity ratio is observable with an increasing bandwidth.
In all cases, the mixed strategy outperforms the other two strategies.
Considering the average resulting request rates, it is observable that the mixed strategy greatly benefits from an increasing source bandwidth.
This indicates the superior replication behavior of the mixed strategy.
As mentioned before, EDF and LDF experience an increases streaming performance with higher source bandwidths, yet experiencing only a small reductions in the request rates and thus costs in terms of traffic caused by the strategies.

%

\newpage
\chapter{Related Work}
\label{sec:related-work}
The problem of P2P live video streaming has been studied extensively in the recent past, both from  the theoretical as well as practical perspective.
Playback continuity, among others, received considerable attention for due
reasons. Zhou \etal ~\cite{zhou2007simple,Zhou:P2Pmodel}
proposed a simple model
 for the analysis of buffer probabilities based on mean-field heuristics.
They brought to light several characteristics of the two most popular chunk selection strategies, \viz, LDF, and EDF.
Moreover, they proposed a mixed strategy as a compromise between LDF and EDF.
Adamu \etal~\cite{adamu2011discrete} made an attempt to analyze  buffer
probabilities in the context of a discrete Markov chain. Zhao \etal~\cite{zhao2009exploring} developed a population model to study different chunk selection strategies.
They modeled  P2P system as a \emph{density-dependent Markov jump process} and made interesting observations about  optimal strategies in different scenarios.

It would be unfair not to mention the contributions that, although made outside
 live video streaming domain, have been impactful in our understanding of the P2P communication systems in general. Hajek \etal~\cite{hajek2010missing,zhu2012stability}
highlighted interesting aspects of
 stability of a P2P system and lay down powerful results on continuous time Markov chain formulation.

The works mentioned above have, in many ways, been insightful in the homogeneous setup: when everyone plays the same strategy, everyone is equally capable etc.
Therefore, they do not answer questions pertaining to heterogeneous setups, such as "what happens if only some of the peers are greedy?".
Also, the influence of the graph structure, to the best of our knowledge, has not been studied so far.
Our endeavour in this article has been to meticulously capture these two important aspects in a principled way.
Infection models (see \cite{liggett2013stochastic,durrett2007random,pastor2002epidemic,eugster2004epidemic}) have proven to be useful in the study of spread of epidemic in random networks.
Our modeling of the system as a contact process in the context of P2P live video streaming to describe the dynamics in terms of \emph{master equations} is also a first, to the best of our knowledge.


Besides the above discussion of works that focus on the theoretical analysis and modelling, an number of more practical works in the area should be named that contributed to our today's understanding of mesh/pull-based streaming systems in general.

Zhang \etal~\cite{zhang07} argue that with an adequate configuration,
pull-based streaming can achieve a high bandwidth utilization.
They estimate lower bound for delivery ratio and complement their simulation study with steady-state mathematical analysis of simple sender-requester topologies.
They identify the source bandwidth and group size to be an important parameter that influences the performance of a scheduling mechanism and that there exists a tradeoff between control overhead and streaming delay in such systems.
As an answer, they propose a hybrid push/pull-based mechanisms to break this tradeoff.

Liang \etal~\cite{liang08} studied the importance of different mechanisms used
in P2P streaming systems and focus on scheduling strategies as important element.
They compare different strategies using testbed experiments and show argue that a previously proposed mechanisms proposed by the same authors in~\cite{guo08} shows the best performance.
Besides they identify key factors that influence the performance of scheduling strategies: the scheduling at the source node, the source bandwidth, the buffer size, and the out-degree of individual nodes.
They also argue that scheduling only plays a role for the system performance in case the overall resource index is low.
In our studies, however, we found that clear differences between scheduling strategies exist, even in settings with a high resource index.
We explain this observation by the fact that content bottlenecks are still likely to be caused by commonly used lightweight scheduling strategies such as EDF and LDF.

In~\cite{guo08}, the same authors propose the above mentioned scheduling approach.
Similar to the objective of this technical report, the proposed strategy tries to leverage peer heterogeneity.
Yet, this only happens implicitly as peers maximize their bandwidth utilization.
For this, the authors propose a rather complex scheduling approach based on multiple queues that are to be managed by each individual peer.
Besides, the approach is partially based on a push delivery approach when relaying video chunks instead of a pull approach as studied in this technical report.
With this approach, the authors show that they can achieve a nearly $100\%$ bandwidth uitilization across peers.
Yet, their proposed mechanisms assumes a fully connected mesh among peers and that each peer is directly connected to the source as well.
This greatly limits the applicability of the approach to larger setups.
In contrast to this approach, we focus on pure pull-based strategies, do not assume a fully connected mesh, and do not focus on maximizing bandwidth utilization only.

Zhang \etal~\cite{zhang07} show that pull-based streaming can achieve high bandwidth utilisation and estimate a lower bound for the delivery ratio, based on simulations and a steady-state analysis of simple sender-requester topologies.
Liang \etal~\cite{liang08} discuss scheduling as key mechanism for P2P streaming and name scheduling and bandwidth of the source, the buffer sizes, and peer degrees as additional factors.
Besides, they argue that scheduling plays a role only at a resource index, whereas we observe clear differences due to content bottlenecks.
In~\cite{guo08}, a scheduling strategy is proposed, implicitly leverages heterogeneity by maximising bandwidth utilisation of peers.
The strategy is rather complex as it is based on multiple queues and a partial push delivery.
The authors show a nearly optimal utilisation only for a fully connected mesh, greatly limiting the applicability to realistic setups.
In contrast, we focus on pure pull strategies, do not assume a fully connected mesh, and do not focus on maximising bandwidth only.
In~\cite{picconi08}, a mesh/push-based streaming system is proposed using LRU as scheduling strategy.
While the authors assume a heterogeneous setup, they do not consider degree heterogeneity as done in this technical report but define a fixed maximum degree to all peers.

%


\newpage
\chapter{Discussion}
\label{sec:discussion}
We conclude our technical report with a short discussion. We summarise our contribution and propound interesting perspectives and questions that we find fit for future scientific probe.

In this technical report, we  contributed to building a sound mathematical framework for swarming on random graphs. The intricate dependence of performance on degree was made explicit. The idea of degree-based (strength-based) combination of  primitive scheduling strategies  led to interesting revelations, \viz, boon of heterogeneity, weak peers outperforming strong ones etc. Propelled by these observations, we proposed our mixed strategy \sys{SchedMix}.

Our mathematical framework is a general one.
We believe it can also serve as a foundation in problems other than the one in pursuit, opening up ample opportunities for future exploration.
It would be interesting to accommodate more complicated scheduling strategies into the model.
From practical perspective, the prospect is even broader.
We expect to see impactful application of \sys{SchedMix} in combination with more sophisticated mechanisms.
One straightforward but important step is the application of \sys{SchedMix} in a state-of-the-art hybrid streaming system, where both mesh/pull and multi-tree/push-based mechanisms coexist.
In this context it would be also interesting to understand the impact of other mechanisms, such as the exchange of buffermaps or a streaming of layered media content.
The results presented in this technical report are encouraging in that \sys{SchedMix} could be used as simple, yet powerful alternative to complex scheduling strategies in the growing number of scenarios where peer heterogeneity is inevitably given, e.g. streaming scenarios with heterogeneous end-user devices, where, e.g., bandwidth-constraint mobile users meet well-connected and high-capacity home users.

\label{ref:conclusion}

\newpage
\begin{appendix}
	
\chapter{Appendix}

\label{sec:appendix} 

\section{Appendix: Mean field master equations}
 \label{appendix:master-equations}

\begin{proof}[Proof of \cref{lemma:for-mean-dynamics}]
	Fix $u \in \mathcal{T}, k \in \mathbb{N}$. It follows, 
	\begin{align*}
		& 	\sum_{Z \in  \mathbb{N}_0^{|\mathcal{T}| \times \mathbb{N}} } z_u^{k} \sum_{\substack{Y : \sum_{ \mathbb{S} v= u} y_v^{l} = z_u^{l} \\ \forall u,v \in \mathcal{T}, l \in 
				\mathbb{N}}} \probOf{Y}  \\
		&= \sum_{z_u^{k}} \sum_{ \substack{y_v^{k} :\\ \sum_{\mathbb{S} v= u} y_v^{k} = z_u^{k} }}
		(\sum_{ \mathbb{S} v= u} y_v^{k}) \probOf{\{y_v^{k} \mid  \mathbb{S} v= u \}} \\
		&= \sum_{ \mathbb{S} v= u } \Eof{z_v^{k}}
	\end{align*}
\end{proof}

\begin{proof}[Proof of \cref{lemma1}]
	\begin{enumerate}
		\item Notice that, for $u,v \in \mathcal{T} : \mathbb{S}v=u$ and $u_{i+1}=1 \implies v_i=1$. Therefore, 
		\begin{eqnarray*}
			\sum_{u \in \mathcal{T}: u_{i+1}=1} \sum_{v \in \mathcal{T}: \mathbb{S}v=u} w_v^{k} &=\sum_{v \in \mathcal{T}: v_i=1} w_v^{k} =&   p_k(i)\\
		\end{eqnarray*}
		\item We simplify the left hand side and omit terms whenever they turn out to be $0$. 
		\begin{align*}
			& \sum_{u \in \mathcal{T}: u_{i}=1}  \sum_{j \in \mathcal{F}} \left[   \lambda^{k}(u-e_j,u)  -\lambda^{k}(u,u+e_j) \right] \\
			& =  \sum_{u \in \mathcal{T}: u_{i}=1}  \left[   \lambda^{k}(u-e_{i},u)  -\lambda^{k}(u,u+e_{i})  \right] 
			 +   \sum_{u \in \mathcal{T}: u_{i}=1}  \sum_{j \in \mathcal{F} \setminus \{i\}} \left[  \lambda^{k}(u-e_j,u)  -\lambda^{k}(u,u+e_j) \right] \\
			&=  \sum_{u \in \mathcal{T}: u_{i}=1} \lambda^{k}(u-e_{i},u) 
		 +   \sum_{j \in \mathcal{F} \setminus \{i\}}  \sum_{u \in \mathcal{T}: u_{i}=1}  \left[  \lambda^{k}(u-e_j,u)  -\lambda^{k}(u,u+e_j)  \right]  \\
			&= \sum_{u \in \mathcal{T}: u_{i}=1} \lambda^{k}(u-e_{i},u) \eqpunkt 
		\end{align*}
		This is because, 
		\begin{align*}
			&  \sum_{j \in \mathcal{F} \setminus \{i\}}  \sum_{u \in \mathcal{T}: u_{i}=1}  ( \lambda^{k}(u-e_j,u)  -\lambda^{k}(u,u+e_j) )  \\
			&=  \sum_{j \in \mathcal{F} \setminus \{i\}} \Bigg[ \sum_{u \in \mathcal{T}: u_{i}=1, u_j=1}  ( \lambda^{k}(u-e_j,u)  -\lambda^{k}(u,u+e_j) ) 
				+  \sum_{u \in \mathcal{T}: u_{i}=1, u_j=0}  ( \lambda^{k}(u-e_j,u)  -\lambda^{k}(u,u+e_j) )  \Bigg]\\
			&=  \sum_{j \in \mathcal{F} \setminus \{i\}} \Bigg[ \sum_{u \in \mathcal{T}: u_{i}=1, u_j=1} \lambda^{k}(u-e_j,u) 
			 -  \sum_{u \in \mathcal{T}: u_{i}=1, u_j=0}  \lambda^{k}(u,u+e_j)  \Bigg] \\
			&=  \sum_{j \in \mathcal{F} \setminus \{i\}} \Bigg[ \sum_{u \in \mathcal{T}: u_{i}=1} \bigg( \lambda^{k}(u-e_j,u) 
			- \lambda^{k}(u-e_j,u) \bigg)  \Bigg]  \\
			&=0 \eqpunkt
		\end{align*}
		Such a rearrangement of summands is possible because $ u \in \mathcal{T}: u_{i}=1, u_j=1 \implies v=u-e_j \in \mathcal{T} : v_{i}=1, v_j=0$. 	This completes the proof.
	\end{enumerate}
\end{proof}

\newpage
 \section{Appendix: Marginal probabilities} 
 \label{appendix:chunkSelectionFunction}

\begin{proof}[Proof of \cref{result:LDFchunkSelectionFunction}]
	\label{proof:LDFchunkSelectionFunction}
	\begin{enumerate}
		\item From \Cref{equation:LDFchunckSelectionFunction}, we have 
		
				\begin{align*}
	  s_k(i+1)-s_k(i)	
	 = & \left[   1- p_k(1)\right] \prod_{j=1}^{i} \left[   p_k(j) + (1- p_k(j))(1-k  \varsigma \theta_j) \right] \\
	& 	- \left[   1- p_k(1)\right] \prod_{j=1}^{i-1}  \left[  p_k(j) + (1- p_k(j))(1-k  \varsigma \theta_j) \right] \\
	 	= & \left[   1- p_k(1)\right] \prod_{j=1}^{i-1} (p_k(j) + (1- p_k(j))(1-k  \varsigma \theta_j))  \\
		& (p_k(i)+(1- p_k(i))(1-k \varsigma \theta_i) -1 ) \\
		= & s_k(i) (-k\theta_i +k \varsigma \theta_i p_k(i)) \\
	 	=&  -k  \varsigma \theta_i (1- p_k(i)) s_k(i) \\
		= &- (p_k(i+1)- p_k(i)) \\
		\Rightarrow \sum_{j=1}^{i-1} ( s_k(j+1) - s_k(j) )  = &-  \sum_{j=1}^{i-1} (p_k(j+1)- p_k(j) ) \\
		\Rightarrow s_k(i) = & s_k(1)+ p_k(1)- p_k(i) 
		\end{align*}
		Substituting $s_k(1)= 1- p_k(1)$, we have $s_k(i)=1- p_k(i)$. 
		\item Follows directly by substituting $s_k(i)$ in \Cref{equation:marginalProbRecursion}.
	\end{enumerate}
\end{proof}

\begin{proof}[Proof of \cref{result:EDFchunkSelectionFunction}]
	\label{proof:EDFchunkSelectionFunction}
	\begin{enumerate}
		\item From \Cref{equation:EDFchunckSelectionFunction}, we have 
%
	\begin{align*}
		s_k(i+1)-s_k(i) 
		= &\left[   1- p_k(1)\right] \prod_{j=i+2}^{n-1}  \left[ p_k(j) + (1- p_k(j))(1-k  \varsigma \theta_j) \right]\\
		&- \left[   1- p_k(1)\right] \prod_{j=i+1}^{n-1}  \left[ p_k(j) + (1- p_k(j))(1-k  \varsigma \theta_j))   \right]  \\
		= & \left[   1- p_k(1)\right]  \prod_{j=i+2}^{n-1}  \left[  p_k(j) + (1- p_k(j))(1-k  \varsigma \theta_j)) \right] \\
		&   \left[   1- p_k(i+1)-(1- p_k(i+1))(1-k\varsigma \theta_{i+1}) \right]\\
		= & s_k(i+1)  \left[   k\theta_{i+1} - k \varsigma \theta_{i+1} p_k(i+1) \right] \\
		=& k  \varsigma \theta_{i+1} \left[  1- p_k(i+1) \right] s_k(i+1) \\
		= & (p_k(i+2)- p_k(i+1)) \\
		\end{align*}		
		
	So, 	
		\begin{align*}
\sum_{j=i}^{n-2} ( s_k(j+1) - s_k(j) ) 
		& =  \sum_{j=i}^{n-2} (p_k(j+2)- p_k(j+1) ) \\
		\Rightarrow s_k(i) &= s_k(n-1)- p_k(n) +p_k(i+1) 
		\end{align*}
		Substituting $s_k(n-1)= 1-p_k(1)$, we have 
		\begin{equation*}
		s_k(i)=1- p_k(1)- p_k(n) + p_k(i+1) \eqpunkt
		\end{equation*}
		\item Follows directly by substituting $s_k(i)$ in \Cref{equation:marginalProbRecursion}.
	\end{enumerate}
\end{proof}


\begin{proof}[Proof of \cref{result:comparisonLDF}]
	\label{proof:comparisonLDF}
Dividing the two differential equations, we get 
\begin{equation*}
\frac{\, dy_1}{\,dy_2} = \frac{k_1  (1-y_1)^2}{k_2 (1-y_2)^2} \eqkomma
\end{equation*}
which, along with the boundary condition $y_2=\frac{1}{M}$ when $y_1=\frac{1}{M}$, gives exact solution 
\begin{equation*}
y_2=\frac{1- (Ck_1  +r)(1-y_1)}{1-Ck_1(1-y_1)} \eqkomma
\end{equation*}
where $r=\frac{k_1}{k_2}$ and $C=\frac{M}{M-1} (\frac{1}{k_1}-\frac{1}{k_2})$. When $M\rightarrow \infty$, the above simplifies to
\begin{equation*}
y_2= \frac{y_1}{1-(1-r)(1-y_1)} \eqpunkt
\end{equation*}
\end{proof}

\begin{proof}[Proof of \cref{result:LDFbuffersizeRequirement}]
\label{proof:LDFbuffersizeRequirement}
Inserting \cref{eq:y1vsy2relationPureLDF} into \cref{eq:pureLDF_ODE}, we get 
\begin{equation*}
\frac{\, dy_1}{\, dx} = \frac{k_1 \varsigma (q_1(r+(1-r)y_1)+q_2)y_1(1-y_1)^2}{(r+(1-r)y_1)} \eqpunkt
\end{equation*}

Writing 
\begin{equation*}
	\begin{split}
&\frac{(r+(1-r)y_1)}  {k_1 \varsigma (q_1(r+(1-r)y_1)+q_2)y_1(1-y_1)^2} \\
&= \frac{A}{y_1} + \frac{B}{(1-y_1)} 
 +\frac{C}{(1-y_1)^2} 
+ \frac{D}{q_1(r+(1-r)y_1)+q_2} \eqkomma
\end{split}
\end{equation*}
and using the boundary condition $y_1=\frac{1}{M}$ when $x=1$, we obtain the desired result. 
\end{proof}

\begin{proof}[Derivation of \cref{eq:y2vsMixed}]
From \Cref{eq:mixed_ODE}, making use of the approximation, we have the following differential equation
\begin{equation*}
\frac{\, dy_1}{\,dy_2} = \frac{r (1-y_1) (y_1-p(1) +\epsilon_1)}{(1-y_2)^2} \eqkomma
\end{equation*}	
which can be exactly solved to get the desired result
\begin{equation*}
y_2= \frac{\frac{1}{r(1-p(1)+\epsilon_1)} \ln \left( \frac{y_1-p(1) + \epsilon_1}{1-y_1} \right) -C- 1}{\frac{1}{r(1-p(1)+\epsilon_1)} \ln \left( \frac{y_1-p(1) + \epsilon_1}{1-y_1} \right) -C} \eqkomma
\end{equation*}
 where $C= \frac{1}{(1-p(1)+\epsilon_1)} \ln \left( \frac{\epsilon_1}{1-p(1)} \right) - \frac{1}{1-p(1)} $.
 	\label{derivation:y2vsy1Mixed}
\end{proof}
	\end{appendix}
\newpage

\section*{Acknowledgment}
This work has been funded by the German Research Foundation~(DFG) as part of project C03 within the Collaborative Research Center~(CRC) 1053 -- MAKI.

\bibliographystyle{IEEEtranS}
\bibliography{main}

\begin{thebibliography}{10}
\providecommand{\url}[1]{#1}
\csname url@samestyle\endcsname
\providecommand{\newblock}{\relax}
\providecommand{\bibinfo}[2]{#2}
\providecommand{\BIBentrySTDinterwordspacing}{\spaceskip=0pt\relax}
\providecommand{\BIBentryALTinterwordstretchfactor}{4}
\providecommand{\BIBentryALTinterwordspacing}{\spaceskip=\fontdimen2\font plus
\BIBentryALTinterwordstretchfactor\fontdimen3\font minus
  \fontdimen4\font\relax}
\providecommand{\BIBforeignlanguage}[2]{{%
\expandafter\ifx\csname l@#1\endcsname\relax
\typeout{** WARNING: IEEEtranS.bst: No hyphenation pattern has been}%
\typeout{** loaded for the language `#1'. Using the pattern for}%
\typeout{** the default language instead.}%
\else
\language=\csname l@#1\endcsname
\fi
#2}}
\providecommand{\BIBdecl}{\relax}
\BIBdecl

\bibitem{adamu2011discrete}
A.~Adamu, Y.~Gaidamaka, and A.~Samuylov, ``{Discrete Markov Chain Model for
  Analyzing Probability Measures of P2P Streaming Network},'' in \emph{NEW2AN},
  ser. LNCS.\hskip 1em plus 0.5em minus 0.4em\relax Springer, 2011, vol. 6869.

\bibitem{BarabasiAlbertScaling}
A.-L. Barab{\'{a}}si and R.~Albert, ``{Emergence of Scaling in Random
  Networks},'' \emph{Science}, vol. 286, no. 5439, 1999.

\bibitem{barvinok2009asymptotic}
A.~Barvinok, ``{Asymptotic Estimates for the Number of Contingency Tables,
  Integer Flows, and Volumes of Transportation Polytopes},''
  \emph{International Mathematics Research Notices}, vol. 2009, no.~2, 2009.

\bibitem{behzad1967no}
M.~Behzad and G.~Chartrand, ``{No Graph is Perfect},'' \emph{American
  Mathematical Monthly}, 1967.

\bibitem{cisco14}
Cisco, ``{Cisco Visual Networking Index: Forecast and Methodology, 2013 --
  2018},'' Tech. Rep., 2014.

\bibitem{igraph}
\BIBentryALTinterwordspacing
G.~Csardi and T.~Nepusz, ``The igraph software package for complex network
  research,'' \emph{InterJournal}, vol. Complex Systems, p. 1695, 2006.
  [Online]. Available: \url{http://igraph.org}
\BIBentrySTDinterwordspacing

\bibitem{deering89}
\BIBentryALTinterwordspacing
S.~E. Deering, ``{Host Extensions for IP Multicasting},'' RFC 1112, Aug 1989.
  [Online]. Available: \url{https://tools.ietf.org/rfc/rfc1112}
\BIBentrySTDinterwordspacing

\bibitem{diot00}
C.~Diot, B.~Levine, B.~Lyles, H.~Kassem, and D.~Balensiefen, ``{Deployment
  Issues for the IP Multicast Service and Architecture},'' \emph{IEEE Network},
  vol.~14, no.~1, 2000.

\bibitem{durrett2007random}
R.~Durrett, \emph{Random Graph Dynamics}.\hskip 1em plus 0.5em minus
  0.4em\relax Cambridge University Press.

\bibitem{eugster2004epidemic}
P.~Eugster, R.~Guerraoui, A.-M. Kermarrec, and L.~Massouli{\'e}, ``{Epidemic
  Information Dissemination in Distributed Systems},'' \emph{IEEE Computer},
  vol.~37, no.~5, 2004.

\bibitem{froemmgen15}
A.~Fr\"ommgen, B.~Richerzhagen, J.~R\"uckert, D.~Hausheer, R.~Steinmetz, and
  A.~Buchmann, ``{Towards the Description and Execution of Transitions in
  Networked Systems},'' in \emph{{AIMS}}, 2015.

\bibitem{guo08}
Y.~Guo, C.~Liang, and Y.~Liu, ``{AQCS: Adaptive Queue-based Chunk Scheduling
  for P2P Live Streaming},'' in \emph{IFIP NETWORKING}, 2008.

\bibitem{hajek2010missing}
B.~Hajek and J.~Zhu, ``{The Missing Piece Syndrome in Peer-to-Peer
  Communication},'' in \emph{IEEE ISIT}, 2010.

\bibitem{kemeny1960finite}
J.~G. Kemeny and J.~L. Snell, \emph{Finite Markov Chains}.\hskip 1em plus 0.5em
  minus 0.4em\relax van Nostrand, Princeton, NJ, 1960.

\bibitem{KhudaBukhsh2016P2P}
W.~R. KhudaBukhsh, J.~R{\"u}ckert, J.~Wulfheide, D.~Hausheer, and H.~Koeppl,
  ``Analysing and leveraging client heterogeneity in swarming-based live
  streaming,'' in \emph{2016 IFIP Networking Conference (IFIP Networking) and
  Workshops}, May 2016, pp. 386--394.

\bibitem{khudabukhsh2018approximate}
W.~R. KhudaBukhsh, A.~Auddy, Y.~Disser, and H.~Koeppl, ``Approximate
  lumpability for markovian agent-based models using local symmetries,''
  \emph{arXiv preprint arXiv:1804.00910}, 2018.

\bibitem{krishnappa15}
D.~Krishnappa, M.~Zink, and R.~Sitaraman, ``{Optimizing the Video Transcoding
  Workflow in CDNs},'' in \emph{ACM MM}, 2015.

\bibitem{liang08}
C.~Liang, Y.~Guo, and Y.~Liu, ``{Is Random Scheduling Sufficient in P2P Video
  Streaming?}'' in \emph{IEEE ICDCS}, 2008.

\bibitem{liggett2013stochastic}
T.~M. Liggett, \emph{Stochastic Interacting Systems: Contact, Voter and
  Exclusion Processes}.\hskip 1em plus 0.5em minus 0.4em\relax Springer, 1999.

\bibitem{liu08}
Y.~Liu, Y.~Guo, and C.~Liang, ``{A Survey on Peer-to-Peer Video Streaming
  Systems},'' \emph{{Peer-to-Peer Networking and Applications}}, vol.~1, 2008.

\bibitem{nesterov1994interior}
Y.~Nesterov and A.~Nemirovskii, \emph{Interior-point Polynomial Algorithms in
  Convex Programming}.\hskip 1em plus 0.5em minus 0.4em\relax SIAM, 1994,
  vol.~13.

\bibitem{nisan2007algorithmic}
N.~Nisan, T.~Roughgarden, {\'{E}}.~Tardos, and V.~V. Vazirani,
  \emph{{Algorithmic Game Theory}}.\hskip 1em plus 0.5em minus 0.4em\relax
  Cambridge University Press, 2007.

\bibitem{oecd14}
{Organisation for Economic Co-operation and Development}, ``{OECD Broadband
  Report},'' Tech. Rep., 2014.

\bibitem{pastor2002epidemic}
R.~Pastor-Satorras and A.~Vespignani, ``{Epidemic Dynamics in Finite Size
  Scale-free Networks},'' \emph{Physical Review E}, vol.~65, no.~3, 2002.

\bibitem{picconi08}
F.~Picconi and L.~Massouli{\'e}, ``{Is There a Future for Mesh-based Live Video
  Streaming?}'' in \emph{IEEE P2P}, 2008.

\bibitem{rejaie14}
R.~Rejaie and N.~Magharei, ``{On Performance Evaluation of Swarm-based Live
  Peer-to-Peer Streaming Applications},'' \emph{Springer Multimedia Systems},
  vol.~20, no.~4, 2014.

\bibitem{richerzhagen15}
B.~Richerzhagen, D.~Stingl, J.~R\"{u}ckert, and R.~Steinmetz, ``{Simonstrator:
  Simulation and Prototyping Platform for Distributed Mobile Applications},''
  in \emph{{ICST/ACM SIMUtools}}, 2015.

\bibitem{rueckert15b}
J.~R\"uckert, B.~Richerzhagen, E.~Lidanski, R.~Steinmetz, and D.~Hausheer,
  ``{TopT: Supporting Flash Crowd Events in Hybrid Overlay-based Live
  Streaming},'' in \emph{{IFIP NETWORKING}}, 2015.

\bibitem{sitaraman14}
R.~K. Sitaraman, M.~Kasbekar, W.~Lichtenstein, and M.~Jain, ``{Overlay
  Networks: An Akamai Perspective},'' in \emph{{Advanced Content Delivery,
  Streaming, and Cloud Services}}.\hskip 1em plus 0.5em minus 0.4em\relax John
  Wiley \& Sons, 2014.

\bibitem{stingl2011}
D.~Stingl, C.~Gross, J.~R\"{u}ckert, L.~Nobach, A.~Kovacevic, and R.~Steinmetz,
  ``{PeerfactSim.KOM: A Simulation Framework for Peer-to-Peer Systems},'' in
  \emph{IEEE HPCS}, 2011.

\bibitem{strogatz2014nonlinear}
S.~H. Strogatz, \emph{{Nonlinear Dynamics and Chaos: With Applications to
  Physics, Biology, Chemistry, and Engineering}}, 2014.

\bibitem{wang10}
F.~Wang, Y.~Xiong, and J.~Liu, ``{mTreebone: A Collaborative Tree-Mesh Overlay
  Network for Multicast Video Streaming},'' \emph{IEEE TPDS}, vol.~21, no.~3,
  2010.

\bibitem{WattsStrogatzSmallWorlds}
D.~J. Watts and S.~H. Strogatz, ``{Collective Dynamics of `Small-world'
  Networks},'' \emph{Nature}, vol. 393, 1998.

\bibitem{wichtlhuber14a}
M.~Wichtlhuber, B.~Richerzhagen, J.~R\"uckert, and D.~Hausheer, ``{TRANSIT:
  Supporting Transitions in Peer-to-Peer Live Video Streaming},'' in
  \emph{{IFIP NETWORKING}}, 2014.

\bibitem{ying2010asymptotic}
L.~Ying, R.~Srikant, and S.~Shakkottai, ``{The Asymptotic Behavior of Minimum
  Buffer Size Requirements in Large P2P Streaming Networks},'' in \emph{IEEE
  Information Theory and Applications Workshop (ITA)}, 2010.

\bibitem{zhang07}
M.~Zhang, Q.~Zhang, L.~Sun, and S.~Yang, ``{Understanding the Power of
  Pull-based Streaming Protocol: Can We Do Better?}'' \emph{IEEE JSAC},
  vol.~25, no.~9, 2007.

\bibitem{zhang12}
X.~Zhang and H.~Hassanein, ``{A Survey of Peer-to-Peer Live Video Streaming
  Schemes - An Algorithmic Perspective},'' \emph{Computer Networks}, vol.~56,
  no.~15, 2012.

\bibitem{zhao2009exploring}
B.~Zhao, J.~Lui, and D.~Chiu, ``{Exploring the Optimal Chunk Selection Policy
  for Data-driven P2P Streaming Systems},'' in \emph{IEEE P2P}, 2009.

\bibitem{zhao2013}
M.~Zhao, P.~Aditya, A.~Chen, Y.~Lin, A.~Haeberlen, P.~Druschel, B.~Maggs,
  B.~Wishon, and M.~Ponec, ``{Peer-Assisted Content Distribution in Akamai
  NetSession},'' in \emph{ACM IMC}, 2013.

\bibitem{zhou2007simple}
Y.~Zhou, D.~M. Chiu \emph{et~al.}, ``{A Simple Model for Analyzing P2P
  Streaming Protocols},'' in \emph{IEEE ICNP}, 2007.

\bibitem{Zhou:P2Pmodel}
Y.~Zhou, D.-M. Chiu, and J.~Lui, ``{A Simple Model for Chunk-scheduling
  Strategies in P2P Streaming},'' \emph{IEEE/ACM TON}, vol.~19, no.~1, 2011.

\bibitem{zhu2012stability}
J.~Zhu and B.~Hajek, ``{Stability of a Peer-to-Peer Communication System},''
  \emph{IEEE Transactions on Information Theory}, vol.~58, no.~7, 2012.

\end{thebibliography}

\end{document}